\documentclass[a4paper,onecolumn,superscriptaddress,11pt]{quantumarticle}
\pdfoutput=1
\usepackage{amsmath,amsthm,amsfonts}
\usepackage[english]{babel}
\usepackage{booktabs}
\usepackage{braket}
\usepackage{etoolbox}
\usepackage[T1]{fontenc}
\usepackage[hidelinks]{hyperref}
\usepackage[utf8]{inputenc}
\usepackage{lipsum}
\usepackage[numbers,sort&compress]{natbib}
\usepackage[usenames,dvipsnames]{xcolor}

\usepackage{tikz}
\usetikzlibrary{calc}

\usepackage{soul}

\usepackage{enumitem}
\usepackage{float} 
\usepackage{etoolbox}
\usepackage[babel=true,tracking=true]{microtype}      
\usepackage{ulem}

\author{Stuart Hadfield $^{1,2,3,*}$, Zhihui Wang $^{1,2}$, Bryan O'Gorman $^{1,4,5}$, Eleanor G. Rieffel $^{1}$, Davide Venturelli $^{1,2}$ and Rupak Biswas $^{1}$}

\affiliation{\noindent
$^{1}$ \quad 
Quantum Artificial Intelligence Laboratory (QuAIL), NASA Ames Research Center, Moffett Field, CA\\%
$^{2}$ \quad USRA Research Institute for Advanced Computer Science (RIACS), Mountain View, CA 94043, USA\\
$^{3}$ \quad Department of Computer Science, Columbia University, New York, NY 10027, USA\\
$^{4}$ \quad Stinger Ghaffarian Technologies, Inc.,  Greenbelt, MD 20770, USA\\
$^{5}$ \quad Berkeley Quantum Information and Computation Center and Departments of Chemistry and Computer Science, University of California, Berkeley, CA 94720, USA \\
$^*$  \quad Correspondence: stuart.hadfield@nasa.gov\\}

\date{\today}
\title{From the Quantum Approximate \\ Optimization Algorithm to a \\ Quantum Alternating Operator Ansatz}

\newcommand{\reals}{{\mathbb{R}}}
\newcommand{\integers}{{\mathbb{Z}}}

\newcommand{\naturals}{{\mathbb{N}}}

\newcommand{\colorParity}{\mathrm{CP}}

\newcommand{\OPT}{\mathrm{OPT}}

\newcommand{\QAOA}{\mathrm{QAOA}} % class of circuits of this form, as well as algorithm name
\newcommand{\QAOAcirc}{Q} % For unitary from a full QAOA circuit, the product of 2p ops.
\newcommand{\domain}{F}
\newcommand{\domainQ}{\mathcal{F}}
\newcommand{\objFunc}{f}
\newcommand{\objHam}{H_{\objFunc}}

\newcommand{\phaseFunc}{g}
\newcommand{\phaseHam}{H_{\mathrm{P}}}
\newcommand{\mixHam}{H_{\mathrm{M}}}
\newcommand{\mixUnitary}{U_{\mathrm{M}}}
\newcommand{\stdDriver}{H_B}
\newcommand{\initial}{s}

\newcommand{\hamDist}{\mathrm{Ham}}

\newcommand\phaseUnitary[1][]{
  \ifstrempty{#1}{
    U_{\mathrm{P}}
  }{
    U_{\mathrm{P},#1}
  }
}

\newcommand\Ham[2][]{
  \ifstrempty{#1}{
    H_{#2}
  }{
    H_{#1, #2}
  }
}

\newcommand\unitary[2][]{
  \ifstrempty{#1}{
    U_{#2}
  }{
    U_{#1, #2}
  }
}

\newcommand{\rnv}[1][r]{#1\text{-}\mathrm{NV}}
\newcommand{\ring}{\mathrm{ring}}
\newcommand{\fc}{\mathrm{FC}}

\newcommand{\timeEv}[1]{e^{-i #1}}
\newcommand{\comp}{(\mathrm{comp})}
\newcommand{\enc}{(\mathrm{enc})}
\newcommand{\parity}{\mathrm{parity}}
\newcommand{\even}{\mathrm{even}}
\newcommand{\odd}{\mathrm{odd}}
\newcommand{\last}{\mathrm{last}}
\newcommand{\identity}{I}
\newcommand{\quditSwap}{\mathrm{XY}}
\newcommand{\binary}{\mathrm{binary}}
\newcommand{\controlX}{\mathrm{CX}}
\newcommand{\control}{\Lambda}

\newcommand{\controlFunc}{\chi}
\newcommand{\proj}{H_{\controlFunc}}

\newcommand{\simMixer}[1]{\mathrm{sim}\text{-}#1}
\newcommand{\seqMixer}[1]{\mathcal{P}\text{-}#1}
\newcommand{\simCX}{\simMixer{\controlX}}
\newcommand{\seqCX}{\seqMixer{\controlX}}

\newcommand{\add}{\mathrm{ADD}}

\newcommand{\nullSwap}{\mathrm{NS}}
\newcommand{\simNullSwap}{\simMixer{\nullSwap}}
\newcommand{\seqNullSwap}{\seqMixer{\nullSwap}}
\newcommand{\controlSwap}{\mathrm{CS}}
\newcommand{\simControlSwap}{\simMixer{\nullSwap}}
\newcommand{\seqControlSwap}{\seqMixer{\nullSwap}}
\newcommand{\permSwap}{\mathrm{PS}}
\newcommand{\simPermSwap}{\simMixer{\permSwap}}

\newcommand{\timeSwap}{\mathrm{TS}}
\newcommand{\timeColor}{\mathrm{TC}}

\newcommand{\NONE}{\mathsf{NONE}}

\newcommand{\neighborFunc}{\mathrm{nbhd}}
\newcommand{\quditDim}{d}

\newcommand{\numVars}{n}
\newcommand{\numConstraints}{m}

\newcommand{\permElmt}{\sigma}

\newcommand{\order}{\boldsymbol{\iota}}
\newcommand{\orderElmt}{\iota}

\newcommand{\graphComp}[1]{\overline{#1}}

\newcommand{\create}{S^{+}}
\newcommand{\annihilate}{S^{-}}

\newcommand{\numColors}{\kappa}
\newcommand{\degree}{D}
\newcommand{\distance}{d}

\newcommand{\strtTime}{s}
\newcommand{\strtTimes}{\mathbf{s}}
\newcommand{\lastStrtTime}{h}
\newcommand{\releaseTime}{r}
\newcommand{\releaseTimes}{\mathbf{r}}
\newcommand{\window}{W}
\newcommand{\deadline}{d}
\newcommand{\deadlines}{\mathbf{\deadline}}
\newcommand{\weight}{w}
\newcommand{\weights}{\mathbf{w}}
\newcommand{\procTime}{p}
\newcommand{\procTimes}{\mathbf{\procTime}}
\newcommand{\horizon}{h}

\newcommand{\NEQ}{\mathsf{NEQ}}
\newcommand{\EQ}{\mathsf{EQ}}
\newcommand{\NOR}{\mathsf{NOR}}
\newcommand{\OR}{\mathsf{OR}}

\newcommand\noNeighbors[2][]{
  \ifstrempty{#1}{
    \NOR(x_{\neighborFunc(#2)})
  }{
    \NOR(x_{\neighborFunc(#1), #2})
  }
}

\newcommand{\quditX}{{\breve{X}}}
\newcommand{\quditZ}{{\breve{Z}}}

\newcommand{\SWAP}{\mathrm{SWAP}}
\newcommand{\XYij}{X_iX_j+Y_iY_j}
\newcommand{\SWAPij}{\SWAP_{i,j}}

\newcommand{\x}{\mathbf{x}}
\newcommand{\y}{\mathbf{y}}

\newcommand{\e}{\epsilon}

\newcommand{\bqa}{\begin{eqnarray}}
\newcommand{\eqa}{\end{eqnarray}}
\newcommand{\ketbra}[2]{\ket{#1}\!\bra{#2}}

\newcommand{\tikzmark}[1]{\tikz[overlay,remember picture] \node (#1) {};}

\begin{document}
\maketitle

\begin{abstract}
The next few years will be exciting as prototype universal quantum processors
emerge, enabling the implementation of a wider variety of algorithms.
Of particular interest are quantum heuristics, which require experimentation
on quantum hardware for their evaluation and which have the potential
to significantly expand the breadth of applications for which quantum 
computers have an established advantage. 
A leading candidate is Farhi et al.'s quantum approximate optimization 
algorithm, which alternates between applying a cost function based
Hamiltonian and a mixing Hamiltonian. 
Here, we extend this framework
to allow alternation between more general families of operators.~The essence of this extension, the quantum alternating operator ansatz,
is the consideration of general parameterized
families of unitaries rather than only those corresponding to the
time evolution under a fixed local Hamiltonian for a time specified by
the parameter. 
This ansatz supports the representation of a larger,
and potentially more useful, set of states than the original formulation,
with potential long-term impact on a broad array of application areas. 
For cases that call for mixing only within a desired subspace, 
refocusing on unitaries rather than Hamiltonians enables
more efficiently implementable mixers 
than was possible in the original framework. 
Such mixers are particularly useful for optimization problems with
hard constraints that must always be satisfied, defining
a feasible subspace, and soft constraints whose violation we wish to
minimize. 
More efficient implementation enables earlier experimental 
exploration of an alternating operator approach, in~the spirit of the quantum
approximate optimization algorithm, to a wide variety of approximate
optimization, exact optimization, and sampling problems. 
In addition to
introducing the quantum alternating operator ansatz, we lay out design criteria
for mixing operators, detail~mappings for eight problems, 
and provide a compendium with brief descriptions of mappings for a diverse
array of problems.
\end{abstract}

\section{Introduction}

Today, challenging computational problems
arising in the practical world are frequently tackled by heuristic algorithms.
These algorithms are empirically shown to be effective, 
% by running them on characteristic sets of problems.
but %Please check meaning has been retained.
they have not been analytically proven to be the best
approach, or even to
% proven analytically to outperform 
outperform the best approach of the previous year.
Until recently, empirical investigation of quantum algorithms has been 
limited to tiny problems, given the typically exponential overhead of 
simulating quantum algorithms on classical processors. 
As prototype quantum hardware emerges which enables experimentation beyond
what is reachable by even the world's largest supercomputers, we 
come into a new era for quantum heuristic~algorithms. 

A key question is: ``What are good quantum heuristic algorithms to try?'' 
A leading candidate is Farhi et al.'s quantum approximate optimization
algorithm, %~\cite{Farhi2014}, 
a quantum gate-model meta-heuristic which alternates between applying 
unitaries drawn from two families, 
a cost function based unitary family
$\phaseUnitary(\gamma) = e^{-i\gamma\objHam}$ and
a family of mixing unitaries 
$\mixUnitary(\beta) = e^{-i\beta\stdDriver}$,
for some fixed cost function based Hamiltonian $\objHam$
and some fixed mixing Hamiltonian $\stdDriver$.
Here, we formally describe a \textit{quantum alternating operator ansatz} (QAOA), %Is the italic necessary?
extending the approach of Farhi et al.~\cite{Farhi2014} to allow alternation
between more general families of operators.
This ansatz supports the representation of a much more varied, %larger
and potentially more useful, set of states than the original formulation.  
Our extension is particularly 
useful for situations in which the feasible subspace is smaller than the
full space, such as when the optimization is over solutions that must 
satisfy hard constraints.
Intuitively, mixing operators that restrict the search to the feasible 
subspace should result in better-performing algorithms.
Our expansion includes families of mixing operators $\mixUnitary(\beta)$ that %are not 
cannot be expressed, as a family, as 
%$\mixUnitary(\beta) = e^{-i\gamma\stdDriver}$
$e^{-i\beta\stdDriver}$ 
for a fixed mixing Hamiltonian $\stdDriver$.
As we shall see, expanding the design space of families of one-parameter mixing
operators allowed enables the ansatz to support
more efficiently implementable mixers 
than was possible in the original framework. 
More efficient implementation enables earlier experimental
exploration of an~alternating operator approach, in the spirit of the quantum
approximate optimization algorithm, \mbox{to a wide} variety of approximate
optimization, exact optimization, and sampling problems.

% The essence of this extension is the consideration of general parameterized
% families of unitaries  rather than only those corresponding to the
% time-evolution of a fixed local Hamiltonian for a time specified by
% the parameter. 
% For cases that call for mixing only within a desired subspace, 
% refocusing on unitaries rather than Hamiltonians enables
% more efficiently implementable mixers 
% than was possible in the original framework. 
% Such mixers are particularly useful for optimization problems with
% hard constraints that must always be satisfied, defining
% a feasible subspace, and soft constraints whose violation we wish to
% minimize. 

We carefully construct a framework for this ansatz, laying out design criteria
for families of mixing operators. We detail QAOA mappings of several
optimization problems, and provide a~compendium of
mappings for a diverse array of problems. 
These mapping range from the
relatively simple, which could be implemented on near-term devices,
to complex mappings with significant resource requirements. This paper
is meant as a starting point for a research program.
Improved mappings and compilations, especially for some of the more complex
problems, are a promising area for future work. 
Architectural codesign could be used to enable experimentation of
QAOA approaches to some problems earlier than would be possible otherwise.

We reworked the original acronym so that ``QAOA'' continues to apply to both
prior work and future work to be done in this more general
framework.  More generally, the reworked acronym refers to a set of states
representable in a certain form, and so can be used without confusion in contexts other than approximate optimization, e.g., exact optimization and sampling.
(Incidentally, this reworking also removes the redundancy from the now commonly-used phrase ``QAOA algorithm''.)

After describing the framework for the ansatz, we show explicit 
mappings to quantum circuits 
and resource estimates 
for a 
%number 
%variety 
%of 
diverse set of 
%quite different 
problems, designing phase separation and mixing operators appropriate for each
problem. 
%By design, we explore a variety of
The resulting mappings and techniques employed are nontrivial and serve as prototypes for a much wider variety of problems and applications; we include brief summaries of mappings for many other well-known optimization problems as an appendix. 

%Here, 
%After describing the framework for the ansatz, we map a number of
%problems, designing phase-separation and mixing operators appropriate for each
%problem. 
We comment here on the relation between these mappings and
those for non-gate-model quantum computing, such as quantum 
annealing (QA). 
Because current quantum annealers have a fixed driver (the mixing Hamiltonian
in the QA setting), all problem dependence must
be captured in the cost Hamiltonian on such devices.
The general strategy is to
incorporate the hard constraints as penalty terms in the cost function
and then convert the cost function to a cost Hamiltonian~\cite{biswas2017nasa,rieffel2015case,LucasIsingNP,hadfield2018representation}. 
However, this approach means that the algorithm must search a much larger
space than if the evolution were confined to feasible configurations, making
the search less efficient than if it were possible to constrain the evolution. 
This issue, and other drawbacks, led 
Hen and Spedalieri~\cite{Hen2016quantum} and Hen and Sarandy~\cite{Hen2016driver}
to suggest a different approach for adiabatic quantum optimization (AQO),
in which the standard driver is replaced by an alternative driver that
confines the evolution to the feasible subspace. Their approach 
resembles a restricted class, H-QAOA (defined below), of
QAOA algorithms. 
While some of our mapppings, e.g., H-QAOA mappings of graph coloring, graph
partitioning, and not-all-equal 3-SAT, %Please define SAT.
are close to those in References~\cite{Hen2016quantum,Hen2016driver}, other mappings we describe, 
including for these problems, are quite different and take advantage 
of the more general families of mixers supported by this ansatz. 
% \redtext{[SH: phrasing of this sentence could be improved] }
Indeed, while QAOA mappings are different from quantum annealing mappings, 
with most of the design effort going into the mixing operator rather than
the cost function based phase separator, QAOA algorithms, like QA and AQO, 
but unlike most other quantum algorithms, are relatively easy for 
people familiar with classical computer science but not quantum computing
to design, as we illustrate in this paper.

%In Section~\ref{sec:background}.
\textls[-10]{In the following section, we %summarize 
overview the relevant background results. 
In Section~\ref{sec:ansatz}, we %carefully 
construct a framework for this ansatz,
laying out design criteria for families of mixing operators. 
Sections~\ref{sec:strings} and~\ref{sec:permutations} detail QAOA
mappings and compilations for several optimization problems, 
illustrating design techniques and a variety of mixers. 
Section~\ref{sec:strings} considers four problems
in which the configuration space of interest is strings: MaxIndependentSet,
Max-$k$-ColorableSubgraph, Max-$k$-ColorableInducedSubgraph, and 
MinGraphColoring. Section~\ref{sec:permutations} considers four problems
in which the configuration space of interest is orderings (or permutations):
%MinTravelingSalesperson, 
%TravelingSalesperson, 
The traveling salesperson, 
and three versions of single
machine scheduling (SMS), also called job sequencing.  
Section~\ref{sec:conclusions} concludes with a discussion
of many open questions and directions for future work. We 
provide a compendium of mappings and compilations
for a diverse array of problems in Appendix~\ref{sec:compendium} and provide resource estimates for their implementation. For the benefit of the reader, we include a glossary of important terminology used in the paper, and a review of some useful elementary quantum operations as Appendices~\ref{sec:glossary} and~\ref{sec:elemOps}, respectively. } 

\section{Background}   \label{sec:background}

Over the last few decades, researchers have discovered several stunning
instances of quantum algorithms that provably outperform the best
existing classical algorithms and, in some cases, the best possible classical
algorithm~\cite{RPbook}.
For most problems, however, it is currently unknown whether quantum computing
can provide an advantage, and if so, how to design quantum algorithms that
realize such advantages. Today, challenging computational problems
arising in the practical world are frequently tackled by heuristic algorithms,
which by definition have not been analytically proven to be the best
approach, or even proven analytically to outperform the best approach 
of the previous year.

For several years now, special-purpose quantum hardware has been used to explore one quantum heuristic algorithm, quantum annealing. 
Emerging gate-model processors will enable investigation of a much broader
array of quantum heuristics beyond quantum annealing.
Within the last year, IBM has made available publicly through the cloud a
gate-model chip with 5 and 16 superconducting qubits~\cite{IBM_Q}, and it recently announced an upgrade 
to a $20$-qubit chip. 
Likewise, Google~\cite{Boixo16} and Rigetti Computing~\cite{Sete16} 
anticipate to provide processors with 40--100 superconducting qubits 
within a year~\cite{Google2017nature}.
Many academic groups, including at TU Delft and at UC Berkeley, have made similar
efforts.  
In addition to superconducting architectures, ion~\cite{debnath2016demonstration} and
neutral atom based~\cite{saffman2016quantum} devices are also reaching the
scale at which intermediate-size experiments would be 
feasible~\cite{Google2017nature}.
Gate-model quantum computing expands the empirical evaluation of
quantum heuristics applications beyond optimization of classical functions, 
as well as enabling a broader array of approaches to optimization~\cite{zahedinejad2017combinatorial}.

While limited exploration of quantum heuristics beyond 
quantum annealing
has been possible through 
small-scale classical simulation,
the exponential overhead in such
simulations has limited their usefulness. 
The next decade will see a blossoming of quantum heuristics as a
broader and more flexible array of quantum computational hardware becomes
available.
The immediate question is: What experiments should we prioritize
that will give us insight into quantum heuristics?  One leading candidate is
the quantum approximate optimization algorithm (QAOA), for which a number of
tantalizing related results have been 
obtained~\cite{Farhi2014b,Farhi2016,Shabani16,Jiang17,Wecker2016training,Wang17,Venturelli17} 
since Farhi  et al.'s initial paper~\cite{Farhi2014}.
In QAOA, a phase-separation operator, usually the problem Hamiltonian that encodes the cost function of the optimization
problem, and a mixing Hamiltonian are applied in alternation.
The class $\QAOA_p$ consists of level-$p$ QAOA circuits,
in which there are $p$ iterations of applying a classical Hamiltonian (derived
from the cost function) and a mixing Hamiltonian.
The $2p$ parameters of the algorithm specify the durations for which each of these two Hamiltonians are applied.

\textls[-20]{Prior work suggests the power and flexibility of QAOA circuits.
Farhi et al.~\cite{Farhi2014b} exhibited a~$\QAOA_1$ algorithm that beat the
existing best approximation bound for efficient classical algorithms for
the problem E3Lin2, only to inspire a better classical
algorithm~\cite{barak2015beating}.
Jiang et al.~\cite{Jiang17} demonstrated that the class of QAOA circuits
is powerful enough to obtain the $\Theta(\sqrt{2^n})$ query complexity
on Grover's problem and also provided the
first algorithm within the QAOA framework
to show a~quantum advantage for a finite number of iterations greater than two.
Farhi and Harrow~\cite{Farhi2016} proved that,
under reasonable complexity assumptions, the output distribution of
even $\QAOA_1$ circuits cannot be efficiently sampled classically.
Yang et al.~\cite{Shabani16} proved that for evolution under a Hamiltonian that
is the weighted sum of Hamiltonian terms, with the weights allowed to
vary in time, the optimal control is (essentially always) bang-bang, i.e., 
constant magnitude, of either the maximum or minimum allowed weight,
for each of the terms in the Hamiltonian at any given time.
Their work implies that QAOA circuits with the right parameters
are optimal among Hamiltonians of the form 
$H(s) = \big(1-f(s)\big)H_B + f(s) H_C$,
where $f(s)$ is a real function in the range $[0,1]$.
It remains an open question whether QAOA provides a quantum
advantage over classical algorithms for approximate optimization, 
either in terms of the quality of approximate solution returned, or the speed of achieving such an approximation. }

This paper generalizes our initial results on quantum approximate optimization for problems with hard and soft constraints \cite{hadfield2017quantum}. 
Since the preprint of these two papers,
%We comment on some recent results related to QAOA that have appeared after this work. 
the approach we proposed to deal with constrained optimization problems
has been applied to 
a benchmarking study on graph-coloring problems (in preparation) %(forthcoming %(to appear) %~\cite{Wang19XY}
and a protein folding optimization problem \cite{fingerhuth2018quantum}. 
QAOA also provides a viable platform to study quantum circuit compilation to realistic architectures~\cite{Farhi2017,Venturelli17,lechner2018quantum}. 
Applications and extensions of QAOA beyond optimization include 
state preparation \cite{ho2018efficient} and machine learning \cite{verdon2017quantum, otterbach2017unsupervised}. 
A different approach in the setting of quantum walks to QAOA for constrained problems 
has recently been proposed \cite{marsh2019quantum}, 
and very recently, Lloyd showed that the QAOA framework with a carefully constructed cost Hamiltonian can be made universal for quantum computation \cite{lloyd2018quantum}.

%Should this subsection be 2.1?  YES

%\subsection{QAOA framework}
\subsection{The Original Quantum Approximate Optimization Algorithm}

We now give an overview of the original quantum approximation optimization algorithm 
proposed %by Farhi, Goldstone and Gutmann 
in Reference~\cite{Farhi2014}. % for constraint satisfaction problems over $n$-bit strings. 
%

%for %an unconstrained optimization problem 
%a constraint satisfaction problem over $n$-bit strings, 
%represented by the computational basis states of $n$ qubits. 
%The application to more general optimization problems was discussed only briefly in \cite{Farhi2014}; we elaborate on this in the next section.
%, where, inspired by the original algorithm, we give a general framework,
%Our more general framework of this paper  %, the quantum alternating operator ansatz, 
%may be similarly applied to optimization problems with constraints by substituting the parameterized state below with a quantum alternating operator ansatz state.  
%and more genera applications 

Consider an unconstrained optimization problem on $n$-bit strings
we seek to approximate. %for example the MaxCut or Max-$2$-SAT problems. 
Given a problem instance, %the algorithm 
the algorithm %quantum approximate optimization algorithm 
is specified by %one initial state,
two Hamiltonians $\phaseHam$ and $\mixHam$, and~$2p$ real parameters  $\gamma_1,\dots,\gamma_p,\beta_1,\dots,\beta_p$.
The main details are the following:
\begin{itemize}
\item The \emph{phase Hamiltonian} $\phaseHam$ encodes the cost function~$f$ to be optimized, 
i.e.,  acts diagonally on $n$-qubit computational basis states as:
$$\phaseHam \ket{\mathbf y}= f(\mathbf y) \ket{\mathbf y}.$$ 
%\vskip 2pc
\item The \emph{mixing Hamiltonian} $\mixHam$ is the transverse field Hamiltonian:
$$ \mixHam = \sum_{j=1}^n X_j,$$
where $X_j$ is the Pauli $X$ operator acting on the $j$th qubit. (The Pauli $X$ operator acts as a bit flip, i.e., $X\ket{0}=\ket{1}$ and $X\ket{1}=\ket{0}$.) 
%\vskip 2pc
\item The initial state is selected to be the equal superposition state of all possible solutions:
$$ \ket{s} = \frac{1}{\sqrt{2^n}} \sum_x \ket{x}\;,$$
which is also the ground-state of $-H_M$ and is used similarly in AQO \cite{Farhi2014}. 
%We discard this restriction and allow for much more general initial states in our design criteria of Section \ref{sec:designCrit} below. 
%
\item %The algorithm consists of alternating 
A \emph{parameterized quantum state} is created by alternately applying Hamiltonians $\phaseHam$ and $\mixHam$ for~$p$ rounds,
where the duration in round $j$ is specified by the parameters $\gamma_j$ and $\beta_j$, respectively:
%A parameterized quantum state is created by applying alternating evolution under Hamiltonian $\phaseHam$ for a fixed time $\gamma_j$, followed by Hamiltonian $\mixHam$ for time $\beta_j$, 
%for $j=1,\dots,p$, 
%, followed by $\phaseHam$ for time $\gamma_2$, and so on. 
%This creates the state
$$\ket{\boldsymbol \beta, \boldsymbol \gamma}
=
e^{-i \beta_p \mixHam}e^{-i \gamma_p \phaseHam} \dots e^{-i \beta_2 \mixHam}e^{-i \gamma_2 \phaseHam}e^{-i \beta_1 \mixHam}e^{-i \gamma_1 \phaseHam}
\ket{\initial}.$$
\item %Typically, for optimization applications, after the $2p$ alternations 
A computational basis measurement is performed on the state, which returns a candidate solution~$\mathbf y$ with probability 
$ |\bra{\mathbf y}\ket{\boldsymbol \beta, \boldsymbol \gamma}|^2.$ Repeating the above state preparation and measurement, 
the expected value of the cost function over the returned solution samples 
%with respect to the 
is given by:
$$ \langle f\rangle = \bra{\boldsymbol \beta, \boldsymbol \gamma} \phaseHam \ket{\boldsymbol \beta, \boldsymbol \gamma},$$   
which can be statistically estimated from the samples produced. 
(For a constraint satisfaction problem with $m$ constraints, fom Chebyshev's inequality it follows that an outcome achieving at least $ \langle f \rangle - 1$ will be obtained with probability at least $1 - 1/m$ after $O(m^2)$ repetitions.)

\item The above %procedure 
steps may then be repeated altogether, with updated sets of time parameters, as part of a classical optimization loop (such as gradient descent or other approaches) used to %find the best possible parameters
optimize the algorithm parameters with respect to an objective such as $\langle f\rangle$.

\item The best problem solution found overall is returned. 
\end{itemize}

A key to success for the algorithm is the selection or discovery of good values for the %$2p$ real 
parameters $\gamma_1,\dots,\gamma_p,\beta_1, \dots, \beta_p$, which result in good approximate solutions. 
In some cases, where the analysis is tractable, such angles may be found
analytically~\cite{Jiang17,Wang17}. 
Parameter setting strategies for QAOA and for the general class of variational quantum 
algorithms remains an active area of research \cite{guerreschi2017practical,mcclean2018barren}.

% In the remainder of this paper, we show how generalizing QAOA from the Hamiltonian to the unitary operator picture enables easy adaptation to large classes of constrained optimization problems, in particular we focus on constructions with low resource requirements. Better understanding the performance of these constructions for approximate optimization, relative to the performance of the best classical algorithms, remains an important direction of future research. 
% Indeed, rich theories of the classical complexity of approximation are known; see  \cite{Ausiello2012complexity} for a thorough introduction. We list the known complexity and hardness of approximation results for a variety of prototypical optimization problems in Appendix~\ref{sec:compendium}.}

%Finally, we remark on the advantages of emerging quantum gate-model devices over alternative hardware designs such as quantum annealers, both near-term and more generally. 
 %}
 
We now turn to our generalized QAOA framework, which is the main subject 
of the paper.

\section{The Quantum Alternating Operator Ansatz (QAOA)}\label{sec:ansatz}

Here, we formally describe the quantum alternating operator ansatz,  
extending the approach of Farhi et al.~\cite{Farhi2014}.
QAOA, in our sense, encompasses a more general class of quantum states that may be
algorithmically accessible and useful. 
We focus here on the application of QAOA to approximate optimization, though it
may also be used in exact optimization~\cite{Jiang17,Wecker2016training}
and sampling~\cite{Farhi2016}.

% This sentence almost verbatim from Papadimitriou's 'Combinatorial Optimization'
An instance of an \emph{optimization problem} is a pair $(\domain, \objFunc)$,
where $\domain$ is the \emph{domain} (set of feasible points) and 
$\objFunc: F \rightarrow \mathbb R$ 
is the \emph{objective function} to be optimized (minimized or maximized).  
Let $\domainQ$ be the Hilbert space of dimension $|\domain|$, whose standard
basis we take to be 
$\left\{\Ket{\mathbf{x}} : \mathbf{x} \in \domain\right\}$.
Generalizing Reference~\cite{Farhi2014}, a $\QAOA$ circuit is characterized by  
two parameterized families of operators on $\domainQ$: 
\begin{itemize}
\item A family of 
\emph{phase-separation operators} $\phaseUnitary(\gamma)$ that depends on the
\emph{objective function} $\objFunc$, and;
\item A family of \emph{mixing operators} $\mixUnitary(\beta)$ that depends on the domain and its structure, 
\end{itemize}
where $\beta$ and $\gamma$ are real parameters.
Specifically, a $\QAOA_p$ circuit consists of $p$ alternating applications of operators from these two families:
\begin{equation}
\QAOAcirc_p(\boldsymbol \beta, \boldsymbol \gamma)
=
\mixUnitary(\beta_p) \phaseUnitary(\gamma_p)
\cdots 
\mixUnitary(\beta_1) \phaseUnitary(\gamma_1).
\end{equation}

This quantum alternating operator ansatz (QAOA) consists of the states representable as the application of such a circuit to a suitably simple initial state $\ket{\initial}$:
\begin{equation}
\ket{\boldsymbol \beta, \boldsymbol \gamma}
=
\QAOAcirc_p(\boldsymbol \beta, \boldsymbol \gamma)
\ket{\initial}.
\end{equation}

We show an overall quantum circuit schematic for a QAOA mapping in Figure \ref{QAOAnsatz_overall} below.
For a given optimization problem, a {\it QAOA mapping\/} of a problem consists
of a family of phase-separation operators, a family of mixing operators, and a
starting state.
The circuits for the original quantum approximate optimization algorithm fit
within this paradigm, with unitaries of the form~$\timeEv{\gamma\phaseHam}$
and~$\timeEv{\beta\mixHam}$, with parameters~$\gamma$ and~$\beta$ indicating
the time for which a fixed Hamiltonian is applied. 

\begin{figure}[H]
\centering
\includegraphics[width=12cm]{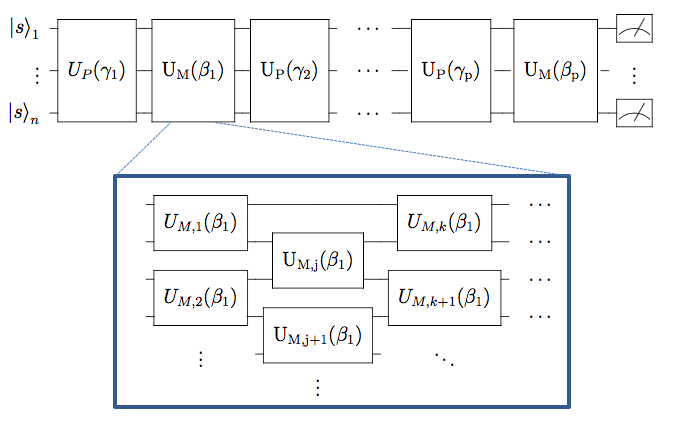}
%\label{fig:MaxIndepSetcircuit}
\caption{The quantum alternating operator ansatz (QAOA$_p$) quantum circuit schematic. 
Here, an encoding to qubits for a given problem domain is assumed. 
The box shows an example decomposition of a QAOA mixing operator family 
$U_M(\beta)$ into a sequence of partial mixers $U_{M,\alpha}(\beta)$. 
% In general, such mixing operators do 
In this ansatz, a one-parameter family of mixing operators does 
not in general correspond to time evolution under a
fixed mixing Hamiltonian $H_M$. The construction of this paper includes
different orderings of the partial mixers, resulting in a variety of
inequivalent mixing operators with different implementation costs. Though not
shown in the figure,  phase and mixing operators will often include ancilla
qubits to facilitate computation and simple compilation to one- and two-qubit
gates. The circuit shown indicates measurement at the end of the algorithm; in
general, a quantum alternating operator ansatz circuit may be instead embedded as part
of a larger quantum algorithm. Likewise, different initial states may be used
which may be constructed by design or the output of another quantum subroutine.
}   \label{QAOAnsatz_overall}
\end{figure}

The domain will usually be expressed as the feasible subset of a larger
\emph{configuration space}, specified by a set of problem \emph{constraints}.  
For implementation on expected near-term quantum hardware, each configuration
space will need to be \emph{encoded} into a subspace of a Hilbert space
of a multiqubit system, with the domain corresponding to a  
\emph{feasible subspace} 
of the configuration space.
For each domain, there are many possible mixing operators.
As we will see, 
using more general one-parameter families of unitaries enables
more efficiently implementable mixers that preserve the feasible subspace.
\mbox{Given a domain,} an encoding of its configuration space, a phase separator, and a mixer,
there are a~variety of \emph{compilations} of the phase separator and mixer to circuits that act on qubits.

%\vskip 1pc 

%\vskip 1pc 

For any function $f$, not just an objective (cost) function, we define
$\objHam$ to be the quantum Hamiltonian that acts as $f$ on basis states as: 
\begin{equation}     \label{eq:functionHam}
    \objHam \ket{\mathbf{x}} = \objFunc(\mathbf{x}) \ket{\mathbf{x}}.
\end{equation}

In prior work, the domain $\domain$ was
the set of all $n$-bit strings, 
$\phaseUnitary(\gamma) = e^{-i\gamma\objHam}$, and
$\mixUnitary(\beta) = e^{-i\beta\stdDriver}$. 
Furthermore, with just one exception, the mixing Hamiltonian was
$\stdDriver = \sum_{j=1}^n X_j$.
We used the notation $X_j$, $Y_j$, $Z_j$ to indicate the Pauli matrices $X$, $Y$, and $Z$ acting on the $j$th qubit.
The corresponding parameterized unitaries are denoted by $X_j(\theta) = \timeEv{\theta X_j}$ and similarly for $Y_j$ and $Z_j$.
The one exception is Section~VIII of~Reference \cite{Farhi2014}, which discusses a variant for
the maximum independent set problem, in which~$\domain$ is the set of
bitstrings 
corresponding to the independent sets of a graph, the phase separator depends 
on the cost function as above, and the mixing operator is 
$\mixUnitary(\beta) = e^{-i\beta H_B}$, where
$H_B$ is such that:
\begin{equation}
\braket{\x  | H_B | \y} = 
\begin{cases}
1, & \x, \y \in \domain \text{ and } \hamDist(\x,\y) = 1,\\
0, & \text{otherwise},
\end{cases}
\end{equation}
which connects feasible 
qubit computational basis states with unit Hamming distance $(\hamDist)$. 
%, i.e., 
%states which differ by a single bitflip. 
%independent sets which differ by a single vertex. 
Section~VIII of~Reference \cite{Farhi2014} does not discuss the implementability
of $\mixUnitary(\beta)$. 
A closely related generalization of QAOA for problems with hard constraints based on quantum walks has recently been proposed \cite{marsh2019quantum}. However, row-computable feasibility oracles are required to enable mixing between feasible states, which are likely to be more expensive to implement in practice than the approach of this paper.

We extended and formalized the approach of Section~VIII of Reference~\cite{Farhi2014}
with an eye to implementability, both in the short and long term.
We also built on a theory developed for adiabatic quantum optimization (AQO) by 
Hen and Spedalieri~\cite{Hen2016quantum} and 
Hen and Sarandy~\cite{Hen2016driver}, 
though the gate-model setting of QAOA leads to different implementation considerations than those for AQO.\@
For example, Hen et al.~identified driver Hamiltonians of the form
$\mixHam = \sum_{j,k} H_{j,k}$, where $H_{j,k} = X_j X_k + Y_j Y_k$, 
as useful in the AQO setting for a variety of optimization problems with hard
and soft constraints; 
such mixers restrict the mixing to the feasible
subspace defined by the hard constraints. 
Analogously, the unitary 
$\mixUnitary = e^{-i\beta\mixHam}$
meets our criteria, discussed in Section~\ref{sec:designCrit}, 
for good mixing for a variety of optimization problems, including those
considered in References~\cite{Hen2016quantum,Hen2016driver}.
Since $H_{j,k}$ and $H_{i,l}$ do not commute when
$|\{j,k\}\cap\{i,l\}| = 1$, compiling $\mixUnitary$ to two-qubit gates
is nontrivial. 
One could Trotterize, but it may be more efficient and
just as effective to use an alternative mixing operator, such as
$\mixUnitary = e^{-i\beta H_{S_r}}\cdots e^{-i\beta H_{S_2}}e^{-i\beta H_{S_1}}$, where 
the pairs of qubits have been partitioned into $r$ subsets ${\{S_i\}}_i$ containing only 
disjoint pairs, motivating in part our more general ansatz.

We define as ``Hamiltonian-based QAOA'' (H-QAOA) the class of QAOA circuits in
which both the phase separator family  $\phaseUnitary(\gamma) =  e^{-i \gamma
\phaseHam}$ and the mixing operator family $\mixUnitary(\beta) = e^{-i \beta
\mixHam}$ correspond to time evolution under some Hamiltonians $\phaseHam$ and
$\mixHam$, respectively.  (In the example mappings to follow, we consider only phase separators 
$\phaseUnitary (\gamma) =\sum_{\x} e^{-i\gamma \phaseFunc(\x)} \ketbra{\x}{\x}$ that correspond to classical functions and thus also correspond to time evolution under some (potentially nonlocal) Hamiltonians, though more general types of phase separators may be considered).
We further define ``local Hamiltonian-based  QAOA'' (LH-QAOA) as the subclass of H-QAOA in which the Hamiltonian~$\mixHam$ is a sum of (polynomially many) local terms.

Before discussing design criteria, we briefly mention that there are
obvious generalizations in which $\phaseUnitary$ and $\mixUnitary$ are
taken from families parameterized by more than a single parameter.  
For example, in Reference~\cite{Farhi2017}, a different parameter for every term in the 
Hamiltonian is considered. 
In this paper, we only consider the case of one-dimensional families,
given that it is a sufficiently rich area of study, 
with the task of finding good parameters $\gamma_1, \ldots, \gamma_p$,
and $\beta_1, \ldots, \beta_p$ already challenging enough due to the 
curse of dimensionality~\cite{Wang17}.
A larger parameter space may support more effective circuits but 
increases the difficulty of finding such circuits by opening up the
design space and making the parameter setting more difficult.

We remark that the quantum gate-model setting offers several advantages over Hamiltonian-based algorithms such as AQO and quantum annealing. 
Higher order ($k$-local) interactions may be compiled down to two-local gates, and
compilations using SWAP gates \cite{Venturelli17,Booth18} %Please define SWAP.
% and other techniques \cite{}
enable the implementation of
quantum operations between qubits that are non-neighboring 
in the physical hardware; % graph;
indeed, locality and connectivity are both well-known bottlenecks for physical quantum annealing devices.
In the longer~term, once mature quantum hardware has been built,
quantum error correction can be applied to robustly implement QAOA.

%Should this be 3.1? YES

\subsection{Design Criteria}\label{sec:designCrit}

Here, we briefly specify design criteria for the three components of
a QAOA mapping of a problem. 
We expect that as exploration of QAOA proceeds, these design criteria
will be strengthened and will depend on the context in which
the ansatz is used. For example, when the aim is a polynomial-time
quantum circuit, the components should have more stringent bounds
on their complexity; without such bounds, the ansatz is not useful
as a model for a strict subset of states producible via
polynomially-sized quantum circuits. 
On the other hand, when the computation is expected to grow exponentially, a~simple polynomial bound on the depth of these operators might be reasonable. 
One example might be for exact optimization of the problems considered here; for these problems, the worst case algorithmic complexity is exponential, but it is worth exploring whether QAOA might outperform classical heuristics in expanding the tractable range for some problems.
%\newline
\vskip 1pc
%\paragraph{Initial state.} 
\noindent \textbf{Initial state.} \quad We require that the initial state $\ket{\initial}$ be
\emph{trivial} to implement, by which we mean that it can be created by
a constant-depth 
(in the size of the problem) quantum circuit from the
$\ket{0\dots 0}$ state.
Here, we often take as our initial state a single feasible solution, 
usually implementable by a depth-$1$ circuit consisting of single-qubit
bit-flip operations $X$. Because in such a case the initial phase operator
only applies a global phase, we may want to consider the algorithm as
starting with a single-mixing operator $\mixUnitary(\beta_0)$ to the initial
state as a first step. In the quantum approximate optimization algorithm,
the standard starting state $\ket{+\cdots+}$ 
is obtained by a depth-$1$ circuit that applies 
a Hadamard $H$ gate to each of the qubits in the $\ket{0\dots 0}$ state. 

This criterion could be relaxed to logarithmic depth if needed.
It should not be relaxed too much:
Relaxing the criterion to polynomial depth would 
obviate the usefulness of the
ansatz as a model for a strict subset of states producible via
polynomially-sized quantum circuits.  
Algorithms with more complex initial states should be considered
hybrid algorithms, with an initialization part and a QAOA part. 
Such algorithms are of interest in cases 
when one expects the computation
to grow exponentially, such as is the case for exact optimization
for many of the problems here, but might still outperform classical
heuristics in expanding the tractable range.
\vskip 1pc  %\newline
%\paragraph{Phase-separation unitaries.} 
\noindent \textbf{Phase-separation unitaries.}   \quad  We require %that each member of
 the family of 
phase-separation operators to 
be diagonal in the computational basis. In almost all cases, we take%the
~$\phaseUnitary(\gamma) = e^{-i \gamma \objHam}$, where $f$ is the %problem 
objective function. 
\vskip 1pc %\newline
%\paragraph{Mixing unitaries (or ``mixers'').} 
\noindent \textbf{Mixing unitaries (or ``mixers'').} \quad We require the family of 
mixing operators $\mixUnitary(\beta)$ to:
\begin{itemize}
\item Preserve the feasible subspace: For all values of the parameter $\beta$, the resulting unitary
takes feasible states to feasible states, and;
\item Provide transitions between all pairs of states corresponding to feasible 
points. More concretely, for any pair of feasible 
computational-basis 
states $\x, \y \in \domain$,
there is some parameter value~$\beta^*$ 
and some positive integer~$r$
such that the corresponding mixer
connects those two states: 
\mbox{$\left|\Braket{\x | \mixUnitary^r(\beta^*) | \y}\right| > 0$.}
\end{itemize}

In some cases, we may want to relax some of these criteria. For example,
if a QAOA circuit is being used as a subroutine within a hybrid 
quantum-classical algorithm, or in a broader quantum algorithm, 
we may use starting states informed by previous runs and thus allow mixing 
operators that mix less. 

\textls[-20]{This framework can be used in many different contexts. Depending on 
the context, different measures of success are appropriate. 
As indicated by the name, the original motivation for \mbox{Farhi  et al.'s}~work
was to develop a quantum approximation algorithm, one for which rigorous 
bounds on the approximation ratio can be proven \cite{Farhi2014,Farhi2014b}.
The same style of algorithm was then applied to exact optimization 
\cite{Wecker2016training} and sampling~\cite{Farhi2016}, 
which have different measures of success. 
In certain cases, rigorous performance guarantees can be provided in
these contexts, e.g.,~for the Grover problem in Reference~\cite{Jiang17}.
Alternatively, it can be applied as a heuristic approach for any of exact
optimization, approximate optimization, or sampling. In these cases, the 
measure of success is not in terms of rigorous analytical bounds, but
rather empirical typical time-to-solution or approximation ratio or
sample quality within a given time. 
Our approach facilitates low-resource constructions that support
empirical evaluation of QAOA as a heuristic for a variety of 
combinatorial optimization problems, and is agnostic as to which
success criterion is being used for~evaluation.}

\section{QAOA Mappings: Strings}\label{sec:strings}

\textls[-20]{This section describes mappings to QAOA for four problems in
which the underlying configuration space is strings 
with letters taken from some alphabet. 
We~introduce some basic families of mixers and~discuss compilations thereof, 
illustrating their use with MaxColorableSubgraph as an example.
We~then build on these basic mixers to design families of more complicated
mixers, such as controlled versions of these mixers, and illustrate their use
in mappings and circuits for the problems MaxIndependentSet,
MaxColorableInducedSubgraph, and MinGraphColoring as examples.
The mixers we develop in this section, and close variants, are applicable
to a wide variety of problems, as we see in Appendix~\ref{sec:compendium}.}

\subsection{Example: Max-$\numColors$-ColorableSubgraph}\label{sec:maxColSubGraph}
%
%\paragraph{Problem.}
%\vskip 1pc 
\noindent \textbf{Problem.} \quad
Given a graph $G = (V,E)$ with $n$ vertices and $m$ edges, and $\numColors$
colors, maximize the size (number of edges) of a properly vertex-$\numColors$-colorable subgraph.

The domain $\domain$ is the set of colorings $\x$ of $G$, an assignment 
of a color to each vertex.
(Note that here and throughout, the term ``colorings'' includes \emph{improper} colorings.)
The~domain $\domain$ can be represented as the set of length $n$ strings
over an alphabet of $\numColors$ characters, 
$\mathbf{x} = x_1x_2\dots x_n$, where $x_i\in [\numColors]$. 
The~objective function 
$\objFunc: {[\numColors]}^n \to {\naturals}$ 
counts the number of properly colored edges in a coloring: 
\begin{equation}\label{eq:maxColorableSubgraph-objFunc}
\objFunc(\x) = \sum_{\{u,v\} \in E} \NEQ(x_u,x_v).
\end{equation}

We then built up machinery to define mixing operators for a QAOA approach
to this problem.
Since some mixing operators are more naturally expressed in one 
encoding rather than another, we~found
it useful to describe different mixing operators in different encodings, 
though we emphasize that doing so is merely for convenience; all mixing 
operators are encoding-independent, so the descriptions may be translated 
from one encoding to another. 
The domain $\domain$ is naturally expressed as strings of $\quditDim$-dits,
a~$\quditDim$-valued generalization of bits. 
For the present problem, $\quditDim = \numColors$.
In addition to discussing colorings as strings of dits, we used a
``one-hot'' encoding into $\numVars\numColors$ bits, with $x_{i,c}$ 
indicating whether or not vertex $i$ is assigned color $c$.

Figure \ref{fig:n4graphMap} below shows a mapping to qubits in the one-hot encoding for Max-$\numColors$-ColorableSubgraph. We explain this and other possible mappings generally over the remainder of the section. 

%\vskip 2pc 
\begin{figure}[H]
\centering
\includegraphics[width=7cm]{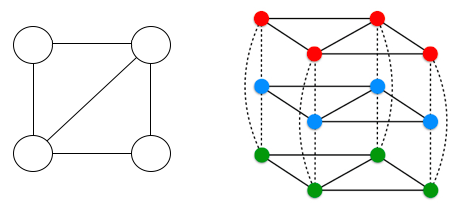}
%\label{fig:MaxIndepSetcircuit}
\caption{Example: Quantum alternating operator ansatz mapping for Max-$\numColors$-ColorableSubgraph with $\numColors=3$ in the one-hot encoding. The $4$-node graph on the left is mapped to $12$ qubits on the right, one vertical layer for each color. The solid lines show pairs of qubits acted on by the phase operator, which checks if adjacent vertices have the same color. The dashed lines show the qubits acted on by the mixing operator, which mixes the possible colors of each vertex independently. } 
\label{fig:n4graphMap}
\end{figure}
\vskip 1pc

\subsubsection{Single Qudit Mixing Operators}\label{sec:mixOpsMaxColSubgraph}

We focused initially on designing \emph{partial mixers}, component operators
that will be used to construct full mixing operators. For this mapping
of the maxColorableSubgraph problem, the partial mixers are 
operators acting on a qudit with dimension $d = \numColors$ that mix between the 
colors associated to a single vertex~$v$. As we see in subsequent sections,
this is a particularly simple case of partial mixers which are often
more complicated multiqubit-controlled operators.
Once we have defined these single-qudit partial
mixing operators, we put them together to create a full mixer for the problem.

We began by considering the following family of single-qudit mixing operators
expressed in terms of qudits, and qudit operators, and then 
considered encodings and compilations to qubit-based architectures,
which inspired us to consider other families of single-qubit 
mixing operators. 
See~Appendix~\ref{sec:quditReview} and References~\cite{gottesman2001encoding,bartlett2002quantum} 
for a review of qudit operators, including the generalized Pauli operators~$\quditX$ and~$\quditZ$.
%%%\newline 
%\paragraph{$r$-nearby-values single-qudit mixer.} 
\vskip 1pc 
\noindent \textbf{$r$-nearby-values single-qudit mixer.} \quad
Let 
$\unitary{\rnv}(\beta) = \timeEv{\beta \Ham{\rnv}}$,
where 
$\Ham{\rnv} = \sum_{i=1}^r \left( \quditX^i + {(\quditX^{\dagger})}^i\right)$,
which acts on a single qudit, with 
$\quditX = \sum_{a=0}^{d-1} \ket{a + 1} \bra{a}$.
We identified two special cases by name:
The ``single-qudit ring mixer'' for $r=1$, $\Ham{\ring} = \Ham{\rnv[1]}$
and the ``fully-connected'' mixer for \mbox{$r=\quditDim-1$}, $\Ham{\fc} = \Ham{\rnv[(\quditDim-1)]}$.
Whenever we introduced a Hamiltonian, we also implicitly introduced 
its corresponding family of unitaries, as with $\Ham{\rnv}$ and 
$\unitary{\rnv}(\beta)$.

The single-qudit ring mixer is %our bread and butter. 
a cornerstone of many of the mixing constructions that we discuss. 
We concentrated on qubit encodings thereof, given the various 
projections of hardware with at least $40$ qubits that will be available 
in the next year or two~\cite{Sete16,Google2017nature},
though it could alternatively be implemented directly using a qudit-based
architecture.
We explored two natural encodings of a qudit into qubits:
(1) The \emph{one-hot} encoding using $\quditDim$ qubits, in which each qudit basis 
state $\ket{a}$, $a=0,1,\dots,d-1$, is encoded as  ${\ket{0}}^{\otimes a} \otimes \ket{1} \otimes 
\ket{0}^{\otimes {\quditDim-1-a}}$; and
(2) the \emph{binary} encoding using $\lceil \log_2 \quditDim \rceil$ qubits,
in which each qudit state $\ket{a}$ is encoded as the qubit computational basis state 
%corresponding to 
labeled by 
the binary representation of the integer $a$. The one-hot
encoding uses more qubits but 
supports a simpler compilation of the single-qudit ring mixer for general $d$,
both in the sense of it being much easier to write down a compilation,
and in the sense that it uses many fewer two-qubit gates;
in the one-hot encoding, %two-qubit mixing operators
$2$-local mixing interactions suffice, whereas binary encoding requires $\lceil \log_2 d \rceil$-local Hamiltonian terms,
with the corresponding unitary requiring further compilation to
two-qubit gates.

In the one-hot encoding, the single-qudit ring mixer is encoded as a qubit 
unitary $\unitary{\ring}^{\enc}$ corresponding to the qubit Hamiltonian:
\begin{equation}  \label{eq:HXY}
\Ham{\ring}^{\enc} = \sum_{a=0}^{\quditDim-1} \left(X_a X_{a+1} + Y_{a} Y_{a+1}\right),
\end{equation}
which acts on the %whole 
qubit Hilbert space in a way that preserves 
the Hamming weight of computational 
basis states and acts 
as $\unitary{\ring}$ on the encoded subspace spanned by
unit Hamming weight computational basis states. 
%\redtext{The \lq\lq XY Model\rq\rq\ Hamiltonian (\ref{eq:HXY}) is well-known from physics; 
%see Appendix~\ref{sec:SWAPvXY} for review its relation to the 

Although $\Ham{\ring}^{\enc}$  is 2-local, its terms do not mutually commute.
There are several implementation options.
First, hardware that natively implements the multiqubit gate
$\unitary{\ring}^{\enc}$ directly may be plausible (much as quantum annealers already
support the simultaneous application of a Hamiltonian to a large
set of qubits), but most proposals for universal quantum processors are
based on two-qubit gates, so a compilation to such gates
is desirable both for applicability to hardware likely to be
available in the near term and
for error-correction and fault tolerance in the longer term.  
Second, we could use constructions used in quantum many-body physics to compile
$\unitary{\ring}^{\enc}$ into a circuit of 2-local gates~\cite{verstraete2009quantum}.  
Third, the multiqubit gate $\unitary{\ring}^{\enc}$ could be implemented
approximately via Trotterization or other Hamiltonian simulation algorithms.
A different approach, and the alternative we
explore most extensively here, is to implement a different unitary rather than
$\unitary{\rnv}$, one related to $\unitary{\rnv}$, and sharing its desirable
mixing properties, as encapsulated in the design criteria of 
Section~\ref{sec:designCrit},
which is easier to implement. 
The form of the circuit obtained by Trotterization is suggestive. We
considered sequentially applying unitaries corresponding to subsets of terms 
in the Hamiltonian, each subset chosen in such a way that the corresponding unitary is
readily implementable. 
This reasoning mirrors the relation of H-QAOA circuits
to Trotterized AQO.\@
We give a few examples of mixers obtained in this way.
%%%\newline
%\paragraph{Parity single-qudit ring mixer.} 
\vskip 1pc 
\noindent \textbf{Parity single-qudit ring mixer.} \quad
Still in the one-hot encoding, we partitioned the $d$ 
terms $\timeEv{\beta\left(X_{a} X_{a+1} + Y_{a} Y_{a+1}\right)}$ by
the parity of their indices. Let:
\begin{equation}
\unitary{\parity}(\beta) = 
\unitary{\last}(\beta)
\unitary{\even}(\beta)
\unitary{\odd}(\beta),
\end{equation}
where:
\begin{align}
\unitary{\odd}(\beta)
&= 
\prod_{a \text{ odd, } a\ne n}\timeEv{\beta\left(X_{a} X_{a+1} + Y_{a} Y_{a+1}\right)},
&
\unitary{\even}(\beta)
&=
\prod_{a \text{ even}}\timeEv{\beta\left(X_{a} X_{a+1} + Y_{a} Y_{a+1}\right)},
\\
\unitary{\last}(\beta)
&=
\begin{cases}
\timeEv{\beta\left(X_{d} X_{1} + Y_{d} Y_{1}\right)},
& \quditDim \text{ odd},\\
\identity, & \quditDim \text{ even}.
\end{cases}
\end{align}

Such ``XY'' gates are natively implemented on certain
superconducting processors~\cite{chow2010quantum}.

It is easy to see that the parity single-qudit ring mixer preserves the Hamming weight and hence meets the first criterion of keeping the evolution within the feasible single-qudit subspace.  
%of the graph consisting of a single vertex. 
To see that it
also meets the second, providing transitions between all feasible
computational basis states, it is useful to consider the quantum swap gate, 
which behaves in exactly the same way as the XY gate on the subspace 
spanned by %$\left\{\ket{0,1}, \ket{1,0}\right\}$.
$\left\{\ket{0_i1_j}, \ket{1_i0_j}\right\}$.
The swap gate 
$\SWAP_{i,j} = \frac{1}{2} (I + X_i X_j + Y_i Y_j + Z_i Z_j)$
is both unitary and Hermitian, and thus: 
\begin{equation}  \label{eq:expSWAP}
e^{i\theta \SWAP_{i,j}} = \cos(\theta) I + i\sin(\theta) \SWAP_{i,j}.
\end{equation}

For $0 < \beta < \pi/2$, each term in the parity single-qudit 
ring mixer is a superposition of a swap gate and the identity. 
A single application of $\unitary{\parity}(\beta)$ will have nonzero
transition amplitudes only between pairs of colors with indices no more 
than two apart. 
Nevertheless, this mixer meets the second criteria because all possible
orderings of $d$ swap gates appear in $\lceil \frac{d}{2} \rceil$ repeats 
of the parity operator for any $0 < \beta < \pi/2$, thus providing nonzero
amplitude transitions between all feasible computational basis states
for a single-vertex graph.

When $\quditDim$ is an integer power of two, there is also a straightforward,
though more resource-intensive, compilation of the parity single-qudit 
ring mixer using the binary encoding.  
Applying a Pauli $X$ gate to the least-significant
qubit acts as $\Ham{\even}$ on the encoded subspace.
Incrementing the register by one, applying the Pauli gate to the least-significant qubit, and finally decrementing the register by one overall acts as $\Ham{\odd}$.
Therefore, we can implement
$\unitary{\parity}$ by incrementing the register by one, applying an $X(\beta) = \timeEv{\beta X}$
to the least-significant qubit, decrementing the register by one, and then
again applying an $X(\beta)$ gate to the least-significant qubit.
For an $l$-bit register, each incrementing and decrementing operation can be
written as a series of $l$ multiply-controlled $X$ gates, with the numbers of
control ranging from $0$ to $l-1$.
%
%\paragraph{Repeated parity single-qudit ring mixers.} 
\vskip 1pc 
\noindent \textbf{Repeated parity single-qudit ring mixers.} \quad As we mentioned above, 
a single application of $\unitary{\parity}(\beta)$ will have nonzero
transition amplitudes only between pairs of colors with indices no more 
than two apart, which suggests that it may be useful to repeat
the parity mixer within one mixing step. 
%
%\paragraph{Partition single-qudit ring mixers.} 
\vskip 1pc 
\noindent \textbf{Partition single-qudit ring mixers.} \quad We now generalize the
above construction for the parity single-qudit ring mixer to more general
partition mixers.
For a given ordered partition $\mathcal{P} = (P_1, \ldots, P_p)$ of the terms
of $\Ham{\rnv}$ such that all pairs of terms within a $P_i$ act on disjoint 
states of the
qudit,
let: \begin{equation}
\unitary{\rnv[\mathcal{P}\text{-}r]}(\beta) = 
    \unitary{P_p \text{-}\quditSwap}(\beta)
\cdots
    \unitary{P_1 \text{-}\quditSwap} (\beta),
\end{equation}
where:
\begin{equation}
\unitary{P \text{-}\quditSwap}(\beta)
=
\prod_{\{a, b\} \in P} 
\timeEv{\beta \left(\ket{a}\bra{b} + \ket{b}\bra{a}\right)}.
\end{equation}

By construction, 
in the one-hot encoding,
the terms of $\unitary{P \text{-}\mathrm{XY}}$ commute
(because they act on disjoint pairs of qubits), and so the ordering does not
matter; all can be implemented in parallel.  
We~call $\unitary{\rnv[\mathcal{P}\text{-}r]}$ the
``partition $\mathcal{P}$'' $r$-nearby-values single-qudit ring mixer, and
$\unitary{\rnv}$ the ``simultaneous'' $r$-nearby-values single-qudit ring 
mixer to distinguish the latter from the
former. The latter is member of H-QAOA, while the former is not.

Even more generally, for a set of single-qudit ring mixers $\{\Ham{\alpha}\}$ indexed by some $\alpha$ and an ordered partition
$\mathcal{P} = (P_1, \ldots, P_p)$ thereof, in which the single-qudit 
mixers within each part mutually commute,
we defined a simultaneous version, $\timeEv{\beta \sum_{\alpha} \Ham{\alpha}}$ and a $\mathcal{P}$-partitioned version, 
$\prod_{i=1}^{P} \left[\prod_{\alpha \in P_i} \timeEv{\beta \Ham{\alpha}}\right]$, 
where~the order of the product over the elements of each part does not matter
because they commute, and the order of the product over the parts of the
partition is given by their ordering within the ordered partition
$\mathcal{P}$.  
%\paragraph{Binary single-qudit mixer for $d = 2^l$.}
\vskip 1pc 
\noindent \textbf{Binary single-qudit mixer for $d = 2^l$.} \quad
We now return briefly to the binary encoding, and describe a
different single-qudit mixer.
An alternative to the $r$-nearby values single-qudit  
mixer, which is easily implementable using the binary encoding when $\quditDim
= 2^l$ is a power of two, is the ``simple binary'' single-qudit mixer:
\begin{equation}
\Ham{\binary}^{\enc}
=
\sum_{i=1}^l X_i,
\end{equation}
where $X_i$ acts on the $i$th qubit in the binary encoding of the qudit.
Since the ordering of the colors was arbitrary to begin with, it
does not much matter whether the Hamiltonian mixes nearby values in the ordering
or mixes the colors in a different way, in this case to colors with indices 
whose binary representations have Hamming distance $1$.

When $d$ is not a power of $2$, a straightforward generalization of the
binary single-qudit mixer experiences difficulty in meeting the first of the design
criteria, since swapping one of the bit values in the binary
representation may take the evolution out of the feasible subspace.
While requiring $d$ to be a power of $2$ restricts its general
applicability, the binary single-qudit mixer could be useful in some interesting
cases, such as $4$-coloring.  
For $2$-coloring (a problem equivalent to MaxCut), 
the full binary single-qudit mixer is simply the standard mixer 
$X$.  
We use this encoding in Section~\ref{sec:smst2} to handle slack 
variables in a single-machine scheduling problem, a case in which there is flexibility in the upper range of the integer to be encoded, allowing us to round up to the nearest power of two when needed. 

\subsubsection{Full QAOA Mapping}
Having introduced several partial mixers for single qudits, we now show a
complete QAOA circuit for %a general instance of 
MaxColorableSubgraph with $\numVars$
vertices, $\numConstraints$ edges, and $\numColors$ colors, compiled to 2-local
gates on qubits.  Using the one-hot encoding, we require $\numVars \numColors$
qubits.
%\paragraph{Mixing operator.} 
\vskip 1pc 
\noindent \textbf{Mixing operator.} \quad We used as the full mixer a parity ring mixer
made up of parity single-qudit ring mixers, one for each of the qudits 
corresponding to each vertex:
\begin{equation}
\mixUnitary = \prod_{v=1}^{\numVars} \unitary[v]{\parity}^{\enc}.
\end{equation}

The single qubit mixers act on different qubits and can be applied in parallel.
The overall parity mixing operator required a depth-2 or depth-3 circuit 
(for even and odd $\numColors$, respectively) of $\numVars \numColors$ gates.  
Other single-qudit mixers we defined above, including $r$ repeats of the 
parity ring mixer, other partitioned mixers, or the binary mixer, could be used 
in place of the parity single-qudit mixer in this construction.
All of these unitary mixers by construction meet our first criterion 
for a mixer: Keeping the evolution in the feasible subspace. 
Further, each of these mixers, after at most 
$\lceil \frac{\numColors}{2} \rceil$ 
repeats, provides nonzero amplitude transitions between all colors at 
a given vertex, with the product providing transitions between any two 
feasible states.
%\paragraph{Phase-separation operator.} 
\vskip 1pc 
\noindent \textls[-10]{\textbf{Phase-separation operator.} \quad The objective function can be written in 
 classical one-hot encoding~as:}
\begin{equation} \label{eq:costgc1}
m - \sum_{\{ u,v \} \in E} \sum_{a=1}^k x_{u,a}x_{v,a},
\end{equation}
where $x_{v,a} = 1$, indicating that vertex $v$ has been assigned color $a$.
To obtain a phase-separation Hamiltonian, we substituted
$(I-Z)/2$ for each binary variable %(as detailed in \cite{hadfield2018representation}), 
to obtain:
\begin{equation}
\phaseHam' =
\frac{4-\numColors}{4}m\identity
+ \frac14 \sum_{\{u,v\} \in E} \sum_{a=1}^{\numColors}
\left(Z_{u, a} + Z_{v, a} - Z_{u,a}Z_{v,a}\right).
\end{equation}

The constant term affects only a physically-irrelevant global phase,
and since we are only concerned about the feasible subspace,
we can disregard each sum $\sum_{a=1}^{\numColors}Z_{u, a}$ of all 
$\numColors$ single $Z$ operators corresponding to a single qudit, 
since they multiply each of the $d$ Hamming weight $1$ elements corresponding to the $d$
single qudit values by the same constant, resulting in a global phase. 
Removing those terms and rescaling, the phase separator now has the simpler form:
\begin{equation}\label{eq:maxColSubgraph-phaseHam-enc}
\phaseHam^{\enc} = 
\sum_{\{u, v\} \in E} \sum_{a = 1}^{\numColors}
Z_{u, a} Z_{v, a},
\end{equation}
where $Z_{v, a}$ acts on the $a$th qubit in the one-hot encoding of the $v$th qudit, corresponding to coloring vertex $v$ with color $a$.
The phase separator requires a circuit containing $\numConstraints \numColors$
two-qubit gates with depth at most $\degree_{G} + 1$, where $\degree_{G}$ is
the maximum degree over all vertices in the instance graph $G$.

%g = sum_{u,v} sum_a Z_{u,a} Z_v,a}
%=sum_{u, v} sum_a (1 - 2 * x_{u,a}) (1 - 2 * x_{v, a})
%= sum_{u, v} sum_a 1
%- 2 * sum_{u, v} sum_a x_{u, a}
%- 2 * sum_{u, v} sum_a x_{v, a}
%+ 4 * sum_{u, v} sum_a  x_{u, a} x_{v, a}
%= m * k - 2 * m - 2 * m + 4 * sum_{u, v} sum_a x_{u, a} x_{v, a}
%= (k - 4) * m + 4 FOO
%
%f = m - FOO 
%=> g = (k - 4) * m + 4 * (m - f) = k * m - 4 * f
Translated back to acting on qudits, Equation~\eqref{eq:maxColSubgraph-phaseHam-enc} 
acts as $\phaseHam = \Ham{\phaseFunc}$ 
(as defined in Equation~\eqref{eq:functionHam}), 
where $\phaseFunc(\x) = \numColors \numConstraints - 4 \objFunc(\x)$.
We refer to this function $\phaseFunc$ as the ``phase function'', which will typically be an affine transformation of the objective function, which corresponds simply to a physically irrelevant global phase and a rescaling of the parameter.
Defining $\phaseHam$ using such a phase function allows us to write a simpler encoded version $\phaseHam^{\enc}$ that corresponds exactly to $\phaseHam$, without qualification, on the encoded subspace.
%
%\paragraph{Initial state.} 
\vskip 1pc 
\noindent \textbf{Initial state.} \quad Any encoded coloring can be generated by a depth-$1$ circuit of %no more
%than 
at most $\numVars$ single-qubit $X$ gates.  A reasonable initial state is one in
which all vertices are assigned the same color. 
Alternatively, we could start with any other feasible state, or
the initial state could be obtained by applying one or more rounds of the 
mixer to a single feasible state, so that the algorithm begins with
a superposition of feasible states.

The circuit depth and gate count for the full algorithm  will increase 
when compiling to realistic near-term hardware with architectures that have
nearest neighbor topological constraints limiting between which pairs of
physical qubits two-qubit gates can be applied. See Reference~\cite{Venturelli17} for 
one approach for compiling to realistic hardware with such constraints.

Further investigation is needed to understand which mixers and initial states, 
for a given resource budget, result in more or less effective
algorithms, and whether some have an advantage with respect
to finding good parameters  
$\gamma_1, \dots, \gamma_p$, and $\beta_1, \dots, \beta_p$
or being robust to error.

\subsection{Example: MaxIndependentSet}\label{sec:MaxIndSet}
%\paragraph{Problem.} 
\noindent \textbf{Problem.} \quad
Given a graph $G=(V,E)$, with $|V|=\numVars$ and $|E|=\numConstraints$, 
find the largest subset $V' \subset V$ of mutually non-adjacent vertices. 

This problem was discussed in Section~VII of Reference~\cite{Farhi2014} as
a ``variant'' of the quantum approximate optimization algorithm introduced
in that paper. 
To handle this problem, Farhi et al.\ suggested restricting
the evolution to what we are calling the feasible subspace of the
overall Hilbert space, the subspace spanned by computational basis elements
corresponding to independent sets, through modification of what we are 
calling the mixing operator. 
We made the construction of the H-QAOA
mixer Farhi  et al.\ defined more explicit, and introduced partitioned mixers
that have implementation advantages over the H-QAOA, or simultaneous, mixer
defined in Farhi et al.

The configuration space is the set of $n$-bit strings, representing 
subsets  $V' \subset V$ of vertices, where $i \in V'$ if and only if $x_i = 1$.
The domain $F$ is represented by the subset of all $n$-bit strings 
corresponding to 
independent sets of $G$. 
In contrast to the domain for MaxColorableSubgraph, this domain is 
dependent on the problem instance, not just on the size of the problem.
Because the configuration is already bit-based, some aspects of mapping
this problem to QAOA are simpler, but the partitioned mixing operators 
are more complicated in that they require controlled operation.

To support the discussion of controlled operators, we used the 
notation $\control_{\y} (Q)$ to indicate a unitary \emph{target operator} $Q$ 
applied to a set of target qubits controlled by the state $\y$ of
a set of control qubits: 
\begin{equation}
\label{eq:ctrlQ}
\control_{\y}(Q) = 
\sum_{\y' \neq \y} \ket{\y'}\bra{\y'} \otimes \identity
+
\ket{\y} \bra{\y} \otimes Q.
\end{equation}

More generally, we used $\control_{\controlFunc}(Q)$ when the operation
was controlled on a predicate $\controlFunc$:
\begin{equation}
\label{eq:ctrlQ2}
\control_{\controlFunc}(Q) = 
\sum_{\y : \lnot \controlFunc(\y)} \ket{\y}\bra{\y} \otimes \identity + 
\sum_{\y : \controlFunc(\y)} \ket{\y}\bra{\y} \otimes Q.
\end{equation}

Whether the subscript of $\control$ is a string or predicate will be
clear from context.
For a Hamiltonian $H_Q$ such that $Q = \timeEv{H_Q}$, we can write the controlled unitary Equation~\eqref{eq:ctrlQ2} as:
\begin{equation}
\label{eq:chi_ctrl_q}
\Lambda_{\controlFunc} (Q)
=
\timeEv{\proj \otimes H_Q}.
\end{equation}

We refer to $\proj \otimes H_Q$ as the \emph{controlled Hamiltonian}, $\controlFunc$-controlled-$H_Q$.
Note that the Hamiltonian $\proj$ that acts as the predicate $\controlFunc$ on computational basis states, in the sense of Equation~\eqref{eq:functionHam}, is precisely the projector $\proj = \sum_{\x : \controlFunc(\x) = 1} \ketbra{\x}{\x}$ that projects onto the subspace on which the predicate is $1$.  
%As with $\control$, we also use $H_{\y}$ for a bitstring $\y$ to indicate the projector for the implied conjunction of the bits in $\y$.
We used this relation to connect corresponding controlled Hamiltonians and controlled unitaries.
In particular, when we want to apply a phase only on the part of the Hilbert space picked out by a predicate $\controlFunc$, we can write:
\begin{equation}
\control_{\controlFunc}
\left(e^{-i \theta}\right)
=
e^{-i \theta H_{\controlFunc}},
\end{equation}
%which corresponds to the target operator being the trivial operator $e^{-i \theta}$ acting on the 1-dimensional Hilbert space corresponding to 0 target qubits.
where we have adapted the control notation of Equation~\eqref{eq:chi_ctrl_q} to mean applying the operator $Q = e^{-i\theta}$ to zero target qubits. 
We often compile controlled unitaries (both phase separators and mixers) by using ancilla qubits to intermediate the control, e.g., for a single ancilla qubit (initialized at $\ket{0}$ and returned thereto):
\begin{equation}
{\left(
\control_{\controlFunc}
\left(
Q
\right)
\right)}^{\comp}
=
\control_{\controlFunc}
\left(
X_{\mathrm{anc}}
\right)
\control_{x_{\mathrm{anc}}}
\left( Q \right)
\control_{\controlFunc}
\left(
X_{\mathrm{anc}}
\right).
\end{equation}

We %elaborate on
explore in detail the construction of such controlled Hamiltonians and unitaries in Reference~\cite{hadfield2018representation}.

\subsubsection{Partial Mixing Operator at Each Vertex} 
Given an independent set $V'$, we can add a vertex $w \notin V'$ to $V'$ while
maintaining feasibility only if none of its neighboring vertices
$\neighborFunc(w)$ are
already in $V'$. On the other hand, we can always remove any vertex $w \in V'$
without affecting feasibility. 
Hence, a bit-flip operation at a vertex, controlled by its neighbors (adjacent vertices), suffices both to remove and add vertices while maintaining the independence property.
These classical moves inspire
the controlled-bit-flip partial mixing operators. 

In general, for a string $\y$ and a set of indices $V$, let $\y_V = {\left(y_{v_i}\right)}_{i=1}^{|V|}$ be the substring of $\y$ in lexicographical order of the indices.
In particular, let 
$\x_{\neighborFunc(v)} = {\left(x_w\right)}_{w \in \neighborFunc(v)}$. (The ordering of the characters within the substring is arbitrary, because we only use this as the argument to predicates that are symmetric under permutation of the arguments.)
For each vertex, we defined the partial mixer as a multiply-controlled $X$ operator: 
\begin{align}
\begin{split}
\Ham{\controlX, v}
& =
X_v 
H_{\noNeighbors{v}} \\
& =
2^{-\degree_v}
X_v \prod_{w \in \neighborFunc(v)} (\identity + Z_w)
,
\end{split}
\end{align}
with corresponding partial mixing unitary, 
a multiply-controlled-$X(\beta)$ single-qubit rotation:
\begin{align}\label{eqn:driverIndepSet}
\begin{split}
\unitary{\controlX, v}(\beta)
& =
\timeEv{\beta \Ham{\controlX, v}} \\
& =
\control_{\noNeighbors{v}}
\left(\timeEv{\beta X_v}\right) \\
& =
\control_{\noNeighbors{v}}
\left(X_v(\beta)\right),
\end{split}
\end{align}
where $X_v(\beta)$ is the single-qubit operator
 $X_v(\beta) = \timeEv{\beta X_v}$.
Since $X_v$ is both Hermitian and unitary, $\timeEv{\beta X_v}$ 
is a linear combination of the identity and $X_v$ for $0 < \beta < \pi/2$.

\subsubsection{Full QAOA Mapping}
%\paragraph{Mixing operators.} 
\noindent \textbf{Mixing operators.} \quad
Let $\Ham{\controlX} = \sum_{i=1}^{\numVars} \Ham{\controlX, v_i}$.
We defined two distinct types of mixers:
\begin{itemize}
\item The \emph{simultaneous} controlled-$X$ mixer, 
$\unitary{\simCX}(\beta) = \timeEv{\beta \Ham{\controlX}}$, and;
\item A class of \emph{partitioned} controlled-$X(\beta)$ mixers, 
$\unitary{\seqCX}(\beta) = \prod_{i=1}^{|{\mathcal P}|}\prod_{v\in P_i} \unitary{\controlX,v}$,
\end{itemize}
where $\mathcal{P}$ is an ordered partition of the partial mixers, in which each part contains mutually commuting partial mixers.
Since the partial mixers often do not commute,
different ordered partitions often result in different mixers.
By design, both the simultaneous and partitioned mixers restrict evolution to the feasible 
subspace. With respect to the second design criterion, 
there is nonzero transition amplitude from the $\ket{0}^{\otimes n}$
state corresponding to the empty set to all other independent sets;
for $0 < \beta < \pi/2$, we got terms corresponding to products of 
the individual control-bit-flip operators for all subsets of vertices,
including those corresponding to independent sets. (For those
subsets $S$ not corresponding to independent sets, the product will result
in a independent set $V' \in S$ that does not include vertices in $S$ which 
have neighbors whose controlled-bit-flip preceded them in the partition order;
thus, different ordered partition affects the amount of nonzero amplitude
in the states corresponding to independent sets.) Two applications of
any such partitioned mixer result in nonzero amplitude between any two feasible 
states. An interesting question is how different ordered partitions
affect the ease with which good parameters can be found and the 
quality of the solutions obtained. 

Partitioned mixers are generally easier to compile than the simultaneous mixer,
since the partitioned mixer is a product of multiqubit-controlled-{\sc not}
operators (generalized Toffoli gates) on at most $\degree_G +1$ qubits. 
Altogether, this construction uses $\numVars$ partial mixers, which can then be
compiled into single- and two-qubit gates. For many graphs, partitions in which
each set contains multiple commuting partial mixers exist, reducing the depth.
%
%\paragraph{Phase-separation operator.} 
\vskip 1pc 
\noindent \textbf{Phase-separation operator.} \quad
The objective function is the size of the
independent set, or $f(\x) = \sum_{i=1}^{\numVars} x_i$, 
which we could translate into a phase-separating Hamiltonian
via substitution of $(I-Z)/2$ for
each binary variable. 
Instead, we used affine transformation of the 
objective function $\phaseFunc(\x) = n - 2\objFunc(\x)$, which, when translated, yields
a phase separation operator of a simpler form:
% g = sum_i Z_i
% = sum_i (1 - 2 x_i)
% = n - 2 sum_i = n - 2 f
\begin{equation}
\phaseUnitary(\gamma) = 
\timeEv{\gamma \Ham{\phaseFunc}}
=
\prod_{i=1}^n \timeEv{\gamma Z_i},
\end{equation}
which is simply a depth-1 circuit of $n$ single-qubit Z-rotations.
%
%\paragraph{Initial state.} 
\vskip 1pc 
\noindent \textbf{Initial state.} \quad
A reasonable initial state is the trivial state
$\ket{\initial} = \ket{0}^{\otimes n}$ corresponding to the empty set.

\subsection{Example: MaxColorableInducedSubgraph}\label{sec:maxColIndSubGraph}\label{sec:induced}
%
%\paragraph{Problem.} 
\noindent \textbf{Problem.} \quad
Given $\numColors$ colors, and a graph $G = (V,E)$ with $\numVars$ vertices and $\numConstraints$ edges, 
find the largest induced subgraph that can be properly $\numColors$-colored.

The \emph{induced subgraph} of a graph $G= (V,E)$ for a subset of vertices
$W \subset V$ is the graph $H = (W, E_W)$, where  
$E_W = \{\{v, w\} \in E: v, w \in W\}$.
The configuration space is the set ${[\numColors+1]}^{\numVars}$ of $(\numColors+1)$-dit strings of length $\numVars$, corresponding to partial $\numColors$-colorings of the graph:
$x_v = 0$ indicates that vertex $v$ is uncolored and $x_v = c> 0$ indicates that the vertex has color $c$.
The induced subgraph is defined by the colored vertices.
The domain is the set of \emph{proper} partial colorings, those in which two colored vertices that are adjacent in $G$ have different colors.
The objective function 
$\objFunc: {[\numColors + 1]}^{\numVars} \to \naturals$
is the number of vertices that are colored:
\begin{equation}
\objFunc(\x) = \sum_v \NEQ(x_v,0).
\end{equation}

\subsubsection{Controlled Null-Swap Mixer at a Vertex}

The controlled null-swap partial mixer we defined has elements of the mixers we
saw for the previous two problems, combining the control by vertex
neighbors from MaxIndependentSet and the color swap from MaxColorableSubgraph.
Here, however, we made substantial use of the uncolored state, and at each vertex only
considered swapping a color with uncolored status.
An uncolored vertex can be assigned color~$c$, maintaining feasibility, 
as long as none of its neighbors are colored~$c$, whereas uncoloring a vertex always preserves feasibility. This suggests a mixer may be obtained by swaps between each color and the uncolored state, controlled for each vertex by the colors of its neighboring vertices to ensure feasibility. 
This reasoning in terms of classical moves inspires, 
for problems containing $\NEQ$ constraints, the 
controlled null-swap partial mixing Hamiltonian:
\begin{equation}
\label{eq:HnullSwap}
\begin{split}
\Ham{\nullSwap, {v, a}}
&=
\left(
\ket{a}\bra{0}_v + \ket{0}\bra{a}_v
\right)
\Ham{\NONE(\x_{\neighborFunc(v)},a)}
\\&=
\left(
\ket{a}\bra{0}_v + \ket{0}\bra{a}_v
\right)
\prod_{w \in \neighborFunc(v)}
\left(\identity_w - \ket{a}\bra{a}_w\right),
\end{split}
\end{equation}
with corresponding controlled null-swap-rotation mixing unitary:
\begin{equation}
\label{eq:UnullSwap}
\unitary{\nullSwap, {v, a}}(\beta)
=
\control_{\NONE(\x_{\neighborFunc(v)}, a)}
\left(
\timeEv{\beta 
\left(\ket{a}\bra{0}_v + \ket{0}\bra{a}_v\right)}
+
\sum_{b \notin \{0, a\}}
\ket{b}\bra{b}_v
\right),
\end{equation}
where:
\begin{equation}\label{eq:NONE_Def}
\NONE(\y, A) = 
\bigwedge_{\substack{y \in \y \\ a \in A}} 
\NEQ(y, a)
=
\bigwedge_{a \in A}
\bigwedge_{i = 1}^{|\y|}
\NEQ(y, a)
=
\bigwedge_{a \in A}
\bigwedge_{i \in \left[|\y|\right]}
\NEQ(y, a)
\end{equation}
is shorthand for none of the variables in $\y$ having value any of the values
in $A$; when $A=\{a\}$ is a~singleton set, we write simply $\NONE(\y, a) =
\NONE(\y, \{a\})$.

\subsubsection{Full QAOA Mapping}
%\paragraph{Mixing operators.} 
\noindent \textbf{Mixing operators.} \quad
Define:
\begin{equation}
\Ham{\nullSwap} = 
\sum_{i=1}^{\numVars}
\sum_{a=1}^{\numColors}
\Ham{\nullSwap, {i, a}}.
\end{equation}

We defined two distinct types of mixers:
\begin{itemize}
\item The \emph{simultaneous} controlled null-swap mixer, 
$\unitary{\simNullSwap}(\beta) = \timeEv{\beta \Ham{\nullSwap}}$, and; 
\item A family of \emph{partitioned} controlled null-swap mixers,
$\unitary{\seqNullSwap}(\beta) = 
\prod_{a=1}^{\numColors}
\prod_{i=1}^{|{\mathcal P}|}
\prod_{v\in P_i}
\unitary{\nullSwap, {v, a}}$.
\end{itemize}

Again, we have a variety of partitioned mixers, each specified by an ordered
partition $\mathcal P$ of the vertices such that for each color the terms 
corresponding to the vertices in the partition commute. 
We~segregated the colors into separate stages,
but other orderings are possible.

We used the one-hot encoding of Section~\ref{sec:maxColSubGraph}, but with
additional variables $x_{v, 0}$ for the uncolored states:
The binary variables for each vertex $v$ 
are $x_{v, 0}, x_{v,1}, \dots,x_{v,k}$. This encoding uses
 $\numVars (\numColors + 1)$ computational qubits.  
In this encoding, a single partial mixer has the form:
\begin{equation}\label{eq:ctrlNullSwapOneHot}
\begin{split}
\unitary{\nullSwap, {v, a}}^{\comp}(\beta)
&=
%%%%%\control_{\noNeighbors[v]{a}}
\control_{\NOR(\bar{\x}_{\neighborFunc(v), a})}
\left(
\timeEv{\beta \left(X_{v, 0} X_{v, a} + Y_{v, a} Y_{v, 0}\right)}
\right)
,
\end{split}
\end{equation}
where 
$\bar{\x}_{\neighborFunc(v), a} = 
{\left(x_{w,a}\right)}_{w \in \neighborFunc(v)}$.
Reasoning similar to that we used for the mixers discussed for the
MaxIndependentSet and MaxColorableSubgraph
problems shows that this mixer has nonzero transition amplitude between any
feasible computational-basis state
and the trivial state corresponding to the empty set as the induced subgraph.
Two applications of this mixer give nonzero transition amplitudes between
any two feasible computational-basis states.

To ease compilation, each partial mixer can be implemented as:
\begin{equation}   \label{eq: controlledNSancilla}
\unitary{\nullSwap, {v, a}}^{\comp}(\beta)
=
%%%%\control_{\noNeighbors[v]{a}}
\control_{\NOR(\bar{\x}_{\neighborFunc(v), a})}
\left(X_{\mathrm{anc}}\right)
\control_{x_{\mathrm{anc}}}
\left(
\timeEv{\beta \left(X_{v, 0} X_{v, a} + Y_{v, a} Y_{v, 0}\right)}
\right)
%%%\control_{\noNeighbors[v]{a}}
\control_{\NOR(\bar{\x}_{\neighborFunc(v), a})}
\left(X_{\mathrm{anc}}\right),
\end{equation}
where the control is intermediated by an ancilla qubit, which is initialized and returns to the zero state.
Altogether, this construction uses $\numColors\numVars$ partial
mixers, which can then be compiled into single- and two-qubit gates. 
For many graphs, partitions in which each set contains multiple 
commuting partial mixers exist, reducing the depth.
%\paragraph{Phase-separation operators.} 
\vskip 1pc 
\noindent \textbf{Phase-separation operator.} \quad
We can translate the objective function
to a Hamiltonian as usual, or~translate a linear modification of the
objective function to obtain a simpler form. 
% f = n - sum_i x_{i, 0}
% g = sum_i Z_{v, 0}
% = sum_i (1 - 2 x_{i, 0})
% = n - 2 f
The phase separator 
function $\phaseFunc(\x) = \numVars - 2 \objFunc(\x)$ yields %corresponds to 
the simple %compiled 
phase separator Hamiltonian:
\begin{equation}\label{eq:costgc2}
\phaseHam^{\comp} = \sum_v Z_{v,0},
\end{equation}
for which the corresponding unitary operator can be implemented using a depth-1 circuit of $\numVars$~single-qubit $Z$-rotations.
%\paragraph{Initial state.} 
\vskip 1pc 
\noindent \textbf{Initial state.} \quad
\textls[-20]{A reasonable initial state is 
$\ket{\initial} = {\left( \ket{1} \otimes {\ket{0}}^{\otimes \numColors}\right)}^{\otimes \numVars}$, 
corresponding to  all vertices uncolored.}

\subsection{Example: MinGraphColoring}\label{sec:chromatic-number}
%\paragraph{Problem.} 
\noindent \textbf{Problem.} \quad Given a graph $G=(V,E)$, find the minimal number of colors $k^*$ required to properly color it.

A graph that can be $\numColors$-colored but not $(\numColors-1)$-colored is said to have \textit{chromatic number}~$\numColors$. 
We took as our configuration space the set of $\numColors$-dit strings 
of length $\numVars$, where $\numColors = \degree_G+2$. 
The domain $\domain$ is the set of \emph{proper} colorings,
many of which will use fewer than $\numColors$ colors.
With $\degree_G+2$ colors, as we explain next, it is possible to get from any
proper coloring to any other by local moves while staying in the feasible
subspace, a property we made use of in designing mixing operators. 
We comment that it may be advantageous
to take use a larger number of colors since that may promote mixing, but
the tradeoffs there would need to be determined in a future investigation.

It is easy to see that any graph can be colored with $\degree_G+1$ colors. 
To see that $\numColors = \degree_G+2$ suffices to get between any
two $\degree_G+1$ colorings, first recognize that given a $\degree_G+2$
coloring, one can always obtain a $\degree_G+1$ coloring by simply 
choosing a color and recoloring each vertex currently colored in that color
with one of the other colors, since at least one of those colors will
not be used by its neighbors. This move is local, in that it depends only
on the neighborhood of the vertex. Now, given two $\degree_G+1$ colorings
$C$ and $C'$, we iterated through colors $c$ to tranform between the two
colorings via local moves while staying in the feasible space. 
Let $S'\subset V$ be the set of vertices 
colored $c$ in $C'$, and let $S\subset S'$ be the set of vertices in $S'$
that are not colored $c$ in $C$. Consider all neighbors of vertices in
$S$. For any neighbor colored $c$, color it with the unused color. We
are now free to color all vertices in $S$ with color $c$. Iterating 
through the $\numColors$ colors provides a means of getting from one 
$\degree_G+1$ coloring to another by local moves that remain in the
feasible space. 

\subsubsection{Partial Mixer at a Vertex}

We used a controlled
version of the mixer in Section~\ref{sec:maxColSubGraph} that allows a
vertex to change colors only when doing so would not result in an 
improper coloring; we may swap colors $a$ and $b$ at vertex $v$
only if none of its neighbors are colored $a$ or $b$.  
The partial mixer we defined has a similar form to the 
controlled null-swap partial mixer defined in Equations~(\ref{eq:HnullSwap})
and~(\ref{eq:UnullSwap}) but supports color changes between any two colors 
at a vertex, rather than only between colored and uncolored.
Define the controlled-swap partial mixing Hamiltonian:
\begin{equation}
\begin{split}
\label{eq:controlledSwap}
\Ham{\controlSwap, {v, \{a, b\}}}
&=
\left(
\ketbra{a}{b}_v + \ketbra{b}{a}_v
\right)
H_{\NONE(\x_{\neighborFunc(v)}, \{a, b\})}
\\&=
\left(
\ketbra{a}{b}_v + \ketbra{b}{a}_v
\right)
\prod_{w \in \neighborFunc(v)}
\left(\identity_w - \ket{a}\bra{a}_w - \ket{b}\bra{b}_w\right),
\end{split}
\end{equation}
with corresponding controled-swap-rotation mixing unitary:
\begin{equation}
\unitary{\controlSwap, {v, \{a, b\}}}
=
\control_{\NONE(\x_{\neighborFunc(v)}, \{a, b\})}
\left(\timeEv{\beta\left(
\ket{a}\bra{b}_v + \ket{b}\bra{a}_v
\right) } 
+ \sum_{c \notin \{a, b\}} \ket{c}\bra{c}_v
\right),
\end{equation}
where $\NONE(\x, A)$ was defined in Equation~(\ref{eq:NONE_Def}).
These mixers are 
controlled versions of the single
qudit fully-connected mixer of Section~\ref{sec:maxColSubGraph}, 
rather than the single qudit ring mixer, which makes sure that
every possible state is reachable. 

\subsubsection{Full QAOA Mapping}
%\paragraph{Mixing Operator.}
\noindent \textbf{Mixing Operator.} \quad
Let:
\begin{equation}
\Ham{\controlSwap} = 
\sum_{v}
\sum_{a, b}
\Ham{\controlSwap, {v, \{a, b\}}}\;.
\end{equation}
We defined two types of mixers:
\begin{itemize}
\item The \emph{simultaneous} controlled-swap mixer:
\begin{equation}
\unitary{\simControlSwap}(\beta) = 
\timeEv{\beta \Ham{\controlSwap}} \text{, and;}
\end{equation}
\item A family of \emph{partitioned} controlled-swap mixers:
\begin{equation}
\unitary{\seqControlSwap}(\beta)
=
\prod_{a, b}
\prod_{i=1}^{|{\mathcal P}|}
\prod_{v\in P_i}
\unitary{\controlSwap, {v, \{a, b\}}}(\beta)\;.
\end{equation}
\end{itemize}

As before, each partitioned mixer is specified by an 
ordered partition ${\mathcal P}$
of the vertices such that, for each color, the partial mixers
for vertices in one set of the partition all commute with each other.
Altogether, this construction uses $(\numColors - 1)\numColors\numVars/2$ 
partial mixers,
For many graphs, partitions in which each set contains multiple
commuting partial mixers exist, allowing different partial mixers 
to be carried out in parallel, reducing the depth.
%
%\paragraph{Phase-separation operator.}
\vskip 1pc 
\noindent \textbf{Phase-separation operator.} \quad The objective function,
$f: {[\numColors]}^{\numVars} \to \mathbf{Z}_+$, is:
\begin{equation}
f(\mathbf x) = \sum_{a=1}^{\numColors} \OR(\EQ(x_1, a), \ldots, \EQ(x_{\numVars}, a)),\;
\end{equation}
which counts the numbers of colors used.
Let $\phaseFunc(\x) = \numColors - \objFunc(\x)$ be the phase operator
that
counts the number of colors \emph{not} used. 
Let $H_{\NONE(\x, a)}$ be the projector onto the subspace of $\mathcal{H}$ spanned by the states corresponding to strings in ${[\numColors]}^{\numVars}$ that do not contain the character $a$.
We have $\Ham{\phaseFunc} = \sum_{a} H_{\NONE(\x, a)}$,~so:
\begin{equation}
\label{eq:phaseunitary_chromatic}
\phaseUnitary(\gamma)
=
\timeEv{\gamma H_{\phaseFunc}}
=
\prod_{a=1}^{\numColors}
\timeEv{\gamma H_{\NONE(\x, a)}}.
\end{equation}
%\paragraph{Initial state.} 
\vskip 1pc 
\noindent \textbf{Initial state.} \quad For the initial state, we used an easily found
$\degree_G+1$ (or $\degree_G+2$) coloring. 

\subsubsection{Compilation in One-Hot Encoding}
\textls[-20]{We now give partial compilations of the elements of the mapping to 
qubits using the one-hot~encoding.}
%
%{\bf Mixing operators. }
%\paragraph{Mixer.}
\vskip 1pc 
\noindent \textbf{Mixer.} \quad
In the one-hot encoding, the controlled-swap mixing Hamiltonian can
be written as:
\begin{equation}
\begin{split}
\Ham{\controlSwap, {v, \{a, b\}}}^{\comp}
&=
\left(X_{v, a} X_{v, b} + Y_{v, a} Y_{v, b}\right)
\Ham{\noNeighbors[v]{\{a, b\}}}
\\&= 2^{-D_v-1}
\left(
X_{v, a} X_{v, b} + Y_{v, a} Y_{v, b}
\right)
\prod_{w \in \neighborFunc(v)}
\prod_{c \in \{a, b\}}
\left(\identity_{w, c} + Z_{w, c}\right)
\end{split}
\end{equation}
with the corresponding unitary written as:
\begin{equation}
\unitary{\controlSwap, {v, \{a, b\}}}^{\comp}(\beta)
=
\control_{\noNeighbors[v]{\{a, b\}}}
\left(\timeEv{\beta\left(
X_{v, a} X_{v, b} + Y_{v, a} Y_{v, b}
\right) } \right),
\end{equation}
\textls[-20]{where, 
for a string doubly indexed $\y = {\left(y_{i,j}\right)}_{i, j}$, 
$\y_{A, B} = 
{\left({\left(y_{i, j}\right)}_{i \in A}\right)}_{j \in B}$
denotes the substring consisting the characters $y_{i, j}$ for which $i \in A$ and $j \in B$, in lexicographical ordering of the two indices.
In~particular, $\x_{\neighborFunc(v), \{a, b\}}$ indicates the bits corresponding to coloring the neighbors of $v$ either color $a$ or color $b$.
The~$\control_{\noNeighbors[v]{\{a, b\}}}$ dictates
that none of the neighbors of $v$ 
take value $a$ or $b$ for the swap to be performed.
Each $\unitary{\controlSwap, {v, \{a, b\}}}^{\comp}$ is a controlled gate 
with $2\degree_v$ control qubits and two target qubits.
Altogether, the full mixing Hamiltonian can be implemented using
$\numColors(\numColors - 1)\numVars/2$
controlled gates on no more than $\degree_G + 2$~qubits.}
%
%{\bf Phase separator. }
%\paragraph{Phase-separation operators.}
%\paragraph{Phase separator.}
\vskip 1pc 
\noindent \textbf{Phase separator.} \quad
Let $\phaseUnitary[a](\gamma) = \timeEv{\gamma H_{\NONE(\x, a)}}$,
so that the phase separator Equation~\eqref{eq:phaseunitary_chromatic} can be written as
$\phaseUnitary(\gamma) = \prod_{a=1}^{\numColors} \phaseUnitary[a](\gamma)$.
% The operator $\phaseUnitary[a](\gamma)$ can be implemented by applying
% the phase to any qubit not among the control qubits, such as
% the qubit corresponding to variable $x_{1,a+1}$.
% \begin{equation}
% {\phaseUnitary[a]}^{\comp} (\gamma)
% =
% \control_{\bar{\x}_{V, a}}
% X_{1,a+1}.
% \end{equation}
% The phase separator can be implemented using $\numColors$ such 
% operators, each with $\numVars$ control qubits and a single target qubits. The 
% set of qubits on which these operators are distinct except for
% those associated with consecutive colors. Thus, the phase
% separator can be implemented in depth $2$ in these
% $\numVars + 1$-qubit controlled operators.
Each partial phase separator can alternatively be written as:
\begin{equation}
\phaseUnitary[a]
=
\control_{\NONE(\x, a)}
\left(\timeEv{\gamma}\right).
\end{equation}
% Through the use of $\numColors$ ancilla, one for each partial phase
% separator, we could reduce the depth to $1$
% in $\numVars + 1$-qubit controlled operators by intermediating the control via the ancilla.
% We can realize the partial phase separator (up to a negligible global phase) as
% \begin{equation}
% {\phaseUnitary[a]}^{\comp}
% (\gamma)
% =
% \left[
% \control_{\NOR(\x_{V, a})}
% \left(
% X_{\mathrm{anc}}
% \right)
% \right]
% \left(e^{{\red i \gamma Z_{\mathrm{anc}} / 2}}\right)
% \left[
% \control_{\NOR(\x_{V, a})}
% \left(
% X_{\mathrm{anc}}
% \right)
% \right]
% \end{equation}
% using a single control ancilla qubit initialized at $\ket{0}$ (and returned thereto).
% The partial phase separators for different colors now act on 
% distinct sets of qubits, so can be implemented in depth $1$.
%
%{\bf Initial state.}
%\paragraph{Initial state.}
%\vskip 1pc 
\noindent \textbf{Initial state.} \quad
Any coloring can  be prepared in depth $1$ using $\numVars$ 
single-qubit $X$ gates:
\begin{equation}
\ket{\x} = 
\left(\prod_{i=1}^{\numVars} X_{i, x_i}\right) 
\ket{0}^{\otimes \numVars \numColors}.
\end{equation}

\section{QAOA Mappings: Orderings and Schedules}\label{sec:permutations}
Many challenging computational problems have a configuration space that 
is fundamentally the set of orderings, permutations, or schedules of some number of items. 
Here, we introduce the machinery for mapping such problems to QAOA, using 
the traveling salesperson and several single-machine scheduling problems 
as illustrative examples.

\subsection{Example: Traveling Salesperson Problem (TSP)}\label{sec:TSP}

\noindent {\bf Problem.} \quad Given a set of $\numVars$ cities, and distances $\distance:
{[\numVars]}^2 \rightarrow \reals_+$, 
%find the ordering of the cities that minimizes the total distance traveled.
find an ordering of the cities that minimizes the total distance traveled for the corresponding tour. A \textit{tour} visits each city exactly once and returns from the last city to the first.  
Note that %we use the notation  
we defined $[n]=\{1,2,\dots,n\}$ and   $[0,n]=\{0,1,\dots,n\}$. 

While for expository purposes, we call these numbers distances, the mapping
works for any cost function on pairs of cities, whether
or not it forms a metric or not; the distances are not required to be
symmetric, or to satisfy the triangle inequality. 

\subsubsection{Mapping}\label{sec:tsp_mapping}
The configuration space here is the set of all orderings of the cities.
Labeling the cities by $[n]$,
the ordering
$\order=(\orderElmt_1, \orderElmt_2, \ldots, \orderElmt_{\numVars - 1}, \orderElmt_{\numVars})$
indicates traveling from city $\orderElmt_1$ to city $\orderElmt_2$, then on to city $\orderElmt_3$ and so on until finally returning from city 
$\orderElmt_{\numVars}$ back to city $\orderElmt_1$.
The configuration space includes some degeneracy in solutions with respect to cyclic permutations; 
specifically, for any ordering $\order$, the configuration space includes both 
$(\orderElmt_1, \orderElmt_2, \ldots, \orderElmt_{\numVars - 1}, \orderElmt_{\numVars})$ 
and 
$(\orderElmt_2, \orderElmt_3, \ldots, \orderElmt_{\numVars}, \orderElmt_{1})$, 
even though they are essentially the same solution to the TSP.\@
We leave in this degeneracy in the constructions of this section in order to preserve symmetries which make it simpler to construct and present our mixers. Note that, in practice, this degeneracy may be removed by fixing a particular city as the starting point, resulting in an $n-1$ city problem wit slightly simpler cost functions that yields the same solutions, for which it is straightforward to adapt the constructions below. 

As there are no problem constraints, the domain is the same as the
configuration space.  The objective function is:
\begin{equation}
f(\order)
=
\sum_{j = 1}^{\numVars}
\distance_{\orderElmt_j, \orderElmt_{j+1}},
\end{equation}
where we again and throughout employed the convention $\orderElmt_{\numVars+1} := \orderElmt_{1}.$  
%\newline
%
%\paragraph{Ordering swap partial mixing Hamiltonians.}
\vskip 1pc 
\noindent \textbf{Ordering swap partial mixing Hamiltonians.} \quad
Our mixers for orderings will be built from partial mixer Hamiltonians we call 
``value-selective ordering swap mixing Hamiltonians.'' 
Consider $\{\orderElmt_i, \orderElmt_j \} = \{u,v\}$, indicating
that city $u$ (resp.~$v$) is visited at the $i$th (resp.~$j$th) stop
on the tour, or vice versa.  
There are $\binom{\numVars}{2}^2$ 
value-selective ordering swap mixing Hamiltonians, 
$\Ham{\permSwap, {\{i, j\}, \{u, v\}}}$,
which swap the $i$th and $j$th elements in the ordering if and only 
if those elements are the cities $u$ and $v$: 
\begin{equation}
\label{eq:mixer_perm_swap} 
\begin{split}
\lefteqn{\Ham{\permSwap, {\{i, j\}, \{u, v\}}}}
\\&= 
\sum_{\order: \{\orderElmt_i, \orderElmt_j\} = \{u, v\}}
\ketbra{\left(
\orderElmt_1, \ldots, 
    \orderElmt_{i-1}, \tikzmark{v2}v, \ldots
    \orderElmt_{j-1}, \tikzmark{u2}u, \ldots
\orderElmt_{\numVars}
\right)}
{\left(
\orderElmt_1, \ldots, 
    \orderElmt_{i-1}, \tikzmark{u1}u, \ldots
    \orderElmt_{j-1}, \tikzmark{v1}v, \ldots
\orderElmt_{\numVars}
\right)} \;.
\begin{tikzpicture}[overlay,remember picture,out=315,in=225,distance=0.4cm]
\draw[->,shorten >=5pt,shorten <=5pt, out=130, in=40, distance=0.5cm] (u1.north) to (u2.north);
\end{tikzpicture}
\begin{tikzpicture}[overlay,remember picture,out=315,in=225,distance=0.4cm]
\draw[->,shorten >=5pt,shorten <=5pt, out=130, in=40, distance=1.0cm] (v1.north) to (v2.north);
\end{tikzpicture}
\end{split}
\end{equation}

We made extensive use of a special case, the \emph{adjacent} 
ordering swap mixing Hamiltonians:
\begin{equation}
\label{eq:mixer_adj_perm_swap} 
\Ham{\permSwap, {i, \{u, v\}}} = \Ham{\permSwap, {\{i, i+1\}, \{u, v\}}}\;.
\end{equation}

To swap the $i$th and $j$th elements of the ordering regardless of which 
cities those are, we used the 
value-independent ordering swap partial mixing Hamiltonian:
\begin{equation} \label{eq:mixer_indep_perm_swap}
\Ham{\permSwap, \{i, j\}} 
=
\sum_{\{u, v\} \in \binom{[n]}{2}}
\Ham{\permSwap, {\{i, j\}, \{u, v\}}}.
\end{equation}

Of these $\binom{\numVars}{2}$ partial mixing Hamiltonians,
$n$ are adjacent
value-independent ordering swap partial mixing Hamiltonians: 
\begin{equation}\label{eq:mixer_adj_indep_perm_swap}
\begin{split}
%\Ham{\permSwap, \{i, i+1\}} 
\Ham{\permSwap, i} 
&= 
\sum_{\order}
\Ket{\left( 
\orderElmt_1, \ldots
\orderElmt_{i-1}, 
\orderElmt_{i+1}, 
\orderElmt_i, 
\orderElmt_{i+2}, 
\ldots,
\orderElmt_{\numVars} 
\right)}
\Bra{\left(
\orderElmt_1, \ldots, \orderElmt_{\numVars}
\right)},
\end{split}
\end{equation}
which swap the $i$th element with the subsequent one
regardless of which cities those are.

These partial mixers can be combined in several ways to form full mixers,
of which we explore two types.
%
%\paragraph{Simultaneous ordering swap mixer.}
\vskip 1pc 
\noindent \textbf{Simultaneous ordering swap mixer.} \quad
Defining 
$\Ham{\permSwap} 
=
\sum_{i=1}^{\numVars} \Ham{\permSwap,i}$,
we have the ``simultaneous ordering swap mixer'':  
\begin{equation}
\unitary{\simPermSwap}(\beta) = \timeEv{\beta \Ham{\permSwap}}.
\end{equation}

Different subsets and partitions of the partial mixers 
$\unitary{\permSwap, {\{i, j\}, \{u, v\}}} = \timeEv{ \Ham{\permSwap, {\{i, j\}, \{u, v\}}}}$
and different orderings
of a partition yield different partitioned ordering swap mixers.  
The color parity mixers we now defined use the adjacent partial mixer.
Other mixers, using more of %all of 
the  ${\binom{\numVars}{2}}^2$ partial
mixers, are possible, as are repeated versions of the following color-parity 
ordering swap mixer.
%
%\paragraph{Color-parity ordering swap mixer.}
\vskip 1pc 
\noindent \textbf{Color-parity ordering swap mixer.} \quad
Simultaneous ordering swap mixer.
To define the ordered partition, we first defined an ordered partition
on the set of adjacent partial mixers $\unitary{\permSwap, {i, \{u, v\}}}$ 
for a fixed tour position $i$, where the parts of this partition
contains mutually commuting partial mixers. We then partitioned the $i$ to 
obtain a full ordered partition. 
Two partial mixers 
$\unitary{\permSwap, {i, \{u, v\}}}$ and 
$\unitary{\permSwap, {i, \{u', v'\}}}$ 
commute as long as $\{u, v\} \cap \{u', v'\} = \emptyset$.
Partitioning the $\binom{\numVars}{2}$ pairs of cities into $\numColors$ parts
such that each part contains only mutually disjoint pairs 
is equivalent to considering a $\numColors$-edge-coloring
of the complete graph $K_{\numVars}$ and assigning an ordering to the colors.
For odd $\numVars$, $\numColors=\numVars$ suffices, and for 
even $\numVars$, $\numColors = \numVars-1$ suffices~\cite{alexander2008mathematical}.
(Using the geometrical construction based on regular polygons, we can define the canonical partition by placing the vertices at the vertices of the polygon in order, with the last one in the center for even $\numVars$; 
the parts of the partition are then ordered by their lowest element under the lexicographical ordering of the pairs of cities $\{u, v\}$.)
Let 
$\mathcal{P}_{\mathrm{col}} = (P_1, \ldots, P_c, \ldots, P_{\numColors})$ 
be the resulting ordered partition, which we call a ``color partition'' of
the pairs of cities.
For example, for $\numVars=4$, the partition is
${\mathcal P}_{\mathrm{col}} = 
\left(
\left\{ \{1,2\}, \{3, 4\} \right\},
\left\{ \{1,3\}, \{2, 4\} \right\},
\left\{ \{1,4\}, \{2, 3\} \right\}
\right)$.
For different tour positions $i$, two partial unitaries
$\unitary{\permSwap, {i, \{u, v\}}}$ and 
$\unitary{\permSwap, {i', \{u', v'\}}}$ 
commute if $i$ and $i'$ are not consecutive ($|i - i' \mod \numVars| > 1$). 
Thus, for partitioning the positions, we may use the parity partition 
$\mathcal{P}_{\mathrm{par}}$, as defined in Section~\ref{sec:maxColSubGraph}. 
We can thus define the ``color-parity'' ordered partition 
$\mathcal{P}_{\mathrm{CP}} = \mathcal{P}_{\mathrm{col}} \times \mathcal{P}_{\mathrm{par}}$, with the induced lexicographical ordering of the parts.
The part $P_{c, \odd}$ contains all
$\unitary{\permSwap, {i, \{u, v\}}}$
such that $i$ is odd and edge $\{u, v\}$ is colored $c$, i.e.,\ in $P_c$,
and defines the unitary: 
\begin{equation}
\unitary{c, \odd} (\beta)
=
\prod_{(i, \{u, v\}) \in P_{c, \odd}}
\unitary{\permSwap, {i, \{u, v\}}}(\beta),
\end{equation}
where the ordering of the products does not matter because each term commutes.
It is a similar case for 
$P_{c, \even}$ and $\unitary{c, \even}$, and 
$P_{c, \last}$ and $\unitary{c, \last}$.
Thus, we have the full color-parity mixer:
\begin{equation}\label{eq:fullColorParitySwapMixer} 
\unitary{\colorParity} (\beta)
=
\unitary{\mathcal{P}_{\colorParity}\text{-}\permSwap} (\beta)
=
\prod_{P_{c, \pi} \in \mathcal{P}_{\colorParity}}
\unitary{c, \pi},
\end{equation}
where the unitaries $\{\unitary{c, \pi}\}$ are applied in the order they appear in $\mathcal{P}_{\colorParity}$.
The color-parity partition is optimal with respect to the number of parts in the partition 
(exactly so for even $\numVars$ and up to an additive factor of 2 for odd $\numVars$).
By construction, application of this mixer to any feasible state results
in a feasible state, thus satisfying the first design criterion. 
With regard to the second criterion, while 
a single application of this mixer will have nonzero transitions 
only between orderings that swap cities in tour positions no more than 
two apart, repeating
the mixer sufficiently many times results in nonzero transitions between
any two states representing orderings. More precisely, since any
ordering can be obtained from any other with no more than $\frac{n(n-1)}{2}$
adjacent swaps, alternating between odd and even swaps, $\frac{n(n-1)}{2}$
repeats suffice for any $0 < \beta < \pi/2$. 

\subsubsection{Compilation}\label{sec:tsp_compilation}
%
%\paragraph{Encoding orderings.}
\noindent \textbf{Encoding orderings.} \quad 
We encoded orderings in two stages: First into strings, and then into bits 
making use of the encodings from Section~\ref{sec:strings}. Here, we focus on 
a ``direct encoding'' as opposed to
the ``absolute encoding'' that is introduced in Section~\ref{sec:smst}.
Other encodings of orderings are possible, such as the Lehmer code and 
inversion tables. In direct encoding, an ordering  
$\order = \left(\orderElmt_1, \ldots, \orderElmt_{\numVars}\right)$
is encoded directly as a string~${[n]}^{\numVars}$ of integers.
Once in the form of strings, any of the string encodings introduced in Section~\ref{sec:strings} can be applied.
We applied the one-hot encoding with $n^2$ binary variables; 
the binary variable~$x_{j,u}$
indicates whether or not~$\orderElmt_j = u$ in the ordering, in other words,
whether city~$u$ is visited at the~$j$-th stop of the tour. 
%
%\paragraph{Phase separator.}
\vskip 1pc 
\noindent \textbf{Phase separator.} \quad
We used the phase function 
$\phaseFunc(\order) = 
4\objFunc(\order)  - (n-2) 
\sum_{u=1}^{\numVars} 
\sum_{v=1}^{\numVars} d(u, v)$,
which translates to a phase separator encoded as:
\begin{equation}
\label{eq:tsp_phase}
\phaseHam^{\enc}
=
\sum_{i=1}^{\numVars}
\sum_{u=1}^{\numVars}
\sum_{v=1}^{\numVars}
\distance(u, v)
Z_{u, i} Z_{v, i+1}.\;
\end{equation}
The phase separating unitary corresponding to Equation~\eqref{eq:tsp_phase}
imparts a phase determined by the sum of the distances between successive
cities to a state corresponding to a tour.  This unitary can be implemented
using $\numVars^2(\numVars-1)$ two-qubit gates, 
which mutually commute.
Using the same color-parity partition of the terms as for the color-parity
ordering swap mixer, this can be done in depth $2\numColors \leq 2\numVars$.
%
%\paragraph{Mixer.}
\vskip 1pc 
\noindent \textbf{Mixer.} \quad
The individual value-selective ordering swap partial mixer, which
swaps cities $u$, $v$ between tour positions $i$ and $j$, 
is expressed in the one-hot encoding as:
\begin{align}
\unitary{\permSwap, {\{i, j\}, \{u, v\}}}^{\enc} (\beta)
&=
\timeEv{\beta \Ham{\permSwap, {\{i, j\}, \{u, v\}}}}\;, \\
\Ham{\permSwap, {\{i, j\}, \{u, v\}}}^{\enc}
&=
\create_{u, i}
\create_{v, j}
\annihilate_{u, j}
\annihilate_{v, i}
+
\annihilate_{u, i}
\annihilate_{v, j}
\create_{u, j}
    \create_{v, i},\label{eq:H-PS-ij-uv-enc}
\end{align}
where:
\begin{align}
\create &= X + iY = \ket{1}\bra{0},\\
\annihilate &= X - iY = \ket{0}\bra{1}.
\end{align}

The $i$th adjacent value-selective swap partial mixer 
(Equation~\eqref{eq:mixer_adj_perm_swap}) is the special case:
\begin{align}
\label{eq:tsp_adj_hami}
\Ham{\permSwap, {i, \{u, v\}}}^{\enc}
&=
\create_{u, i}
\create_{v, i+1}
\annihilate_{u, i+1}
\annihilate_{v, i}
+
\annihilate_{u, i}
\annihilate_{v, i+1}
\create_{u, i+1}
\create_{v, i}\;.
\end{align}

Each of the two terms of the form $S^+S^+S^-S^-$ in Equation~\eqref{eq:tsp_adj_hami},
can be written as a sum of eight terms, each a product of $4$ Pauli operators 
(e.g.,\ $XXYY$).
The color-parity partitioned ordering swap mixer of
Equation~\eqref{eq:fullColorParitySwapMixer} can be implemented using 
$(\numVars-1) \binom{\numVars}{2}$ of these 4-qubit gates, implementable
in depth $2\kappa \leq 2\numVars$ in these gates. The two-qubit gate
circuit depth is at most $2\kappa$ times the depth of a compilation for such 
$4$-qubit gates.
%
%\paragraph{Initial state.} 
\vskip 1pc 
\noindent \textbf{Initial state.} \quad The initial state, an arbitrary ordering,
 can be prepared from the zero state~$\ket{00\dots0}$ using at most $\numVars$ single-qubit $X$ gates.

\subsection{Example: Single Machine Scheduling (SMS), Minimizing Total Squared Tardiness}\label{sec:smst2}
%\noindent {\bf Problem}  
%$(1|d_j|\sum w_j T_j^2)$. \quad 
\noindent {\bf Problem}. \quad   $(1|d_j|\sum w_j T_j^2).$ 
Given a set of $\numVars$ jobs with processing times $\procTimes$, deadlines 
$\deadlines$, % \in \integers_+$, 
and weights $\weights$, find a schedule minimizing the total weighted squared tardiness $\sum_{j=1}^{\numVars} w_j T_j^2$.  The tardiness of job $j$ with completion time $C_j$ is defined as $T_j=\max\{0,C_j-d_j\}$. Here, we took all quantities to be integers. 
%The tardiness of job $j$ is defined in terms of its completion time $C_j$ as $T_j=\max\{0,C_j-d_j\}$.

The configuration space and domain are the set of all orderings of the jobs.
Given an ordering $\order$ of the jobs in which job $i$ is the $\permElmt_i$-th job to start,
the corresponding schedule $\strtTimes(\order)$ is that in which each job starts as soon as the earlier jobs finish:
$\strtTime_j(\order) = \sum_{i=1}^{\permElmt_j-1} \procTime_{\orderElmt_i}$.

For a job $i$ starting at time $\strtTime_i$, consider the expression:
\begin{equation}
\min_{y_i \in [0, \deadline_i - \procTime_i]}
{\left(
\strtTime_i + \procTime_i - \deadline_i + y_i
\right)}^2
=
\begin{cases}
{\left(\strtTime_i + \procTime_i - \deadline_i\right)}^2, 
& \strtTime_i + \procTime_i > \deadline_i,\\
0, & \text{otherwise}.
\end{cases}
\end{equation}

When the ``slack'' variable $y_i \in [0, \deadline_i - \procTime_i]$ is minimized, this expression is equal to the square of the tardiness of job $i$.
Therefore, we recast SMS as the minimization of: 
\begin{equation}
\objFunc(\order, \y)
=
\sum_{i=1}^{\numVars}
\weight_i
{\left(
\strtTime_i(\order) + \procTime_i - \deadline_i + y_i
\right)}^2
\end{equation}
over the configuration space of orderings $\order$ and 
slack variables %$y_i$. % \in [0, \deadline_i - \procTime_i]$.
$\boldsymbol{y}$.

Using the direct one-hot encoding defined in {Section}~\ref{sec:tsp_compilation},
in which $x_{j, \alpha}$ indicates that job $j$ is the $\alpha$-th to start, this is equivalent to : 
\begin{equation}
\label{eq:smst_obj}
\objFunc(\x, \y)
=
\sum_{i=1}^{\numVars}
\weight_i
{\left(
\strtTime_i(\x) + \procTime_i - \deadline_i + y_i
\right)}^2,
\end{equation}
where: 
\begin{equation}
\strtTime_i(\x)
=
\sum_{\alpha=2}^{\numVars}
x_{i, \alpha}
\sum_{j \neq i}
\procTime_j
\sum_{\beta = 1}^{\alpha - 1}
x_{j, \beta}.
\end{equation}

Note Equation~\eqref{eq:smst_obj} may seem to be quartic;
however,  
the encoding constraints 
$\sum_{i} x_{i,\alpha}= \sum_{\alpha} x_{i, \alpha} = 1$ 
that come with the direct one-hot encoding imply that the quartic terms disappear in the full expansion.
The objective function is thus a cubic pseudo-Boolean function, which corresponds to a 3-local diagonal Hamiltonian for the phase separator.
%
%\paragraph{Mixer and initial state} 
\vskip 1pc 
\noindent \textbf{Mixer and initial state.} \quad We used the same initial state preparation, and the same mixer as in TSP for mixing the ordering, 
in addition to any of the single-qudit mixers from Section~\ref{sec:mixOpsMaxColSubgraph}
for each of the slack variables.
Because the ordering and slack mixers act on separate sets of qubits ($\x$ and $\y$), they can be implemented in parallel.
Note that the only requirement for the upper bound of the range of the slack variable $y_i$ is that it be \emph{at least} $\deadline_i - \procTime_i + 1$.
In particular, it could be $2^{\lceil \log_2 (\deadline_i - \procTime_i + 1)\rceil}$, allowing us to use the binary encoding without modification.

\subsection{SMS, Minimizing Total Tardiness}\label{sec:SMSTotalTardiness}\label{sec:smst}
\noindent {\bf Problem}. \quad $(1|d_j|\sum w_j T_j)$. 
Given a set of jobs with integer processing times $\procTimes$, deadlines $\deadlines$, %\in \integers_+$, 
and weights~$\weights$,
find a schedule minimizing the total weighted tardiness $\sum_{j=1}^{\numVars} w_j T_j$.

\subsubsection{Encoding and Mixer}
The configuration space is the set of orderings of the jobs; the domain is the same.
%
%\paragraph{Absolute and positional encodings.}
\vskip 1pc 
\noindent \textbf{Absolute and positional encodings.} \quad
In the ``absolute'' encoding of the ordering $\order = (\orderElmt_1, \ldots, \orderElmt_{\numVars})$, we assigned each item $i$ a value $\strtTime_i \in [0, \horizon]$, where the ``horizon'' $\horizon$ is a parameter of the encodings, such that for all $i < j$, $\strtTime_{\orderElmt_i} < \strtTime_{\orderElmt_j}$.
In certain cases, there will be an item-specific horizon $\horizon_i$ such that $\strtTime_i \in [0, \horizon_i]$.
Note that in general, the relationship between encoded states and the orderings they encode is not injective, but it will be in the domains to which we apply it.
Once the ordering $\order$ is encoded as a string 
$\strtTimes(\order) \in \times_{i=1}^{\numVars} [0, \horizon_i]$, the resulting string can be encoded using any of the string encodings previously introduced.
We call the special case of the ``absolute'' encoding with $\horizon=\numVars$ the ``positional'' encoding; 
using the one-hot encoding of the resulting strings, the ``direct'' and ``positional'' encodings are the~same.
%
%\paragraph{Time-swap mixer.}
\vskip 1pc 
\noindent \textbf{Time-swap mixer.} \quad
We now introduce a mixer that is specific to the absolute encoding in which there is a single horizon $\horizon$ and each job $i$ has a processing time $\procTime_i$.
Let the horizon be $\horizon = \sum_{i=1}^{\numVars} \procTime_i$.
Each job can start between time $0$ and $\lastStrtTime_i = \horizon - \procTime_i$. (Other optimizations may be made on an instance-specific basis, though we neglect elaborating on these for ease of exposition.)
We used a ``time-swap'' partial mixer that acts on absolutely encoded orderings:
\begin{equation}
\Ham{\timeSwap, {t, \{i, j\}}}
=
\Ket{t+\procTime_j, t}
\Bra{t, t+\procTime_i}_{i, j}
+
\Ket{t, t + \procTime_i}
\Bra{t+\procTime_j, t}_{i, j},
\end{equation}
which swaps jobs $i$ and $j$ when they are scheduled immediately after one another, with the earlier one starting at time $t$.
To swap them regardless of the time at which the earlier one starts, we used:
\vspace{6pt}
\begin{equation}
\Ham{\timeSwap, \{i, j\}}
=
\sum_{t=0}^{\horizon - \procTime_i - \procTime_j}
\Ham{\timeSwap, {t, \{i, j\}}}.
\end{equation}

The \emph{simultaneous} time-swap mixer is constructed as usual:
\begin{equation}
\unitary{\seqMixer{\timeSwap}}(\beta)
=
\timeEv{\beta \Ham{\timeSwap}},
\end{equation}
where 
$\Ham{\timeSwap} = 
\sum_{\{i, j\} \in \binom{[n]}{2}} \Ham{\timeSwap, {\{i, j\}}}$.
Note that while the simultaneous versions of the total time-swap and adjacent permutation mixers are exactly the same, and in particular:
\begin{equation}
\Ham{\timeSwap, {\{i, i'\}}}
=
\sum_{j = 1}^{\numVars -1}
\Ham{\permSwap, {j, \{i, i'\}}},
\end{equation}
the individual partial mixer $\Ham{\timeSwap, {t, \{i, j\}}}$ has no equivalent that acts on the unencoded ordering, because it acts depending on the total processing times of the preceding jobs rather than their number.

Now consider the ``time--color'' ordered partition $\mathcal{P}_{\timeColor} = [0, \lastStrtTime] \times [\kappa]$, 
where, as in {Section}~\ref{sec:tsp_mapping}, 
we used a particular $\numColors$-edge-coloring of the complete graph.
(Again, further optimizations may be made on an instance-specific basis.)
The partition $P_{t, c}$ contains the partial time-swap mixers $\unitary{\timeSwap, {t, \{i, j\}}}$ for which the edge $\{i, j\}$ is colored $c$.
The full ``time--color'' mixer is: 
\begin{equation}
\unitary{\timeSwap}(\beta)
=
\prod_{t=0}^{\lastStrtTime}
\prod_{c=1}^{\numColors}
\unitary{(t, c)\text{-}\timeSwap}(\beta),
\end{equation}
where 
$\unitary{(t, c)\text{-}\timeSwap}$ is the product of the (mutually commuting)
partial mixers in the part $P_{t, c}$.

\subsubsection{Mapping and Compilation}
We used  absolute one-hot encoding, in which the ordering is encoded as a string using  absolute encoding and then the string is further encoded using  one-hot encoding.
Specifically, we encoded an ordering $\order$ into 
$\prod_i (\lastStrtTime_i - 1) \leq \numVars \horizon$ qubits, 
where qubit $(i, t)$ indicates if job $i$ starts at time $t$:
\begin{equation}
\ket{\order} \mapsto 
\bigotimes_{i=1}^{\numVars}
\left[
{\ket{0}}^{\otimes (\strtTime_i - 1)}
\ket{1}
{\ket{0}}^{\otimes (\lastStrtTime_i - \strtTime_i)}
\right].
\end{equation}
%
%\paragraph{Phase separator.}
\vskip 1pc 
\noindent \textbf{Phase separator.} \quad
The objective function is the weighted total tardiness:
\begin{equation}
\label{eq:smstobj}
f(\order)
=
\sum_{i=1}^{\numVars}
\weight_i \max\{0, \deadline_i - \strtTime_i(\order) - \procTime_i\}.
\end{equation}

This yields the encoded phase Hamiltonian:
\begin{equation}
\label{eq:smstHp}
\phaseHam^{\enc} =
\sum_{i=1}^{\numVars}
\weight_i
\sum_{t=\deadline_i - \procTime_i + 1}^{\lastStrtTime_i}
(t + \procTime_i - \deadline_i) Z_{i, t}.
\end{equation}
%
%\paragraph{Mixer.}
\vskip 1pc 
\noindent \textbf{Mixer.} \quad
The partial time-swap mixer:
\begin{equation}
\Ham{\timeSwap, {t, \{i, j\}}}
=
\Ket{t+\procTime_j, t}
\Bra{t, t+\procTime_i}_{i, j} + \Ket{t,t+\procTime_i}
\Bra{t+\procTime_j,t}_{i, j}
\end{equation}
in  absolute one-hot encoding is equivalent to:
\begin{equation}
\Ham{\timeSwap, {t, \{i, j\}}}^{\enc}
=
\create_{i, t + \procTime_j}
\create_{j, t}
\annihilate_{i, t}
\annihilate_{j, t + \procTime_i}
+
\create_{i, t}
\create_{j, t+\procTime_i}
\annihilate_{i, t + \procTime_j}
\annihilate_{j, t}.
\end{equation}

Using the time--color partitioned time-swap mixer, 
this corresponds to a circuit of $\lastStrtTime \binom{\numVars}{2}$ 4-qubit gates in depth $\lastStrtTime \numColors$.
%$\Ham{\timeSwap, {t, \{i, j\}}}$ corresponds to edges $\{i, j\}$ colored $c$ as explained above. 
%
%\paragraph{Initial state.}
\vskip 1pc 
\noindent \textbf{Initial state.} \quad
We used an arbitrary ordering of the jobs as the initial state.

\subsection{SMS, with Release Dates}\label{sec:smswr}
%\vskip 1pc
\noindent {\bf Problem}. \quad $(1|d_j,r_j|*)$.  
Given a set of jobs with integer processing times~$\procTimes$, deadlines~$\deadlines$, 
release times~$\releaseTimes$, and weights~$\weights$,
find a schedule that minimizes some function of the tardiness, 
such that each job starts no earlier than its release time.

We considered SMS problems with release times $r_j$.
We do not specify the objective function here, as any of those used in previous sections are still applicable.
Our focus in this section is to introduce a modification of the time-swap mixer that preserves feasibility with release times. 

Let the horizon $\horizon$ be some upper bound on the maximum completion time, e.g., $\max_j d_j + \sum_j \procTime_j$.
Let $b_j = \horizon + \sum_{i=1}^{j-1} \procTime_i$ be a special ``buffer'' time for job $j$.
Let $\window_j = [\releaseTime_j, \horizon - \procTime_j]$ be the window of times in which job $j$ can start.

Consider the configuration space
$\times_{j=1}^{\numVars} \left[\window_j \cup \{b_j\}\right]$, i.e., schedules $\{\strtTimes\}$ in which a job is scheduled either between its release time and the horizon or at its buffer time slot.
The domain is the subset of the configuration space that satisfies the problem constraint that no two jobs overlap.

\subsubsection{Partial Mixer: Controlled Null-Swap Mixer}
We now introduce a mixer, specifically a controlled null-swap mixer as used for graph coloring in Section~\ref{sec:maxColIndSubGraph}, 
that preserves feasibility and avoids getting ``stuck'':
\begin{equation}
\label{eq:HnullSwapSMSwr}
\Ham{\nullSwap, {i, t}}
=
\left(
\Ket{b_i}\Bra{t}_i
+
\Ket{t}\Ket{b_i}
\right)
\prod_{j \neq i}
\left(
\identity_j - 
\sum_{t' \in \neighborFunc_{i, t} (j)}
\Ket{t'}\Bra{t'}_j
\right),
\end{equation}
which corresponds to the unitary:
\begin{equation}
\label{eq:UnullSwapSMSwr}
\unitary{\nullSwap, {i, t}}(\beta)
=
\Lambda_{\bar{\strtTimes}_{\neighborFunc_{i, t}(j)}}
\left(
\timeEv{\beta \left(\Ket{b_i} \Bra{t}_i + \Ket{t}\Bra{b_i}\right)}
+
\sum_{t' \in \window_i \setminus \{t\}}
\Ket{t'} \Bra{t'}
\right)
\end{equation}
where:
\begin{equation}
\neighborFunc_{i,t}(j) 
=
\left[
\max\{t-\procTime_j+1,\releaseTime_j\},
t+\procTime_i - 1
\right]
\end{equation}
is the temporal ``neighborhood'' of job $i$ at time $t$ with respect to job $j$, i.e., the set of times at which starting job $j$ would conflict with job $i$,
and
$\strtTimes_{\neighborFunc_{i, t}} = 
{\left(
{\left( \strtTime_{j, t'} \right)}_{t' \in \neighborFunc_{i, t}(j)}
\right)}_{j\neq i}$.
The role of the buffer site is similar to the ``uncolored'' option in finding the maximal colorable induced subgraph in Section~\ref{sec:maxColIndSubGraph}.
Such mixing terms enables jobs to move freely in and out $[0,T_0]$ without inducing job overlap, hence enabling exploration of the whole feasible subspace.
Recall the controlled unitary notation $\control_{\bf y}(Q)$ is defined in Equation~\eqref{eq:ctrlQ} in Section~\ref{sec:MaxIndSet}.

\subsubsection{Encoding and Compilation}
Given the $\sum_{i=1}^{\numVars} \left|\window_i\right|$ partial mixers, one for each job $i$ and time $t \in \window_i$, we can define simultaneous and sequential mixers as above.
Using  one-hot encoding, the partial mixer Hamiltonian is:
\begin{equation}
\Ham{\nullSwap, {i, t}}^{\enc}=
\left(\prod_{j\ne i}\prod_{t'\in \neighborFunc_{i,t}(j)}\frac{I+Z_{j,t'}}{2}\right) 
\left(X_{i,t}X_{i,b_i}+Y_{i,t}Y_{i,b_i}\right)\;.
\end{equation}

The controlled null-swap can be further compiled using ancilla qubits, as in Section~\ref{sec:MaxIndSet}.
While the cost of compilation could be bounded based on the degree of the graph, %Please check meaning has been retained.
the overlap of the various partial mixers may be more complicated (with respect to partitioning into disjoint sets) and expensive (with respect to number of gates and ancilla qubits) depending on the SMS instance.
%
%
%{\bf Objective function.}
%\paragraph{Objective function.}
\vskip 1pc 
\noindent \textbf{Objective function.} \quad
As an example objective function, we again considered minimizing the weighted total tardiness, Equation~\eqref{eq:smstobj}.
The one-hot encoded phase Hamiltonian takes the form of Equation~\eqref{eq:smstHp},
with $b_j$ included in the summation range of $t$:
\begin{equation}
\label{eq:smswrHp}
\phaseHam^{\enc} =
\sum_{i=1}^{\numVars}
\weight_i
\left[
(b_{i} - d_i) Z_{i, b_i}
+
\sum_{t=\deadline_i - \procTime_i + 1}^{\lastStrtTime_i}
(t + \procTime_i - \deadline_i) Z_{i, t}, 
\right]
\end{equation}
and the corresponding unitary can be implemented with $\sum_j (h_j-d_j+1)$ single qubit $Z$-rotation gates.
%
%{\bf Initial state.}
%\paragraph{Initial state.}
\vskip 1pc 
\noindent \textbf{Initial state.} \quad
Any feasible schedule can be used as the initial state.
In particular, we used a greedy earliest-release-date schedule.
Assume without loss of generality that the jobs are ordered by their release times, i.e., \ $\releaseTime_1 \leq \releaseTime_2 \leq \cdots \leq \releaseTime_{\numVars}$.
Then set $\strtTime_1 = \releaseTime_1$ and recursively set 
$\strtTime_i = \max\{\releaseTime_i, \strtTime_{i-1} + \procTime_{i-1}\}$,
which is feasible though likely suboptimal.

\subsubsection{Mapping Variants}
In the construction above, each job $j$ is assigned a ``buffer'' time $b_j$, and a phase $b_j - d_j$ applied whenever that job is scheduled at its buffer time.
The factor $b_j - d_j$ is arbitrary.
Rather than considering schedules in which some jobs are at their buffer time, one could consider ``partial schedules'', in which a job is either scheduled at a time between $0$ and $\horizon$ or is in its buffer.
The phase applied when a job is in its buffer need not be $b_j - d_j$ but in fact can be arbitrary, e.g., some common ``buffer phase factor'' $B$.
\mbox{In this case}, we must define a scheme for associating each partial schedule with a canonical complete schedule, e.g., greedily starting the buffered jobs after those that are already scheduled.
In this way, the states corresponding to partial schedules can still be considered as part of the domain.

\section{Conclusions}\label{sec:conclusions}

We introduced a quantum alternating operator ansatz (QAOA), an
extension of Farhi et al.'s quantum approximate optimization algorithm, 
and showed how to apply the ansatz to a diverse set of optimization problems
with hard constraints.  The essence of this extension is the consideration of
general parameterized families of unitaries  rather than only those
corresponding to the time evolution of a local Hamiltonian, which allows it to
represent a larger, and potentially more useful, set of states than the
original formulation.
In particular, refocusing on unitaries rather than Hamiltonians in the specification allows for more efficiently implementable mixing operators.

The original algorithm is already a leading candidate quantum heuristic for exploration on near-term hardware;
our extension makes early testing on emerging gate-model quantum 
processors possible for a wider array of problems at an earlier stage.
After formally introducing the ansatz, and~providing design criteria for
its components, we worked through a number of examples in detail, 
illustrating design techniques and exhibiting a variety of mixing operators.
In the appendix, we~provide a compendium of QAOA mappings 
for over $20$ problems, organized by type of mixer.

While the approach of designing mixing operators to keep the evolution
within the feasible subspace appears quite general, as illustrated by
the wide variety of examples we have worked out, it~is not universally
applicable. 
Many of the problems in Zuckerman~\cite{zuckerman1996unapproximable} have the form of
optimizing a~quantity within a feasible subspace consisting of the 
solutions to an NP-complete problem. %Please define NP.
Not only is an initial starting state
(corresponding to one or more feasible solutions) hard to find, 
designing the mixing operator is also problematic.
Even given a set of feasible solutions to an NP-complete problem, it is
typically computationally difficult to find 
another~\cite{papadimitriou1994complexity}, making it
difficult to design a~mixer that fully explores the feasible subspace.
The situation here is somewhat subtle, with it being easy to show
in the case of SAT that finding a second solution when given a first
remains NP-complete, but for a Hamiltonian cycle on cubic graphs, given a first
solution, a second is easy to find (but not a~third). 
See References ~\cite{yato2003complexity,ueda1996np} for results
on the complexity of ``another solution problems'' (ASPs).  
This difficulty in mapping Zuckerman's problems to QAOA further illustrates
reasons for the difficulty of these approximate optimization problems.

While we have given basic design criteria for initial states, mixing
operators, and phase-separation operators, we have barely scratched
the surface in terms of which possibilities perform better than others. 
For most of the example problems, we discussed multiple mixers, 
coming from different 
partitions and orderings of partial mixers, different choices related
to connectivity, or different numbers of repeats of operators. 
Analytic, numerical, and ultimately experimental work is required to understand 
which of these mixers are most effective, and also to determine potential 
trade-offs with respect to robustness to noise or control error, efficiency of
implementation, effectiveness of the mixing, and difficulty in optimizing the parameters.
Similar questions arise with respect to choosing an initial state. 
An important future research direction is to explore the trade-offs between different mixing operator designs, in particular how a mixer's connectivity relates to its effectiveness in different settings.
Of particular interest is whether or not allocating resources to more complicated mixing operators at the expense of overall QAOA circuit depth results in enhanced performance. 
Similarly, whether $k$-local mixers can be sufficiently advantageous to warrant their increased gate complexity remains to be explored.
How the efficiency of different mixers depends on the underlying mixing graph (e.g., ring vs. fully-connected mixers) is studied for graph coloring problems in %an upcoming study~\cite{Wang19XY}. 
a forthcoming paper. 

%mixer
%Once determined on a mixing Hamiltonian,
%we in the paper introduced different versions (simultaneous vs partitioned, ring vs complete-graph, etc)
%of mixing unitaries.  In particular,
%the connectivity of the graph underlying the mixer may greatly affect the ``mixing" efficiency,
%as studied in Ref.~\cite{Wang19XY} in detail for the graph coloring problem.
%In the mixer designing, one also needs to be cautious that,
%for a general optimization problem,
%When the problem structure is effectively not explored (or not explored enough),
%it is generally expected that the Grover speedup would be the upperbound.
%We would like to point out that, even in this case
%how to achieve the Grover's speedup in different framework is still highly non-trivial:
%for example a different quantum algorithm achieving the same speedup
%may provide advantages in experimental realizations
%or higher adaptivity to other algorithms.

%For problems with hard constraints, 
%the topology of the feasible subspace within the full Hilbert space of the qubits 
%should have a big impact on how complicated the mixers need to be. 
%For example, a non-local mixer may require deep circuits 
%when compiled to local gates,
%which may dominate the complexity of the algorithm 
%and lead to little or no quantum advantage.

Effective parameter setting remains a critical, but mostly open, area
of research. 
While for fixed~$p$, brute-force search
of parameter search space was proposed~\cite{Farhi2014}, it is practical
only for small values of~$p$; as~$p$ increases, the parameter optimization
becomes inefficient due to the curse of dimensionality, 
Guerreschi et al.~\cite{guerreschi2017practical} providing a detailed
analysis. 
In certain simple~\cite{Wang17} or highly symmetric~\cite{Jiang17} cases,
some insight into parameter setting for~$p>1$ has been obtained, but even in 
the simplest cases, understanding good choices of parameters seems
nontrivial~\cite{Wang17}. 
Improved parameter setting protocols may come from adapting techniques from existing control theory and parameter optimization methods, and 
through using insights gained from classical simulation of quantum circuits and experimentation on quantum hardware as it becomes available.
In particular, classical simulation of the quantum circuits can take advantage
of the local structure of the objective function (when it is indeed local)
and the feasibility of classically simulating the measurement of 
$\log(n)$-qubits in a QAOA circuit by adapting results for IQP 
circuits~\cite{bremner2010classical}.%Please define IQP.

To run on near-term quantum hardware, further compilation will be 
required. 
In many cases, we have left the compilation at the level
of multiqubit operators that need to be further compiled to the gate set natively available on the hardware, most likely certain one- and two-qubit gates.
Clearly, issues such as %hardware 
connectivity will be critical to the implementability and success of QAOA circuits. 
Furthermore, near-term hardware will have additional restrictions,
including which qubits each gate can be applied to, the duration and fidelity of the gates, and cross-talk, among others.
This necessitates additional compilation, especially to optimize success probability on pre-fault-tolerant devices.
Other architectures, e.g., ones based on higher-dimensional qudits, may prompt other sorts of compilations as well. 
Recently,
approaches for compiling quantum circuits, including QAOA circuits, to realistic
hardware have been explored~\cite{Venturelli17,Booth18,li2018tackling}, but that research direction remains 
open to innovation. 
An interesting research direction is to explore the power of quantum alternating operator ansatz states built from natively available gates and directly incorporating the connectivity of  the underlying hardware.

\textls[-10]{The robustness of these circuits to realistic error requires further
exploration.
In the near term, 
the question is how robust QAOA circuits are to realistic noise and how best to incorporate resource-efficient techniques to improve the robustness.
In the longer term, 
the question becomes how to best incorporate the full spectrum of error correction techniques into this framework.}

We expect some cross-fertilization between research on QAOA and research on AQO and quantum 
annealing, especially in the $p \gg 0$ regime. 
A particularly 
fruitful area of further study would be to build on the
results of Yang et al.~\cite{Shabani16}. 
They used Pontryagin's minimization 
principle to show that a bang--bang schedule is (essentially) always
optimal, providing support for QAOA, and gate-based approaches, 
generally. 
However, the argument does not provide an efficient means
to find such a schedule. 
Thus, how to find effective schedules remains open in the AQO realm, with it currently being completely open
as to whether allowing non-bang--bang schedules makes the finding
of good schedule parameters easier. 
Similarly, exploiting certain
structural commonalities with VQE %Please define VQE.
is likely to be fruitful.
In the small $p$ regime, we expect further cross-fertilization
between QAOA and other models being considered for early quantum
supremacy experiments, such as random circuits, boson sampling, and 
IQP circuits. 
Results for IQP circuits, whose close
structural similarities with QAOA circuits have already provided insights~\cite{Farhi2016}, and whose advanced status with respect to error analysis
and fault tolerance~\cite{bremner2016average,bremner2016achieving} are 
likely to prove especially useful. %This sentence is incomplete and it is impossible for me to edit without guesswork: Results for IQP circuits...what? Everything that comes after is a subordinate clause and the main clause is missing. Please add a comma at the end after "useful" and finish the sentence.

A number of extensions are possible. 
We briefly mentioned two classes.
The first is hybrid methods. 
As one example, some other algorithm, either classical or quantum,
could be run first in order to provide a good initial state, 
and QAOA then used to improve upon it; this could be repeated several times with different initial states when the other algorithm is stochastic. 
Another example of a hybrid method is a parameter-setting protocol, in which 
a classical algorithm uses the results of measurements during and after runs of a QAOA circuit to iteratively improve the parameters.
A second class of extensions is
versions of QAOA with many parameters.\@
For example, we introduced mixing operators that are repeated applications of some basic mixer; each application could use a different parameter.
The same could be done with mixers consisting of many partial mixers.
Until we have a better grasp of parameter setting in the 
single parameter per mixing operator case, and for the effectiveness of the
different mixers, it makes sense to restrict to the simple case. 
The one
exception is to take advantage of the specific gates natively available,
and to use essentially a VQE approach, as suggested in Reference~\cite{Farhi2017}. 
That~approach makes excellent use of the capabilities of near-term
hardware but may be more limited than single-parameter-per-phase QAOA in what it can tell us about 
parameter setting and the design of scalable quantum heuristics.

The biggest open question is the effectiveness of this approach as a quantum
heuristic, and its potential impact in broadening the array of established
applications of quantum computing. 
In~particular, whether or not this approach---or quantum algorithms generally---can provide advantages for approximate optimization in terms of either approximability or computational efficiency remains a~fundamental yet tantalizing open problem. 
While obtaining further analytic results may
be possible in some cases, in general, we will have to try it out and see. 
Improved simulation techniques for quantum circuits, including potentially 
approaches tailored to QAOA circuits, can provide some insight, but the
ultimate test will be experimentation on quantum hardware itself.

\section{Acknowledgements}
The authors would like to thank the other members of NASA's QuAIL group
for useful discussions about QAOA-related topics, particularly
Salvatore Mandr\`a and Zhang Jiang.
S.H.~was supported by NASA under award NNX12AK33A. He thanks 
Al Aho for additional support and guidance, and  
the Universities Space Research Association
(USRA) and the NASA Ames Research Center for the opportunity to participate in
the USRA Feynman Quantum Computing Academy program that enabled this research. 
Z.W. and D.V. were supported by NASA Academic Mission Services, contract number NNA16BD14C. 
B.O. was supported by a NASA Space Technology Research Fellowship. 
The authors would like to acknowledge support from the NASA Advanced
Exploration Systems program and the NASA Ames Research Center.
The views and conclusions contained herein
are those of the authors and should not be interpreted as necessarily
representing the official policies or endorsements, either expressed or
implied, of the U.S. Government.
The U.S. Government is authorized to reproduce and distribute reprints for
Governmental purpose notwithstanding any copyright annotation thereon.

\appendix

\section{Compendium of Mappings and Compilations}\label{sec:compendium}

We summarize QAOA mappings for a variety of problems. All problems considered
are in NPO~\cite{trevisan2004inapproximability,Ausiello2012complexity}. %Please define NPO.
Problem names are compact versions of the names in 
Ausiello et al.~\cite{Ausiello2012complexity} unless a different reference 
is given. Most of the mappings are new to this paper; for the exceptions,
a reference is given, though in all such cases only H-QAOA mappings have been considered previously.
Approximability results are taken from
Ausiello et al.~\cite{Ausiello2012complexity} unless otherwise mentioned.

We specify partial mixing Hamiltonians for each problem, which can then
be used to define two types of full mixers, simultaneous mixers in H-QAOA  that correspond to time evolution under the sum of the partial mixing Hamiltonians, and
partitioned mixers (which in general are not
in H-QAOA) that correspond to products of unitary operators defined by all
the partial mixers in an~order specified by an ordered partition, possibly
repeated. We will not specify the full mixers, since they are straightforward
to derive from the partial mixers. For each problem, we provide a compilation
with resource counts for at least one mixer or a class of mixers. In most
cases, we mention only one partial mixer and compilation, though as we
have seen, there are many possibilities for partial mixers, and for
compilations of the operators. For the phase operators, we employed affine transformations of the objective functions where possible to eliminate constants and identitied terms as described in the main text. A rigorous approach to constructing phase operators is given in Reference~\cite{hadfield2018representation}. 

The resource counts given are upper bounds. We have not worked to
find optimal compilations. For the simpler problems that have the
potential for implementation in the near term, we give exact resource
bounds when we have such results, rather than only giving the complexity.
We give bounds for the number of computational
qubits required, and also for the number of ancilla qubits, when used.
We give resource counts for the number of arbitrary two-qubit gates 
required to implement a~given operator, merging single qubit operations
into two-qubit gates before or after when possible. We computed the depth
for such circuits, which gives a lower bound on the circuit depth on near-term 
hardware, which will generally be higher due to %topological 
connectivity and other constraints. 
In many cases, we did not compile all the way down to two-qubit gates, 
but instead specified the resource count in terms of multiqubit 
operators which would then yield a two-qubit gate count and depth 
given compilations of those multiqubit mixers.
Throughout,  the initial state is always obtained by applying a~constant-depth circuit to the state $\Ket{0}^{\otimes *}$.

%Recall that 
An algorithm $\mathcal{A}$ achieves an \textit{approximation ratio} $r \leq 1$ if
for all instances $x$ of a maximization problem it satisfies
$\mathcal{A}(x)/\OPT(x) \geq r$, where $\OPT(x)$ is the optimal solution of the
given instance. Both~cases in which $r$ is constant and in which $r$ is a 
function of the problem parameters were considered. For minimization problems, 
the approximation ratio is similarly defined with $r \geq 1$ and 
$\mathcal{A}(x)/\OPT(x) \leq r$. Unfortunately, there are multiple
conventions in the literature, with the approximation ratio given
as $1/r$ for minimization, maximization, or both. 
Fortunately, there is no ambiguity for a given value of $r$ to which convention is being used.
Here, instead of being
internally consistent, we simply took the ratio as stated in the 
reference cited so as to facilitate easy comparison with the literature.
All approximation results below concern efficient classical algorithms. 
Quantifying the %approximation ratio achieved by 
performance of QAOA for these problems remains an important future research direction. 

\textls[-20]{For all problems on a graph $G=(V,E)$, let $|V|=n$, $|E|=m$, 
and $\degree_G$ be the maximum vertex~degree.  }

\subsection{Bit-Flip ($X$) Mixers}
%\input{sections/appendix/compendium/all-states-feasible}
%\noindent 
In this section, we consider problems where all states of the configuration space are feasible, and hence the
original QAOA construction may be used. Specifically, for all problems we can
use the~following:\\
{\bf Variables:} $n$ binary variables.\\
{\bf Initial state:} $\ket{0}^{\otimes n}$ or $\ket{+}^{\otimes n}$.\\
{\bf Mixer:} The standard mixer $\mixUnitary^H(\beta) = e^{-i\beta B}$,\\ where $B=\sum_{j=1}^m X_j$, which can be implemented with depth $1$. 
(Since all terms commute and act on separate qubits, we do not need
to consider partitioned variants of the standard mixer.)\\
{\bf Phase separator:} $\phaseUnitary(\gamma) = \exp[-i\gamma \phaseHam]$,
where we specify $\phaseHam$ for each problem.\\
All problems in this section can be trivially extended to their weighted version 
by multiplying the terms of the phase Hamiltonian by the corresponding weights.

\subsubsection{Maximum Cut}\label{sec:comp:maxcut}
\noindent {\bf Problem:} 
Given a graph $G=(V,E)$, 
find a subset $S\subset V$ such that the number of edges between $S$ and $V\setminus S$ is the largest.\\
{\bf Prior QAOA work:}~%\cite{Farhi2014,Wang17}.\\
Numerical \cite{Farhi2014} and analytical \cite{Wang17} results obtained for $p=1$.\\
{\bf Approximability:}
APX-complete~\cite{papadimitriou1991, khanna1998syntactic}. %Please define APX.
NP-hard to approximate better than  $16/17$~\cite{Hastad2001}. Semidefinite programming achieves $0.8785$~\cite{goemans1995improved}, which is optimal under the unique games conjecture~\cite{khot2007optimal}. 
On~bounded degree graphs with
$D_G \geq 3$, it can be approximated to within
$0.8785 + O (D_G)$~\cite{feige2002improved}, in particular to $0.921$ for $D_G=3$, but remains APX-complete~\cite{papadimitriou1991}. 
\\
{\bf Configuration space, Domain, and Encoding:} ${\{0,1\}}^n$, indicating whether each vertex is in $S$ or not.\\
{\bf Objective:} 
$\max \left|
\left\{
\{u, v\} \in E : u \in S, v \notin S
\right\}
\right|$.\\
{\bf Phase separator}: 
$
\phaseHam = \sum_{\{u,v\}\in E} Z_u Z_v
$.
\\
{\bf Resource count for phase separator:} 
$m$ gates with depth at most $D_G+1$.\\
{\bf Variant:} 
Directed-MaxCut (Max-Directed-Cut), where we seek to maximize the number of the \emph{directed} edges leaving $S$. 
The phase separator is replaced by 
$
\phaseHam=\sum_{(u,v)\in E}  (Z_u - Z_v + Z_u Z_v)
$.

\subsubsection{Max-$\ell$-SAT}
\noindent {\bf Problem:} Given $m$ disjunctive clauses over $n$ Boolean variables $\bf x$, 
where each clause contains at most $\ell \geq 2$ literals,
find a variable assignment that maximizes the number of satisfied clauses. 
Let $\x_{\alpha}$ be the variables involved in clause $\alpha$ and $C_{\alpha}(\x_{\alpha}) = 1$ ($0$) if the state of $\x_{\alpha}$ (resp., does not) satisfies the clause.\\
{\bf Prior QAOA work:} Considered in Reference~\cite{Wecker2016training}.\\
{\bf Approximability:} APX-complete~\cite{papadimitriou1991}. 
The best classical approximation ratio for Max-$2$-Sat 
is $0.940$~\cite{lewin2002improved}, which cannot be improved beyond $0.943$ 
under the unique games conjecture~\cite{khot2007optimal}, or beyond $0.954$ 
unless $P=NP$~\cite{Hastad2001}. 
For Max-$3$-Sat, an efficient $7/8$-approximation~\cite{karloff19977} is known, which is optimal unless $P=NP$~\cite{Hastad2001}. 
Remains APX-complete when any literal appears in at most $3$ clauses~\cite{Ausiello2012complexity}. \\
{\bf Objective:} $\max\sum_{j=\alpha}^m C_{\alpha}(\x_{\alpha})$.\\
{\bf Phase separator:} 
$\phaseHam = \sum_{\alpha=1}^m H_{C_{\alpha}}(\x_{\alpha})$.\\
{\bf Resource count for phase separator:} 
Can be implemented with $\binom{\ell}{k}$ many $k$-local $ZZ \cdots Z$ gates,
for $k=1,\ldots,\ell$,
overall requiring at most $O(m2^\ell)$ two-qubit gates. 

\subsubsection{Min-$\ell$-SAT}
\noindent {\bf Problem:} 
The same as Max-$\ell$-SAT, except with minimization instead of maximization of the number of satisfied clauses.\\
{\bf Approximability:}  APX-complete~\cite{kohli1994minimum}. 
Unless $P=NP$, no classical algorithm can do better than $1.3606$ for MinSAT~\cite{dinur2002importance}, 
or, for Min-$\ell$-SAT, better than $15/14\simeq 1.0714$ for $k=2$ or $7/6\simeq 1.1667$ for arbitrary~$\ell$~\cite{avidor2005approximating}. 
On the other hand, constructive algorithms are known which achieve approximation ratios of $1.1037$, $1.2136$, 
and $1-2^{1-\ell}$ for Min-$2$-SAT, Min-$3$-SAT, and Min-$\ell$-SAT, respectively~\cite{avidor2005approximating,bertsimas1999dependent}. \\
{\bf Mapping:} The same as Max-$\ell$-SAT.

\subsubsection{Max-Not-All-Equal-$\ell$-SAT (NAE-$\ell$-SAT)} 
\noindent {\bf Problem:} 
The same as in Max-$\ell$-SAT, except that in NAE-$\ell$-SAT, 
each clause is satisfied only if all $\ell\geq 3$ variables 
do not all have the same value. \\
{\bf Approximability:} APX-complete~\cite{papadimitriou1991}. 
A classical 
$1.38$-approximation algorithm is known~\cite{andersson1998better}. For $\ell=3$ approximable to $1.138$ but no better than $1.090$~\cite{zwick1998approximation,Ausiello2012complexity}. \\
{\bf Mapping:} Same as Max-$\ell$-SAT but with a slight modification of the phase operator; see, e.g., \cite{hadfield2018representation}.\\
{\bf Variant:} 
Similar considerations apply to Max-$1$-in-$\ell$-Sat, in which
a clause is satisfied when exactly one of its variables is true.

\subsubsection{Set Splitting}
\noindent {\bf Problem:} Given a set $\cal S$ and a collection of subsets ${\{S_j\}}_{j=1}^m$,
seek a partition ${\cal S}={\cal S}_1 \cup ({\cal S}\setminus {\cal S}_1)$ 
that maximizes the number of split sets, i.e.,  subsets $S_j$ 
with elements in both ${\cal S}_1$ and ${\cal S}\setminus {\cal S}_1$. \\
{\bf Approximability:} 
APX-complete~\cite{petrank1994hardness}. 
Can be approximated to $0.7499$~\cite{zhang2004improved}. 
Remains APX-complete if each $S_j$ is restricted to having at most or exactly $k\geq 2$ elements~\cite{lovasz1973coverings}. 
For each $S_j$ having exactly $k \geq 4$ elements, unless $P=NP$, there is no efficient classical algorithm that does essentially better than a random partition~\cite{Hastad2001, guruswami2004inapproximability}.
The generalization MaxHypergraphCut, in which each subset is given a weight and we seek to maximize the total weight of the split sets, can be approximated to $0.7499$~\cite{zhang2004improved}.\\
{\bf Reduction to Not-All-Equal-$\ell$-SAT:}
This problem is a special case of NAE-$\ell$-SAT, where none of the variables are negated.

\subsubsection{E3Lin2}\label{sec:comp:e3lin2}
\noindent {\bf Problem:} 
Given a set of $m$ three-variable equations 
${\cal A}=\{A_j\}$, over $n$ binary variables $\x \in {\{0,1\}}^n$, 
where each equation $A_j$ is of the form
$x_{a_{1,j}} + x_{a_{2,j}} + x_{a_{3,j}}= b_j \mod 2$ % or 
% $x_{a_1} + x_{a_2} + x_{a_3}= b \mod 2$ 
where $a_{1,j},~a_{2,j},~a_{3,j}\in [n]$ and $b_j\in\{0,1\}$,
find an assignment $\x \in {\{0,1\}}^n$ 
that maximizes the number of satisfied equations.
\\
{\bf Prior QAOA work:} Considered in Reference~\cite{Farhi2014b}.\\
{\bf Approximability:} %source: wikipedia
No efficient $1+\epsilon$ classical approximation algorithm unless $P=NP$ \cite{Hastad2001}. 
For the case where each variable appears in at most $D$ equations, an efficient classical algorithm satisfies a $1/2 +  \Omega ( D^{-1/2} )$ fraction of the equations  \cite{barak2015beating}. 
Shown in Reference \cite{Farhi2014b} that QAOA with $p=1$ achieves the same result up to a factor logarithmic in $D$ for this case.\\ 
{\bf Objective:} Maximize the number of satisfied equations.\\
% {\bf Configuration space/Domain/Encoding:} $\{0,1\}^n$.  \\
{\bf Phase Separator}: 
$\phaseHam=\sum_{j=1}^m H_j$
where
$
H_j={(-1)}^{b_j} Z_{a_{1,j}} Z_{a_{2,j}} Z_{a_{3,j}}\;.
$
% where constant
% $c_\phi=\pm 1$ depends on the specific equation $\phi$.
\\
{\bf Resource count for phase separator:} 
Can be implemented with $m$ many 3-qubit gates. 
%with depth at most $\max\{D_j\}+1$, where $D_j$ is the number of equations in which variable $x_j$ appears.
%\egr{I don't believe the depth is correct?  We have 3-qubit gates here. Could someone check?} 

%\input{sections/appendix/compendium/max-indep-set} %file contains multiple problems

\subsection{Controlled-Bit-Flip ($\Lambda_f (X)$) Mixers}
%\noindent 
All problems in this section are 
graph or subset problems with
QAOA mappings that use controlled-bit-flip
mixers on bit strings. 
(See Section~\ref{sec:MaxIndSet}.)
Note that MaxSetPacking (Section~\ref{sec:comp-MaxSetPack}) reduces to MaxIndependentSet (Section~\ref{sec:comp-MaxIndSet}) on the constraint graph;
MinVertexCover (Section~\ref{sec:comp-MinVertexCover}) is a special case of MinSetCover (Section~\ref{sec:comp-MinSetCover}), in which the universe is the set of edges, and each subset in the collection corresponds to the edges adjacent to a single vertex;
and 
MaxClique (Section~\ref{sec:comp-MaxClique}) is simply MaxIndependentSet on the complement graph.
\\
{\bf Configuration space:}
The configuration space of each problem is the subsets $V'$ of some set $V$ (resp., $S'\subset S$), represented by bitstrings $\x\in{\{0, 1\}}^{\numVars}$, with $x_v = 1$ indicating $v \in V'$. \\
{\bf Constraint graph:} An instance of each problem is either specified by a graph or has a natural corresponding constraint graph whose vertices correspond to the variables and with respect to which each variable $v$ %has a neighborhood 
is graph-adjacent to $\neighborFunc(v)$.\\
{\bf Domain:} %The domain is the 
%Subset of the configuration space with elements 
Elements of the configuration space that satisfy some CNF formula %Please define CNF.
(whose clauses correspond to the edges of the problem or the constraint graph, except for MinSetCover) that specifies a required property. \\
{\bf Objective:} Maximize or minimize the subset cardinality.\\
{\bf Mixing rule:} 
Swap an element $v$ in or out of $V'$ if some predicate $\controlFunc (\x_{\neighborFunc(v)})$ is satisfied by the partial state of its neighbors.
(The predicate $\controlFunc$ will depend on the problem.)\\
{\bf Partial mixing Hamiltonian:} 
$\Ham{\controlX, v} =  X_v H_{P_v}$,
which can be used to define both a simultaneous controlled bit-flip mixer
and a class of partitioned controlled bit-flip mixers.~(See Section~\ref{sec:MaxIndSet}.)

\subsubsection{MaxIndependentSet [Section~\ref{sec:MaxIndSet}]}\label{sec:comp-MaxIndSet}
\noindent {\bf Problem:} Given $G=(V,E)$, 
maximize the size of a subset $V' \subset V$ of mutually non-adjacent 
vertices. \\ 
{\bf Prior QAOA work:}~\cite{Farhi2014}. See Section~\ref{sec:MaxIndSet}
for a detailed discussion of the mapping and generalization. \\
{\bf Approximability:} Poly-APX-complete~\cite{bazgan2005completeness}, and has no constant factor approximation
unless $P = NP$.\@
On bounded degree graphs with maximum degree $D_G \geq 3$ can be approximated to $(D_G +2)/3$~\cite{bazgan2005completeness}, but remains APX-complete~\cite{papadimitriou1991}.  \\
%{\bf Configuration space:} 
%All subsets $V'$ of $V$, represented by elements $\x$ of ${\{0, 1\}}^{\numVars}$, with $x_v = 1$ indicating $v \in V'$.\\
{\bf Domain:} Independent sets, 
$\left\{ \x : \bigwedge_{\{u, v\} \in E} \bar{x}_u \lor \bar{x}_v\right\}$.\\
{\bf Objective:} max $|V'| = \sum_{v\in V} x_v$.\\
{\bf Mixing rule:} Swap $v$ in or out of $V'$ if none of its neighbors are in $V'$.\\
{\bf Partial mixing Hamiltonian:} 
$\Ham{\controlX, v} = 
X_v H_{\noNeighbors{v}}$.\\
{\bf Phase separator:} 
$\phaseUnitary(\gamma) = \exp(-i\gamma \sum_{u\in V} Z_{u} )$.\\
{\bf Initial state:} $\ket{\initial} = \ket{0}^{\otimes n}$, i.e., $V' = \emptyset$.\\
{\bf Resource count:}
\begin{itemize}[nolistsep,noitemsep]%[noitemsep,nolistsep]
\item {\bf Controlled-bit-flip mixers:}
$\numVars$ multiqubit-controlled-$X(\beta)$ gates,
each with at most $\degree_G$ controls (exactly $D_v$ controls for each vertex). 
Depth at most $\numVars$ but will be much less for sparsely connected graphs. 
\vskip 4pt
\item {\bf Phase separator:} $n$ single-qubit $Z$-rotations. Depth $1$.
\end{itemize}
{\bf Variants:} Extends easily to weighted-MaxIndependentSet with
objective function $\objFunc= \sum_{i=1}^n w_i x_i$. 

\subsubsection{MaxClique}\label{sec:comp-MaxClique}
\noindent {\bf Problem:} Given $G=(V,E)$, 
maximize the size of a clique in $G$ (a subset $V' \subset V$ that induces a subgraph in which all pairs
of vertices are adjacent). \\
{\bf Approximability:} 
A $O(\numVars / \log^2 \numVars)$-approximation is known \cite{boppana1992approximating}, but 
cannot be efficiently approximated classically better than $O(n^{1-\e})$ for any $\e>0$  
unless $P=NP$~\cite{zuckerman2006linear}. \\
%{\bf Configuration space:}
%All subsets $V'$ of $V$, represented by elements $\x$ of ${\{0, 1\}}^{\numVars}$, with $x_v = 1$ indicating $v \in V'$.\\
{\bf Domain:} Cliques, 
$\left\{ \x : \bigwedge_{\{u, v\} \in E(\graphComp{G})} \bar{x}_u \lor \bar{x}_v\right\}$.
\\
{\bf Objective:} max $|V'| = \sum_{v\in V} x_v$.\\
{\bf Reduction to MaxIndependentSet:} 
Every clique on $G=(V,E)$ gives on independent set on the complement graph
$\graphComp{G}=(V,E(\graphComp{G}))$ where $E(\graphComp{G}) = \binom{V}{2} \setminus E$. 
Therefore, a mapping for MaxClique is given by the mapping for MaxIndependentSet applied to the complement graph $\graphComp{G}$. \\ 
{\bf Resource count:}  The same resources as for MaxIndependentSet, 
except that the controlled bit-flip mixers are now multiqubit-controlled-X gates with at most $n - \degree_G - 1$ controls (exactly $n - \degree_v - 1$ controls for each vertex $v$).
%{\bf Variants:} Extends easily to weighted versions MaxIndependentSet, with
%objective functions $\objFunc= \sum_{i=1}^n w_i x_i$. 

\subsubsection{MinVertexCover}\label{sec:comp-MinVertexCover}
\noindent {\bf Problem:} Given $G=(V,E)$, 
minimize the size of a subset $V' \subset V$ %such that 
that covers $V$, i.e., 
for every $(u,v)\in E$, $u\in V'$ or $v\in V'$. \\
{\bf Prior QAOA work:} A quantum walk generalization of QAOA is proposed for this problem in Reference~\cite{marsh2019quantum}.  \\
{\bf Approximability:} APX-complete~\cite{papadimitriou1991}. Has a $(2-\Theta(1/\sqrt{\log n}))$-approximation~\cite{karakostas2009better} but cannot be approximated better than $1.3606$ unless $P=NP$~\cite{dinur2005hardness}. \\
%{\bf Configuration space:} 
%All subsets $V'$ of $V$, represented by elements $\x$ of ${\{0, 1\}}^{\numVars}$, with $x_v = 1$ indicating $v \in V'$.\\
{\bf Domain:} Vertex covers, 
$\left\{ \x : \bigwedge_{\{u, v\} \in E(\graphComp{G})} x_u \lor x_v\right\}$.\\
{\bf Objective:} min $|V'|$.\\
{\bf Initial state:} $\ket{\initial} = \ket{1}^{\otimes n}$, i.e., $V' = V$.\\
{\bf Mixing rule:} 
Swap $v$ in or out of $V'$ if all of the edges incident to $v$ are covered by $V' \cap \neighborFunc(v)$.\\
{\bf Phase Separator:} $\phaseUnitary(\gamma) = \exp(-i\gamma \sum_{u\in V} Z_{u} )$.\\
{\bf Reduction to MaxIndependentSet:}
A subset $V' \subset V$ is a vertex cover if and only if $V \backslash V'$ is
an independent set, so the problem of finding a minimum vertex cover is 
equivalent to that of finding a maximum independent set. 
While as approximation problems they are not equivalent~\cite{trevisan2004inapproximability}, 
we can use the same mapping as for MaxIndependentSet
with each $\bar{x}_v$ replaced by $x_v$. 
The resource counts are the same as for MaxIndependentSet.\\
{\bf Reduction to MinSat:}  
Maranthe et al.~\cite{marathe1996approximation} give an approximation-preserving reduction to Min-$D_G$-SAT %so the previous mapping for MinSat applies.\\
enabling us to use the QAOA constuction given above for MinSat. 
The resource counts are the same as for MinSat with $m$ variables and $n$ clauses.

\subsubsection{MaxSetPacking}\label{sec:comp-MaxSetPack}
\noindent {\bf Problem:} 
Given a universe $[n]$ and $m$ subsets 
$\mathcal{S} = {\left(S_j\right)}_{j=1}^m$, $S_j \subset [n]$,  
find the maximum cardinality subcollection $\mathcal{S'} \subset \mathcal{S}$ of pairwise disjoint subsets.\\ 
{\bf Approximability:} 
As difficult as MaxClique~\cite{Ausiello2012complexity}. 
Cannot be efficiently approximated within any constant factor unless $P=NP$~\cite{hazan2006complexity}. The best known algorithm gives an $O(\sqrt{n})$-approximation~\cite{halldorsson2000independent}. 
APX-complete with the restriction $|S_j|\leq k$ (Max-$k$-Set-Packing). \\
%{\bf Configuration space:} 
%All subsets $\mathcal{S}'$ of $\mathcal{S}$, represented by elements $\x$ of ${\{0, 1\}}^m$, with %$x_j = 1$ indicating $S_j \in \mathcal{S}'$.\\
{\bf Constraint graph:}
Vertices $V =[m]$ corresponding to elements of $\mathcal{S}$, edges $E$ corresponding to pairs of intersecting subsets.\\
{\bf Domain:} Subcollections containing mutually disjoint subsets, 
$\left\{ \x : \bigwedge_{\{i, j\} \in E} \bar{x}_i \lor \bar{x}_j\right\}$.\\
{\bf Objective:} max $|\mathcal{S}'| = \sum_{j=1}^m x_j$ \\
{\bf Initial state:} $\ket{\initial} = \ket{0}^{\otimes m}$, i.e., \ $\mathcal{S}' = \emptyset$.\\
{\bf Mixing rule:} 
Swap $S_j$ in or out of $\mathcal{S}'$ if $S_j$ is disjoint from the other subsets in $\mathcal{S}'$, i.e.,  $\x_{\neighborFunc(j)} = \mathbf{0}$.\\
{\bf Partial mixing Hamiltonian:} 
$\Ham{\controlX, j}
= 2^{-\degree_j} X_j \prod_{i \in \neighborFunc(j)} (\identity + Z_i)
= X_j \Ham{\noNeighbors{j}}$.\\ 
{\bf Phase separator:} 
$\phaseUnitary(\gamma) = \exp(-i\gamma \sum_{j=1}^m Z_{j} )$.\\
{\bf Resource count:}
\begin{itemize}[noitemsep,nolistsep]
\item {\bf Controlled-bit-flip mixers:} 
Each partial mixer $\Ham{\controlX, j}$ is implemented as a controlled-$R_X$ gate with $\degree_j$ control qubits.
Partial mixer depth at most $m$.
\vskip 4pt
\item {\bf Phase separator:} $m$ single-qubit $Z$-rotations. Depth $1$.
\vskip 4pt
\item {\bf Initial state:} Depth $0$.
\end{itemize}
{\bf Variants}: Also called MaxHypergraphMatching. 
Equivalent to MaxClique under a PTAS-reduction~\cite{Ausiello2012complexity}.

\subsubsection{MinSetCover}\label{sec:comp-MinSetCover}
\noindent {\bf Problem:} Given a universe $[n]$ and $m$ subsets 
$\mathcal{S} = {\left(S_j\right)}_{j=1}^n$, $S_j \subset [n]$,  
find the minimum cardinality subcollection $\mathcal{S}' \subset \mathcal{S}$ of the $S_j$ such that their union recovers $[n]$. \\ 
{\bf Approximability:} There exists a $1 + \ln n$ algorithm~\cite{johnson1973approximation}. 
Cannot be efficiently approximated to $(1-o(1)) \ln n$ unless $P=NP$~\cite{dinur2014analytical}. 
 APX-complete with the restriction $|S_j|\leq k$ (Max-$k$-Set-Cover)~\cite{Ausiello2012complexity}. \\
%{\bf Configuration space:} 
%All subsets $\mathcal{S}'$ of $\mathcal{S}$, represented by elements $\x$ of ${\{0, 1\}}^m$, with $x_j = 1$ indicating $S_j \in \mathcal{S}'$.\\
{\bf Constraint graph:}
Vertices $[m]$ corresponding to subcollections in $\mathcal{S}$; two vertices $\{i, j\}$ are adjacent if and only if their corresponding sets intersect, $S_i \cap S_j \neq \emptyset$.\\
{\bf Domain:} Set covers, $\left\{ \x : \bigcup_{j \in [m] : x_j = 1} S_j = [n]\right\}$.\\
{\bf Objective:} min $|\mathcal{S}'| = \sum_{j=1}^m x_j$.\\
{\bf Initial state:} $\ket{\initial} = \ket{1}^{\otimes m}$, i.e.,  $\mathcal{S}' = \mathcal{S}$.\\
{\bf Mixing rule:} 
Swap set $S_j$ in or out of $\mathcal{S}'$ if $\mathcal{S'} \setminus S_j$ covers $[n]$.\\
{\bf Partial mixing Hamiltonian:} 
\begin{equation}
\Ham{\controlX, j} = 
\left(
\prod_{i=1}^n 
\left(
I - \prod_{\ell \in [m] : \ell \neq j, i \in S_{\ell}} \frac{I+Z_l}{2}
\right)
\right)
X_j 
=
X_j
H_{\chi_{\mathrm{MSC, j}}},
\end{equation}
where
$\chi_{\mathrm{MSC, j}} (\x_{\neighborFunc(j)})= 
\bigwedge_{i \in S_j} 
\bigvee_{\ell \in \neighborFunc(j): i \in S_{\ell}}
x_{\ell}
$.%\\
\vskip 0.3pc
\noindent {\bf Phase separator:} 
$\phaseUnitary(\gamma) = \exp(-i\gamma \sum_{j=1}^m Z_{j} )$.\\
{\bf Resource count:}
\begin{itemize}[noitemsep,nolistsep]
\item {\bf Controlled-bit-flip mixers:} 
For each $\Ham{\controlX, j}$, use $|S_j|$ ancilla qubits.
Use each ancilla qubit $i$ to compute $\bigvee_{\ell \in \neighborFunc(j): i \in S_{\ell}} x_{\ell}$ using a controlled NOT gate with 
$|\{\ell \in \neighborFunc(j): i \in S_{\ell}\}| \leq |\neighborFunc(j)| = \degree_j$ control qubits.
Then implement $\Ham{\controlX, j}$  using a controlled X gate on qubit $j$ with the $|S_j|$ ancilla qubits as the control.
Finally, uncompute the ancilla qubits using the same $|S_j|$ controlled NOT gates as in the first step.
Depth at most $2 \degree_j + 1$ per partial mixer.
\vskip 4pt
\item {\bf Phase separator:} $m$ single-qubit $Z$-rotations. Depth $1$.
\vskip 4pt
\item {\bf Initial state:} Depth $0$.
\end{itemize}
{\bf Variants}: Equivalent to minimum hitting set~\cite{Ausiello2012complexity} and under $L$-reductions equivalent to minimum dominating set (which is a special case of minimum set cover).

\subsection{$XY$ Mixers} 

\noindent In this section, we consider problems where the domain is strings of $d$-dits, $d\geq 3$, with QAOA mappings that use XY mixers.   
%letters taken from some alphabet

\subsubsection{Max-$\numColors$-ColorableSubgraph  [Section~\ref{sec:maxColSubGraph}]}
\noindent {\bf Problem:} Given a graph $G$ 
and $\numColors$ colors, maximize the size (number of edges) of a properly
colored subgraph. \\
{\bf Approximability:}  A random coloring properly colors a fraction $1-1/\numColors$ of edges in expectation. Equivalent to MaxCut for $\numColors=2$. For $\numColors>2$,  semidefinite programming gives a 
$(1 - 1/\numColors + \left(2+o(\numColors)\right) \frac{\ln \numColors}{\numColors^2})-$approximation~\cite{frieze1997improved}, which is optimal up to the $o(\numColors)$ factor under the unique games conjecture~\cite{khot2007optimal}. APX-complete for $\numColors\geq 2$~\cite{papadimitriou1991} and no PTAS exists unless $P=NP$~\cite{petrank1994hardness}. \\
{\bf Configuration space and domain:} 
The set of all colorings with at most $\numColors$ colors, 
$\x \in {[\numColors]}^{\numVars}$.\\
{\bf Objective:} max $\sum_{\{u,v\} \in E} \NEQ(x_u, x_v)$.\\
{\bf Initial state:} All vertices colored with color $1$, $\ket{1}^\numVars$.\\
{\bf Partial Mixing Hamiltonian:} 
$\Ham{\rnv} = \sum_{a=1}^r \left( \quditX^a + {(\quditX^{\dagger})}^a\right)$.\\
{\bf Encoding:} One-hot.
\begin{itemize}[noitemsep,nosep]
\item {\bf Phase Separator:}  $\phaseUnitary (\gamma) = \prod_{\{u, v\} \in E} \prod_{j = 1}^{\numColors} 
\exp\left(-i \gamma Z_{v, j} Z_{v, j} \right)$ 
\vskip 4pt
\item {\bf Resource count:} 
\begin{itemize}[noitemsep,nolistsep]
\item {\bf Number of qubits:} $\numVars \numColors$.
\vskip 4pt
\item {\bf Parity ring mixer:} $\numVars\numColors$ two-qubit ($XY$) gates, 
with depth at most $2$ ($\numColors$ even) or $3$ ($\numColors$ odd).
\vskip 4pt
\item {\bf Phase separator:} $\numVars\numColors$ two-qubit ($ZZ$) gates. 
Depth at most $\degree_G +1$.
\vskip 4pt
\item {\bf Initial state:} $n$ single-qubit $X$ gates. Depth $1$.
\end{itemize}
\end{itemize}
{\bf Alternative encoding (for $\numColors=2^l$):} Binary 
\begin{itemize}[noitemsep,nosep]
\item {\bf Phase separator:} 
$\phaseUnitary(\gamma) = 
\prod_{\{u, v \} \in E}
\Lambda_{\EQ{\mathbf{x}_u, \mathbf{v}}}
\left(\timeEv{\gamma} \right)$
\vskip 4pt
\item {\bf Mixer:} 
$\mixUnitary(\beta)
= \prod_{i=1}^{\numVars}
\timeEv{\beta X_{i, 0}}
{\add_i(1)}^{\dagger} \timeEv{\beta X_{i, 0}} \add_i (1) .$
where $\add_i(z)$ adds $z$ to the register $i$ encoding an integer in binary.
\vskip 4pt
\item {\bf Resource count:}
\begin{itemize}[noitemsep,nosep]
\item {\bf Parity ring mixer:} $2n$ $\add$s, $2n$ single-qubit rotations; completely parallelizable.
\vskip 4pt
\item {\bf Phase separator:}
$\numConstraints$ controlled-phase gates with $2l$ controls; depth $O(\degree_G)$.
\vskip 4pt
\item {\bf Initial state:} $n$ single-qubit $X$ gates in depth $1$.
\end{itemize}
\end{itemize}
{\bf Variants:} Also known as Max-$\numColors$-Cut~\cite{frieze1997improved, khot2007optimal}, sometimes with weighted edges. For the weighted case, the same approximability results apply.

\subsubsection{Graph Partitioning (Minimum Bisection)}\label{sec:comendium:graph_partition}
\noindent {\bf Problem}%
{\bf:} 
Given a graph $G$ such that $\numVars$ is even, 
find a subset $V_0\subset V$ satisfying $|V_0|=n/2$ 
such that the number of edges between $V_0$ and $V\setminus V_0$ is minimized.\\
{\bf Prior AQO work:} Studied in the context of AQO for constrained optimization~\cite{Hen2016quantum}.\\
{\bf Approximability:}  
An efficient $O( \log^{1.5} n)$-approximate algorithm is known~\cite{krauthgamer2006polylogarithmic}. \\
{\bf Configuration space, domain, and encoding:} Bit strings $\bf x$ of Hamming weight $n/2$. \\
{\bf Objective}: $\min\sum_{\{uv\}\in E} \NEQ(x_u, x_v)$.\\
{\bf Initial state:} Any $\bf x$ of Hamming weight $n/2$.\\
{\bf Phase separator}: 
$\phaseUnitary (\gamma) = \exp(-i\gamma \sum_{\{uv\}\in E}  Z_{u}Z_{v}) $.\\
{\bf Partial mixing Hamiltonian:} Qubit ring XY mixer 
$\Ham{\ring}^{\enc} = \sum_{u=1}^n  \left( X_{u}X_{u+1}+Y_{u}Y_{u+1} \right)$.\\
{\bf Resource count:}
\begin{itemize}[noitemsep,nolistsep]
\item {\bf Number of qubits:} $n$.
\vskip 4pt
\item {\bf Parity ring mixer:} $n$ two-qubit (XX+YY) gates, with depth at most 2 ($n$ even) or 3 ($n$ odd).
\vskip 4pt
\item {\bf Phase separator:} $m$ two-qubit (ZZ) gates. Depth at most $\degree_G+1$.
\vskip 4pt
\item {\bf Initial state:} $n/2$ single-qubit $X$ gates. Depth $1$.
\end{itemize}

\subsubsection{Maximum Bisection}
\noindent {\bf Problem:} 
Given a graph $G$ such that $\numVars$ is even, and edge weights $w_j$, 
find a subset $V_0\subset V$ satisfying $|V_0|=n/2$ 
such that the total weight of 
edges crossing between $V_0$ and $V\setminus V_0$ is maximized.\\
{\bf Approximability:} 
A random bisection gives an $0.5$-approximation in expectation, improved to~$0.65$~\cite{frieze1997improved}. \\
{\bf Mapping:} Same as graph partitioning (Section~\ref{sec:comendium:graph_partition}) with weights included in the phase separator.

\subsubsection{Maximum Vertex $\numColors$-Cover}
\noindent {\bf Problem}: Variant of vertex cover optimization problem. 
Given a graph $G$ and an integer $\numColors \leq n$, 
find a subset $V_0 \subset V$ of size $|V_0| = \numColors$ 
such that the number of edges covered by $V_0$ is maximized. \\
{\bf Approximability:} 
It is NP-hard to decide whether or not a fraction $(1-\e)$ of the edges can be $\numColors$-covered~\cite{petrank1994hardness}. \\
{\bf Mapping:} Same as graph partitioning (Section~\ref{sec:comendium:graph_partition}) with the Hamming weight $n/2$ replaced by~$\numColors$.

\subsection{Controlled-$XY$ Mixers}
%\input{sections/appendix/compendium/max-prop-colored-induced-subgraph}

%\noindent
 In this section, we consider problems with QAOA mappings that use controlled-XY mixers. 

%\subsection{Maximum properly colored induced subgraph 
\subsubsection{Max-$\numColors$-ColorableInducedSubgraph [Section~\ref{sec:induced}]}%Is this necessary? if not, can it be delete?the same below.
% \subsubsection{MaxColorableInducedSubgraph}
\noindent {\bf Problem:} Given a graph $G$ and $\numColors$ colors, 
% with $\numVars$ vertices and a number of colors $\numColors$, 
maximize the size of a subset of vertices $V'\subset V$ 
whose induced subgraph is $\numColors$-colorable. \\
{\bf Approximability:} Equivalent to MaxIndependentSet for $k=1$. Both as easy and as hard to approximate as MaxIndependentSet for $k\geq 1$~\cite{panconesi1990quantifiers, Ausiello2012complexity}.
On bounded degree graphs, can be approximated to $(D_G/k +1)/2$, but remains APX-complete~\cite{halldorsson1995approximating}.
\\
% {\bf Configuration space:} ${[\numColors+1]}^{\numVars}$.  Strings of length
% $\numVars$ over an alphabet of size $(\numColors+1)$, the $\numColors$ plus a
% character for uncolored.
%
{\bf Configuration space:} ${[\numColors + 1]}^{\numVars}$. ($0$-th color is
``uncolored'' and represents $v \notin V'$.) \\
{\bf Domain:} $\numColors$-colorable induced subgraphs, 
$\left\{
\x : 
\bigwedge_{\{i, j\} \in E} \left(\EQ(x_i, 0) \lor\EQ(x_j, 0) \lor \NEQ(x_i, x_j) \right)
\right\}$. \\
{\bf Objective:} min $\sum_{i=1}^{\numVars} \NEQ(x_i, 0)$.\\
{\bf Initial state:} All vertices uncolored, $\ket{10\dots0}^{\otimes n}$.\\
{\bf Mixing rule:} 
A vertex can switch between being uncolored and colored $j$ only if none of its neighbours are colored $j$.\\
%A vertex can always be uncolored, but can be colored $j$ (from the uncolored state) only if none of its neighbours are colored $j$.\\
%
{\bf Partial mixing Hamiltonian:} Controlled null-swap mixer at a vertex.
% Simultaneous Mixer: $\mixUnitary^H(\beta) = e^{-i\beta B}, \;\;$ $B=\sum_{u\in V} \sum_{j=1}^k B_{u,j}, \;\;$ $B_{u,j} =   (\prod_{v \in \neighborFunc(u)} \bar{x}_{v,j}) \cdot (X_{u,0}X_{u,j} + Y_{u,0} Y_{u,j}) 
 % =   (X_{u,0}X_{u,j} + Y_{u,0} Y_{u,j}) \prod_{v \in \neighborFunc(u)} \frac{I+Z_{v,j}}{2}$\\
%
%Sequential Mixer: $\mixUnitary(\beta) =  \prod_{u\in V} \prod_{j=1}^{\numColors} \exp(-i \beta (X_{u,0}X_{u,j} + Y_{u,0}Y_{u,j})\prod_{v \in \neighborFunc(u)} \left(I + Z_{v,j}\right)/2)$
% Sequential Mixer: $\mixUnitary(\beta) =  \prod_{u\in V} \prod_{j}^{\numColors} \Lambda_{f_{uj} } ( \exp(-i \beta (X_{u,0}X_{u,j} + Y_{u,0}Y_{u,j}))),\;\;$ $f_{uj} =  \prod_{v \in \neighborFunc(u)} \frac{I + Z_{v,i}}{2}$\\
\\
{\bf Encoding:} One-hot. \\
{\bf Phase separator:}  $\phaseUnitary (\gamma) = \exp(\frac{i}{2}\gamma \sum_{u\in V} \sum_{j=1}^{\numColors} Z_{u,j} )$.\\
{\bf Resource count:}
\begin{itemize}[noitemsep,nolistsep]
\item {\bf Partitioned controlled null-swap mixers:}
$\numVars\numColors$ partial mixers, 
each acting on at most $\degree_G + 1$ qubits.
Depth at most $\numVars\numColors$ but will be much less for sparsely connected graphs.
\vskip 4pt
\item {\bf Phase separator:} $n$ single-qubit $Z$-rotations. Depth $1$.
\vskip 4pt
\item {\bf Initial state:} $\ket{\initial} = {\left( \ket{1} \otimes {\ket{0}}^{\otimes \numColors}\right)}^{\otimes \numVars}$, implemented in depth $1$ with $n$ $X$ gates.
\end{itemize}

\subsubsection{MinGraphColoring [Section~\ref{sec:chromatic-number}]}
\noindent {\bf Problem:} Given a graph $G$, minimize the number
of colors required to properly color it.\\
{\bf Approximability:} The best classical algorithm~\cite{halldorsson1993still}
achieves approximation ratio $O(n \frac{{(\log \log n)}^2}{\log^3 n} )$, and we
cannot do better than %$O(n^{1-\e})$ 
$n^{1-\e}$ for any $\e>0$ unless $P=NP$~\cite{zuckerman2006linear}.
For edge-colorings of multigraphs, there is a ($1.1 + 0.8/\numColors^*$)-approximate algorithm~\cite{nishizeki19901}. 
\\ 
{\bf Configuration space:} ${[\numColors]}^{\numVars}$, $\;\numColors=\degree_G + 2$. \\
{\bf Domain:} Proper $\numColors$-colorings of $G$ (many of which use fewer
than $\numColors$ colors), 
$\left\{
\x : \bigwedge_{\{i, j\} \in E} \NEQ(x_i, x_j)
\right\}$.\\
{\bf Objective:} Minimize number of used colors:
$\sum_{a=1}^{\numColors} \OR(\EQ(x_1, a), \ldots, \EQ(x_{\numVars}, a))$. \\
{\bf Mixing rule:} The color of vertex $u$ may be swapped between colors
$c$ and $c'$ if none of its neighbours are already colored $c$ or $c'$.\\ 
{\bf Partial mixing Hamiltonian:}  Controlled-swap partial mixing Hamiltonian. \\
{\bf Encoding:} One-hot.\\
{\bf Phase separator:} 
$
\prod_{a=1}^{\numColors}
\control_{\OR(\x_{[n],a})}
\left(\timeEv{\gamma}\right)$.\\
{\bf Resource count:} 
\begin{itemize}[noitemsep,nolistsep]
\item {\bf Partitioned controlled-swap mixers:} $\numColors(\numColors - 1)\numVars/2$ 
controlled gates on no more than $\degree_G + 2$ qubits.
\vskip 4pt
\item {\bf Phase separator:}  $\numColors$ partial phase separators acting
on $\numVars + 1$ qubits, one target qubit and $\numVars$ control qubits.
Depth $2$ in partial phase separators, or depth $1$ with the addition
of $\numColors$ ancilla qubits.
\vskip 4pt
\item {\bf Initial state:} Any valid $\numColors$ coloring (can be efficiently 
computed classically).  Can be implemented in depth $1$ using $\numVars$
single-qubit $X$ gates.
\end{itemize}
{\bf Reduction from MinEdgeColoring:} 
In MinEdgeColoring, the objective is to minimize the number of colors needed to color the edges so that no two adjacent edges have the same color.
This is equivalent to MinGraphColoring on the line graph.

\subsubsection{MinCliqueCover}
\noindent {\bf Problem:} Given a graph $G$, we seek the smallest collection of cliques $S_1,\ldots,S_k \subset V$, such that every vertex belongs to at least one clique. \\
{\bf Approximability:} If MaxClique is approximable within $f(n)$ for a given instance, then MinCliqueCover approximable to $O(f(n))$~\cite{Ausiello2012complexity}. 
Not approximable within $n^\e$ for any $\e>0$~\cite{lund1994hardness}. \\
{\bf Reduction to MinGraphColoring:} A partition of the vertices of $G$ is a $k$-clique cover if and only if it is a proper $k$-coloring of the complement graph $G'=(V,E^c)$, and moreover, the smallest clique cover corresponds to the chromatic number of the complement graph. Thus the previous construction~suffices. 

\subsection{Permutation Mixers}

 In this section, we consider problems with QAOA mappings that use the mixers of Section~ \ref{sec:permutations} for domains consisting of ordering, permutations, or schedules. In the SMS problems below, for simplicity, we take all time parameters to be nonnegative integers.

\subsubsection{TravelingSalespersonProblem (TSP) [Section~\ref{sec:TSP}]}  \label{sec:tspEntry} 
\noindent {\bf Problem:} Given a set of $\numVars$ cities and distances $\distance:
{[\numVars]}^2 \rightarrow \reals_+$, % (with $d(i, i) = 0$), 
find an ordering of the cities that minimizes the total distance traveled on the corresponding tour. \\
{\bf Approximability:} \textls[-20]{NPO-complete~\cite{orponen1990approximation}.  
MetricTSP is APX-complete~\cite{papadimitriou1993traveling} and has a $3/2$-approximation~\cite{christofides1976worst}.
The corresponding MaxTSP problem is approximable within $7/5$ for symmetric distance, and $63/38$ if~asymmetric. }
\\
{\bf Configuration space and domain}: Orderings of the cities $\{\order\}$.\\
{\bf Objective:} min $f(\order) = \sum_{j = 1}^{\numVars}\distance_{\orderElmt_j, \orderElmt_{j+1}}$. \\
{\bf Partial mixer:} Partial permutation swap mixer. \\
{\bf Encoding:} Direct one-hot.\\
{\bf Compilation:} 
\begin{itemize}[noitemsep,nolistsep]
\item {\bf Phase separator:}
$\phaseHam^{\enc} = \sum_{i=1}^{\numVars} \sum_{u=1}^{\numVars}
\sum_{v=1}^{\numVars} \distance(u, v) Z_{u, i} Z_{v, i+1}$. 
\vskip 4pt
\item {\bf Partial mixer:} 
$\unitary{\permSwap, {\{i, j\}, \{u, v\}}}^{\enc} (\beta) =
\timeEv{\beta \Ham{\permSwap, {\{i, j\}, \{u, v\}}}}$, where
$\Ham{\permSwap, {\{i, j\}, \{u, v\}}}^{\enc}$
is given in Equation~\eqref{eq:H-PS-ij-uv-enc}.
\end{itemize}
{\bf Initial state:} Arbitrary ordering.\\
{\bf Resource count:} 
\begin{itemize}[noitemsep,nolistsep]
\item {\bf Color-parity permutation swap mixer (Section~\ref{sec:TSP}):} 
At most $(\numVars-1) \binom{\numVars}{2}$ $4$-qubit partial
mixers, in depth at most $2n$.
\vskip 4pt
\item {\bf Phase separator:} $\numVars^2(\numVars-1)$ mutually commuting 
two-qubit gates. 
Depth no more than $\degree_G +1$.
\vskip 4pt
\item {\bf Initial state:} $n$ single-qubit $X$ gates. Depth $1$.
\end{itemize}

\subsubsection{SMS, Minimizing Total Weighted Squared Tardiness [Section~\ref{sec:smst2}]}\label{sec:smstEntry}
\noindent {\bf Problem:}
$(1|d_j|\sum w_j T_j^2)$. 
Given a set of jobs with processing times $\procTimes$, deadlines $\deadlines\in \integers_+$, and weights $\weights$,
find a schedule that minimizes the total weighted squared tardiness $\sum_{j=1}^{\numVars} w_j T_j^2$. \\
{\bf Approximability:} Considered in Reference~\cite{schaller2012minimizing}.\\
{\bf Configuration space and domain:} Orderings of the jobs $\{\mathbf{i}\}$, 
and an integer slack variable $y_j\in[0,d_j-p_j]$ for each job.\\
{\bf Objective:} min 
$
\objFunc(\order, \y)
=
\sum_{j=1}^{\numVars}
\weight_j
{\left(
\strtTime_j(\order) + \procTime_j - \deadline_j + y_j
\right)}^2.
$\\
{\bf Partial mixer:} Partial permutation swap mixer for computational qubits; binary mixer mixer for the slack qubits. \\
{\bf Encoding:} Direct one-hot for the ordering variables; binary for the slack variables.\\
{\bf Compilation:} 
\begin{itemize}[noitemsep,nolistsep]
\item {\bf Phase separator:}
The encoded phase separator is a 3-local Hamiltonian containing
\begin{itemize}[noitemsep,nolistsep]
\item all 1-local terms;
\vskip 4pt
\item all 2-local terms of two computational qubits corresponding to different jobs at different places in the ordering, all 2-local terms of two ancilla qubits corresponding to the same job, and all 2-local terms of one computational qubit and one ancilla qubit except when they correspond to different jobs and the computational qubit corresponds to that job being last in the ordering;
\vskip 4pt
\item all 3-local terms of three computational qubits corresponding to different jobs at different places in the ordering and all 3-local terms containing two computational qubits corresponding to different jobs at different places in the ordering and one ancilla qubit corresponding to the later job.
\end{itemize}
\vskip 5pt
\item{\bf Partial mixer: }
$
\Ham{\permSwap, {\{i, j\}, \{u, v\}}}^{\enc} =
\create_{u, i} \create_{v, j} \annihilate_{u, j} \annihilate_{v, i} +
\annihilate_{u, i} \annihilate_{v, j} 
\create_{u, j} \create_{v, i}
$.
\vskip 5pt
\item{\bf Initial state} Arbitrary ordering. 
\end{itemize}
{\bf Resource count:} 
\begin{itemize}[noitemsep,nolistsep]
\item {\bf Color-parity permutation swap mixer (Section~\ref{sec:TSP}):} 
At most $(\numVars-1) \binom{\numVars}{2}$ $4$-qubit partial
mixers, in depth at most $2n$.
Single-qubit $X$ mixer for slack binary variables can be done in parallel with the permutation swap mixer.
\vskip 4pt
\item {\bf Phase separator:} 
Let $\mu_i = \left\lceil \log_2 (\deadline_i - \procTime_i + 1)\right\rceil$ be the number of bits needed for the slack variable $y_i$, and $\mu = \sum_{i=1}^{\numVars} \mu_i$.
\begin{itemize}[noitemsep,nolistsep]
\item Number of 1-local gates: $\numVars^2 + \mu$.
\vskip 4pt
\item Number of 2-local gates: $2{\binom{\numVars}{2}}^2 + \sum_{i=1}^{\numVars} \binom{\mu_i}{2} + \mu (n^2 - n + 1)$.
\vskip 4pt
\item Number of 3-local gates: $6 {\binom{\numVars}{3}}^2 + 2 \mu {\binom{n}{2}}^2$.
\end{itemize}
\vskip 5pt
\item {\bf Initial state:} $n$ single-qubit $X$ gates. Depth $1$.
\end{itemize}

\subsubsection{SMS, Minimizing Total Weighted Tardiness [Section~\ref{sec:smst}]}\label{sec:smstEntry2}
\noindent {\bf Problem:}
$(1|d_j|\sum w_j T_j)$. 
Given a set of $n$ jobs with processing times $\procTimes$, deadlines $\deadlines \in \integers_+$, and weights $\weights$,
find a schedule that minimizes the total weighted tardiness $\sum_{j=1}^{\numVars} w_j T_j$. \\
{\bf Approximability:}  There exists an $(n-1)$-approximation~\cite{cheng2005single}.
The decision version is strongly NP-hard~\cite{lenstra1977complexity}. \\
{\bf Configuration space and domain:} Orderings of the jobs $\{\mathbf{i}\}$.\\
{\bf Objective:} 
$
f(\order)
=
\sum_{i=1}^{\numVars}
\weight_i \max\{0, \deadline_i - \strtTime_i(\order) - \procTime_i\}.
$\\
{\bf Encoding:} Absolute one-hot.\\
{\bf Mixing rule:} Swap two jobs only if they are scheduled in consecutive order.\\
{\bf Partial mixer:}
Partial time-swap mixer (specific to the absolute encoding).\\
{\bf Initial state:} Arbitrary ordering. \\
{\bf Compilation:} 
\begin{itemize}[noitemsep,nolistsep]
\item {\bf Phase separator:}
$
\phaseHam^{\enc} =
\sum_{i=1}^{\numVars}
\weight_i
\sum_{t=\deadline_i - \procTime_i + 1}^{\lastStrtTime_i}
(t + \procTime_i - \deadline_i) Z_{i, t}.
$
\vskip 4pt
\item{\bf Partial mixer: }
${\Ham{\timeSwap, {t, \{i, j\}}}^{\enc}
=
\create_{i, t + \procTime_j}
\create_{j, t}
\annihilate_{i, t}
\annihilate_{j, t + \procTime_i}
+
\annihilate_{i, t + \procTime_j}
\annihilate_{j, t}.
\create_{i, t}
\create_{j, t+\procTime_i}
}$.
\end{itemize}
{\bf Resource count:} 
\begin{itemize}[noitemsep,nolistsep]
\item {\bf Color--time partitioned time-swap mixer:} 
$\lastStrtTime \binom{\numVars}{2}$ 4-qubit gates in depth $\lastStrtTime \numColors$.
\vskip 4pt
\item {\bf Phase separator:} 
At most $\numVars \lastStrtTime$ single-qubit gates, depth 1.
\vskip 4pt
\item {\bf Initial state:} $n$ single-qubit $X$ gates, depth $1$.
\end{itemize}

\subsubsection{SMS, with Release Dates [Section~\ref{sec:smswr}]}\label{sec:smswrEntry}
\noindent {\bf Problem:}
$(1|d_j,r_j|f)$. 
Given a set of jobs with processing times $\procTimes$, deadlines $\deadlines$, 
release times $\releaseTimes  \in \integers_+ $ and weights $\weights$,
find a schedule that minimizes some (given) function $f$ of the schedule,
e.g.,\ weighted total tardiness such that each job starts no earlier than its release time.\\
{\bf Approximability:} 
For all deadlines being zero, the minimal weighted total tardiness 
(completion times in this special case) is $1.685$-approximable~\cite{goemans2002single}.\\
{\bf Configuration space:} 
$\{\mathbf{s}\} = \times_{j=1}^{\numVars} \left[\window_j \cup \{b_j\}\right]$, 
where $\window_j = [\releaseTime_j, \horizon - \procTime_j]$ is the window of times in which job $j$ can start and $b_j$ is the job's ``buffer'' slot.\\
{\bf Domain:}
Schedules in the configuration space such that no two jobs overlap.\\
{\bf Objective:} 
Maximize (or minimize) a given cost function $f(\mathbf{s})$, e.g., weighted total tardiness 
$f(\mathbf{s}) = \sum_{j=1}^{\numVars} \weight_j T_j$, 
where $T_j= \max\{0, \deadline_i - \strtTime_i - \procTime_i\}$ is the tardiness of job $j$.
\\
{\bf Mixing rule:} 
Swap a job between time $t \in \horizon$ and its buffer slot if no other job is running at time $t$.\\
{\bf Partial mixer:}
Controlled null-swap mixer; see Section~\ref{sec:smswr}.\\
{\bf Initial state:} Greedy earliest-release-date schedule. \\
{\bf Encoding:} One-hot encoding.\\
{\bf Compilation:}
\begin{itemize}[noitemsep,nolistsep]
\item{\bf Partial mixer:}
$
\Ham{\nullSwap, {i, t}}^{\enc}=
\left(\prod_{j\ne i}\prod_{t'\in \neighborFunc_{i,t}(j)}\frac12 (I+Z_{j,t'})\right) 
\left(X_{i,t}X_{i,b_i}+Y_{i,t}Y_{i,b_i}\right).
$
\vskip 4pt
\item {\bf Phase separator (for min weighted total tardiness):}
{See Equation~\eqref{eq:smswrHp}.}
\end{itemize}
{\bf Resource count:} 
\begin{itemize}[noitemsep,nolistsep]
\item {\bf Controlled null-swap mixer:} SMS-instance dependent, see discussion in Section~\ref{sec:smswr}.
\vskip 4pt
\item {\bf Phase separator (for min weighted total tardiness):} 
At most $\sum_j (h-d_j+1)$ single-qubit $Z$ gates. Depth $1$.
\vskip 4pt
\item {\bf Initial state:} $\numVars$ single-qubit $X$ gates. Depth $1$.
\end{itemize}

\section{Glossary of Mapping Terms}  \label{sec:glossary}

Here, we provide a summary the mapping terminology that appears throughout the paper. 

\subsection{Mixers}
We list partial mixers %based on their 
by mixing mechanism in
Appendix~\ref{sec:gl:partialMixer}, and by different ways to partition, used for
partitioned mixers, in Appendix~\ref{sec:gl:partition}

%\subsubsection{\bf Partial Mixing Hamiltonians:}\label{sec:gl:partialMixer}
\subsubsection{Partial Mixing Hamiltonians}\label{sec:gl:partialMixer}

For strings and subsets thereof, we focus on mixing operators that are composed of 
%independent mixing 
operators acting independently on each of the variables.
Specifically, we consider the following single-qudit mixing operators in the absence of constraints:
\begin{itemize}
\item {$r$-nearby-values mixer:} 
$\Ham{\rnv} = \sum_{a=1}^r \left( \quditX^a + {(\quditX^{\dagger})}^a\right)$.
The special cases of $r=1$ and $r=d-1$ are called the ``ring mixer'' and ``fully-connected mixer'', respectively.
\item simple binary mixer:
When $\quditDim = 2^l$ is a power of two: 
$
\Ham{\binary}^{\enc}
=
\sum_{i=1}^l X_i.
$
\item null-swap mixer: For cases when one of the $d$ values corresponds to a ``null'' value (e.g.,\ black or uncolored in graph coloring),
$
H_{\nullSwap}= 
\sum_{a=1}^{d-1} \left(\ket{0}\bra{a} + \ket{a} \bra{0}\right)$.
\end{itemize}

For problems in the ordering/schedules family, we considered swapping-based mixers.
\begin{itemize}
\item Value-selective permutation swap mixer: 
Swaps the $i$th and $j$th elements in the ordering if those elements are $u$ and $v$, see Equation~\eqref{eq:mixer_perm_swap} 
in Section~\ref{sec:TSP}.
\item Value-independent permutation swap mixer: 
Swaps the $i$th and $j$th elements of the ordering regardless of which items those are, see
Equation~\eqref{eq:mixer_indep_perm_swap}
in Section~\ref{sec:TSP}.
\end{itemize}

When there are constraints, we considered modifications of the above mixers that are controlled on not violating the constraints. A few examples:
\begin{itemize}
\item Controlled null-swap mixer: 
\begin{itemize}[noitemsep,nolistsep]
\item Section~\ref{sec:maxColIndSubGraph} for MaxColorableInducedSubgraph, Equation~\eqref{eq:HnullSwap}.
\vskip 4pt
\item
Section~\ref{sec:smswr} for SMS with release dates, Equation~\eqref{eq:HnullSwapSMSwr}.
\end{itemize}

\item Controlled-SWAP mixer:
Section~\ref{sec:chromatic-number} for MinGraphColoring,
Equation~\eqref{eq:controlledSwap}.
\end{itemize}

%\subsubsection{\bf Partitions:}\label{sec:gl:partition}
\subsubsection{Partitions}\label{sec:gl:partition}
The two elementary partitions we use for generating the family of partitioned mixing unitaries~are:
\begin{itemize}
\item Parity-mixer: For Hamiltonian terms of type $\sum_u A_u B_{u+1}$, 
where $u\in[n]$ and $A_u$ and $B_{u+1}$ are operators acting on qubit $u$ and $u+1$, respectively. Partition the index set $\{u\}$ into even and odd subsets. See Section~\ref{sec:maxColSubGraph} for details.
\item Color-mixer: For index pairs $(u,v)\in {[n]\choose 2}$, 
let $\mathcal{P}_{\mathrm{col}} = (P_1, \ldots, P_{\numColors})$ be an ordered partition of the indices $\binom{[\numVars]}{2}$ into $\numColors$ parts such that each part contains only mutually disjoint pairs of indices from $[\numVars]$.
This is equivalent to considering a $\numColors$-edge-coloring of the complete graph $K_{\numVars}$, and assigning an ordering to the colors, so we call $\mathcal{P}_{\mathrm{col}}$ the ``color partition''.
For even $\numVars$, $\numColors = \numVars-1$ suffices, and for odd $\numVars$, $\numColors=\numVars$.
See Section~\ref{sec:TSP} for its use.
\end{itemize}

For mixing Hamiltonians of different coupling connectivity, these partitions
can be combined and tailored as desired.  For example, the color-parity
permutation swap mixer for TSP in Section~\ref{sec:TSP} and 
the time--color partition for SMS $(1|d_j|\sum w_j T_j)$ in Section~\ref{sec:smst}.

\subsection{Encodings}
We considered two types of encodings of strings into qubit space.
 One-hot encoding enables more concise circuits at the expense of requiring more qubits, 
whereas  binary encoding makes the opposite trade-off.
\begin{itemize}
\item One-hot encoding: 
%The state $\ket{a} \in {\mathbb C}^d$ is encoded as 
%${\ket{0}}^{\otimes a} \otimes \ket{1} \otimes \ket{1}^{\otimes {d-1-a}} \in {\mathbb C}^{2^d}$.
The qudit basis states $\ket{a}$, $a=0,\dots,d-1$, are %each 
encoded as the $d$-qubit states   
${\ket{0}}^{\otimes a} \otimes \ket{1} \otimes \ket{1}^{\otimes {d-1-a}}$. 
See Section~\ref{sec:mixOpsMaxColSubgraph}

%\item binary encoding: 
%The state $\ket{a}$ is encoded as $\ket{\mathbf b} \in {\mathbb C}^{2^{l^*}}$, where 
%%$l^* = \lceil \ln(d-1) \rceil$ and 
%$l^* = \lceil \log_2 d \rceil$ and 
%$\mathbf b \in {[2]}^{l^*}$ is the binary representation of $a$.
%See Section~\ref{sec:mixOpsMaxColSubgraph}.

\item Binary encoding: 
Each qudit basis state $\ket{a}$ is encoded as the $\ell$-qubit basis state $\ket{a_2}$, where $a_2$ denotes the binary representation of $a$, and $\ell=\lceil \log_2 d \rceil$. 
See Section~\ref{sec:mixOpsMaxColSubgraph}.

A generalization of the binary encoding is radix encoding, which represents $a$ in base-$r$, with positive  integer $r$. While the binary encoding %bears the advantage in implementability, 
is convenient for qubits, and is hence appealing in terms of implementability, 
it is plausible that for some problems, a more general radix encoding could be a natural choice.
%Radix encoding is a generalization of the binary encoding with the latter bears the advantage of easy implementability.
\end{itemize}

For problems in the ordering/scheduling family, the encoding is composed of two-steps: 
First encoding into strings, and then applying the above encodings into qubit space.
The following encodings are considered for step one:
\begin{itemize}
\item
Direct encoding: An ordering  
$\order = \left(\orderElmt_1, \ldots, \orderElmt_{\numVars}\right)$
is encoded directly as a string ${[n]}^{\numVars}$ of integers. 
It is demonstrated for TSP and SMS $(1|d_j|\sum_j w_j T^2_j)$ in Sections~\ref{sec:TSP} and~\ref{sec:smst2}, respectively.

\item
Absolute encoding: 
To encode the ordering $\order = (\orderElmt_1, \ldots, \orderElmt_{\numVars})$, we assigned each item $i$ a value $\strtTime_i \in [0, \horizon]$, where the ``horizon'' $\horizon$ is a parameter of the encodings, such that for all $i < j$, $\strtTime_{\orderElmt_i} < \strtTime_{\orderElmt_j}$. 
It is demonstrated for SMS $(1|d_j|\sum_j w_j T_j)$ in Section~\ref{sec:smst}.
\end{itemize}

The name of the overall encoding indicates both steps. 
For example: 
\begin{itemize}%[noitemsep,nolistsep]
\item Direct one-hot encoding, see Section~\ref{sec:TSP}.
\item Absolute one-hot encoding, see Section~\ref{sec:smst}.
\end{itemize}

\section{Elementary Operators}    \label{sec:elemOps}
Here, we elaborate on some of the basic quantum operators 
used in the text 
and their properties.
%that are used in the text.
Appendix~\ref{sec:SWAPvXY} explains 
the relationship between the $\SWAP$ gate and the XY-model Hamiltonian, %$XX + YY$,
which were each used as building blocks for encoded mixers in many mappings.
Appendix~\ref{sec:quditReview} contains a brief review of 
generalized Pauli operators for qudits of arbitrary dimension.

\subsection{SWAP and XY Opertors}\label{sec:SWAPvXY}
We examined the relationship between the quantum SWAP operator
$\SWAPij$, which swaps the state of qubit $i$ and $j$, 
and the XY operator, $\frac12(\XYij)$.
In many of our mapping constructions, these operators can be used
interchangeably; an exception is given in the last item below. 

Recall Equation~\eqref{eq:expSWAP} and the discussion of SWAP preceeding it. 
The XY operator acting on qubits $i,j$ 
can be similarly expressed as:
\begin{equation}
XY_{i,j} := \frac12(X_i X_j + Y_i Y_j)
=
\ketbra{1_i0_j}{0_i1_j}+\ketbra{0_i1_j}{1_i0_j}.
\end{equation}

We observed the following connections between the two operators: %Hamiltonians:
\begin{itemize} 
\item In the subspace spanned by $\{\ket{01},\ket{10}\}$, %$\XYij$ 
$XY_{i,j}$ and $\SWAPij$
behave identically.
\item In the subspace $\{\ket{11},\ket{00}\}$, %$\XYij$ 
$XY_{i,j}$ acts as null while $\SWAPij$ acts as an identity.
\item The operators $XY_{i,j}$  %$\XYij$ 
and $\SWAPij$ are both %unitary and 
Hermitian. $\SWAPij$ is unitary.
\item Applied to a multiqubit system, Hamiltonians of the form 
$H_\SWAP=\sum_{i,j}\SWAPij$
or  
$H_\text{XY} = \sum_{i,j} XY_{i,j} $, where either sum may be taken over arbitrary subsets of indices,  each preserve the Hamming weight of computational basis states; 
hence, so do the corresponding unitaries $\exp[-i\beta H_\text{XY}]$ and
$\exp[-i\beta H_\SWAP]$. Although the two operators do not behave identically on 
the full Hilbert space, they can both serve as mixers in situations in 
which Hamming weight is the relevant constraint.  
\item To enforce simultaneous swaps of multiple qubit pairs, in Hamiltonians such as 
$H=\prod_{\{i,j\}}\SWAPij$, each $\SWAPij$ cannot in general be directly replaced by $XY_{i,j}$ due to the second item above. See the TSP problem in Section~\ref{sec:TSP} as an example.
\end{itemize}

\subsection{Generalized Pauli Gates for Qudits}\label{sec:quditReview}
Consider $d$-dimensional qudits. Let $\omega = e^{2\pi i /d}$.
%We have the following operators 
Define the operators: 
\begin{align}
\quditZ &= \sum_{a=0}^{d-1} \omega_d^a \ket{a} \bra{a},
&
\quditX &= \sum_{a=0}^{d-1} \ket{a + 1} \bra{a}, 
&
\end{align}
where all arithmetic is modulo $d$.
As in the main text, we used operator notations ``$\quditZ$'' and ``$\quditX$''
for qudits of arbitrary dimension, 
and reserved ``$X$'' and ``$Z$'' specifically for the qubit case. 
For $d=2$, they~are the same.
Note that for $d>2$, the generalized Pauli operators, while unitary, are not Hermitian.
\mbox{In many cases}, we used the sum of a generalized Pauli operator and its conjugate to generalize its qubit analog, e.g., $\quditX + \quditX^{\dagger}$ to generalize $X$.

Because $\quditX^r=\sum_{a=0}^{d-1}\ketbra{a+r}{a}$,
it and its Hermitian conjugate  $\quditX^r+{(\quditX^\dag)}^r$ together generate
transitions between $\ket a$ and $\ket {a+r}$ for any $a$. 
Section~\ref{sec:mixOpsMaxColSubgraph} uses this operator in
the coloring of a single vertex.

\bibliographystyle{IEEEtran}
\bibliography{bibs/gen-qaoa}
\end{document}